\documentstyle{mn}

\newif\ifAMStwofonts
\ifoldfss
  \ifCUPmtlplainloaded \else
    \NewTextAlphabet{textbfit} {cmbxti10} {}
    \NewTextAlphabet{textbfss} {cmssbx10} {}
    \NewMathAlphabet{mathbfit} {cmbxti10} {} % for math mode
    \NewMathAlphabet{mathbfss} {cmssbx10} {} %  "   "    "
  \fi
  \ifAMStwofonts
    \ifCUPmtlplainloaded \else
      \NewSymbolFont{upmath} {eurm10}
      \NewSymbolFont{AMSa} {msam10}
      \NewMathSymbol{\upi}     {0}{upmath}{19}
      \NewMathSymbol{\umu}     {0}{upmath}{16}
      \NewMathSymbol{\upartial}{0}{upmath}{40}
      \NewMathSymbol{\leqslant}{3}{AMSa}{36}
      \NewMathSymbol{\geqslant}{3}{AMSa}{3E}

    \fi
  \fi
\fi % End of OFSS

\ifnfssone
  \newmathalphabet{\mathit}
  \addtoversion{normal}{\mathit}{cmr}{m}{it}
  \addtoversion{bold}{\mathit}{cmr}{bx}{it}
  \newmathalphabet{\mathbfit} % math mode version of \textbfit{..}
  \addtoversion{normal}{\mathbfit}{cmr}{bx}{it}
  \addtoversion{bold}{\mathbfit}{cmr}{bx}{it}
  \newmathalphabet{\mathbfss} % math mode version of \textbfss{..}
  \addtoversion{normal}{\mathbfss}{cmss}{bx}{n}
  \addtoversion{bold}{\mathbfss}{cmss}{bx}{n}
  \ifAMStwofonts
    \ifCUPmtlplainloaded \else
      \UseAMStwoboldmath
      \makeatletter
      \new@mathgroup\upmath@group
      \define@mathgroup\mv@normal\upmath@group{eur}{m}{n}
      \define@mathgroup\mv@bold\upmath@group{eur}{b}{n}
      \edef\UPM{\hexnumber\upmath@group}
      \new@mathgroup\amsa@group
      \define@mathgroup\mv@normal\amsa@group{msa}{m}{n}
      \define@mathgroup\mv@bold\amsa@group{msa}{m}{n}
      \edef\AMSa{\hexnumber\amsa@group}
      \makeatother
      \mathchardef\upi="0\UPM19
      \mathchardef\umu="0\UPM16
      \mathchardef\upartial="0\UPM40
      \mathchardef\leqslant="3\AMSa36
      \mathchardef\geqslant="3\AMSa3E
    \fi
  \fi
\fi % End of NFSS release 1

\ifnfsstwo
  \DeclareMathAlphabet{\mathbfit}{OT1}{cmr}{bx}{it}
  \SetMathAlphabet\mathbfit{bold}{OT1}{cmr}{bx}{it}
  \DeclareMathAlphabet{\mathbfss}{OT1}{cmss}{bx}{n}
  \SetMathAlphabet\mathbfss{bold}{OT1}{cmss}{bx}{n}
  \ifAMStwofonts
    \ifCUPmtlplainloaded \else
      \DeclareSymbolFont{UPM}{U}{eur}{m}{n}
      \SetSymbolFont{UPM}{bold}{U}{eur}{b}{n}
      \DeclareSymbolFont{AMSa}{U}{msa}{m}{n}
      \DeclareMathSymbol{\upi}{0}{UPM}{"19}
      \DeclareMathSymbol{\umu}{0}{UPM}{"16}
      \DeclareMathSymbol{\upartial}{0}{UPM}{"40}
      \DeclareMathSymbol{\leqslant}{3}{AMSa}{"36}
      \DeclareMathSymbol{\geqslant}{3}{AMSa}{"3E}
    \fi
  \fi
\fi % End of NFSS release 2

\ifCUPmtlplainloaded \else
  \ifAMStwofonts \else % If no AMS fonts
    \def\upi{\pi}
    \def\umu{\mu}
    \def\upartial{\partial}
  \fi
\fi

\title{Optical and infrared observations of the Type~IIP SN~2002hh from day 3 to 397}
\author[M. Pozzo, et al.]
       {M. Pozzo$^{1}$, W.P.S. Meikle$^{1}$, J.T. Rayner$^{2}$, R.D. Joseph$^{2}$, A.V. Filippenko$^{3}$, \newauthor
        R.J. Foley$^{3}$, W. Li$^{3}$, S. Mattila$^{4}$, J. Sollerman$^{4, 5}$\\
        $^1$ Imperial College London, Blackett Laboratory,
        Prince Consort Road, London, SW7 2BW, U.K. \\
        $^2$ Institute for Astronomy, University of Hawaii, 2680 Woodlawn Drive, Honolulu, 
        HI 96822 \\
        $^3$ Department of Astronomy, 601 Campbell Hall, University of California, Berkeley, CA 94720-3411 USA \\
        $^4$ Stockholm Observatory, AlbaNova, Dept. of Astronomy, Stockholm SE 106 91, Sweden \\
        $^5$ DARK cosmology center, NBI, Copenhagen University, Denmark \\}
        
\date{Accepted ... .
      Received .... ;
      in original form 2004 ...}

\pagerange{\pageref{firstpage}--\pageref{lastpage}}
\pubyear{0000}

\begin{document}

\maketitle

\label{firstpage}

\begin{abstract}

We present optical and infrared (IR) observations of the type~II
SN~2002hh from 3 to 397 days after explosion.  The optical
spectroscopic (4--397~d) and photometric (3--278~d) data are
complemented by spectroscopic (137--381~d) and photometric(137--314~d)
data acquired at IR wavelengths.  This is the first time $L$-band
spectra have ever been successfully obtained for asupernova at a 
distance beyond the Local Group.  The $VRI$ light curves in the first 
40 days reveal SN~2002hh to be a SN~IIP (plateau) -- the most common
of all core-collapse supernovae.  SN~2002hh is one of the most
highly extinguished supernovae ever investigated.  To provide a match
between its early-time spectrum and a coeval spectrum of the Type~IIP
SN~1999em, as well as maintaining consistency with K~I interstellar
absorption, we invoke a 2-component extinction model. One component is
due to the combined effect of the interstellar medium of our Milky Way 
Galaxy and the SN host galaxy, while the other component is due to a
``dust pocket'' where the grains have a mean size smaller than in the
interstellar medium. The early-time optical light curves of SNe~1999em and 
2002hh are generally well-matched, as are the radioactive tails of these two 
SNe and SN~1987A.  The late-time similarity of the SN~2002hh optical light
curves to those of SN~1987A, together with measurements of the 
optical/IR luminosity and [Fe~II]~1.257~$\mu$m emission indicate that
$0.07 \pm 0.02$~M$_{\odot}$ of $^{56}$Ni was ejected by SN~2002hh.
However, during the nebular phase the $HKL'$ luminosities of SN~2002hh
exhibit a growing excess with respect to those of SN~1987A. We
attribute much of this excess to an IR echo from a pre-existing, dusty
circumstellar medium.  Based on an IR-echo interpretation of the
near-IR excess, we deduce that the progenitor of SN~2002hh
underwent recent mass loss of $\sim$0.3~M$_{\odot}$.  A detailed
comparison of the late-time optical and near-IR spectra of SNe~1987A and
2002hh is presented. While the overall impression is one of similarity
between the spectra of the two events, there are notable differences. 
The Mg~I~1.503~$\mu$m luminosity of SN~2002hh is a factor of
2.5 greater than in SN~1987A at similar epochs, yet coeval
silicon and calcium lines in SN~2002hh are fainter.  Interpreting
these differences as being due to abundance variations, the overall
abundance trend between SN~1987A and SN~2002hh is not consistent with
explosion model predictions. It appears that during the burning to
intermediate-mass elements, the nucleosynthesis did not progress as
far as might have been expected given the mass of iron ejected.
Evidence for mixing in the ejecta is presented.  Pronounced 
blueshifts seen in the more isolated lines are attributed to
asymmetry in the ejecta.  However, during the timespan of these
observations ($\sim$1~year post-explosion) we find no evidence of dust
condensation in the ejecta such as might have been revealed by an
increasing blueshift and/or attenuation of the red wings of the
emission lines.  Nevertheless, the clear detection of first overtone
CO emission by 200~days and the reddening trend in ($K-L'$)$_0$
suggest that dust formation in the ejecta may occur at later epochs.
From the [O~I]~$\lambda\lambda$6300, 6364~\AA\ doublet luminosity we 
infer a 16--18 M$_{\odot}$ main-sequence progenitor star.  The progenitor 
of SN~2002hh was probably a red supergiant with a substantial, dusty
wind.

\end{abstract}


\section{Introduction}\label{intro}

\begin{figure}
\vspace{8cm} \includegraphics{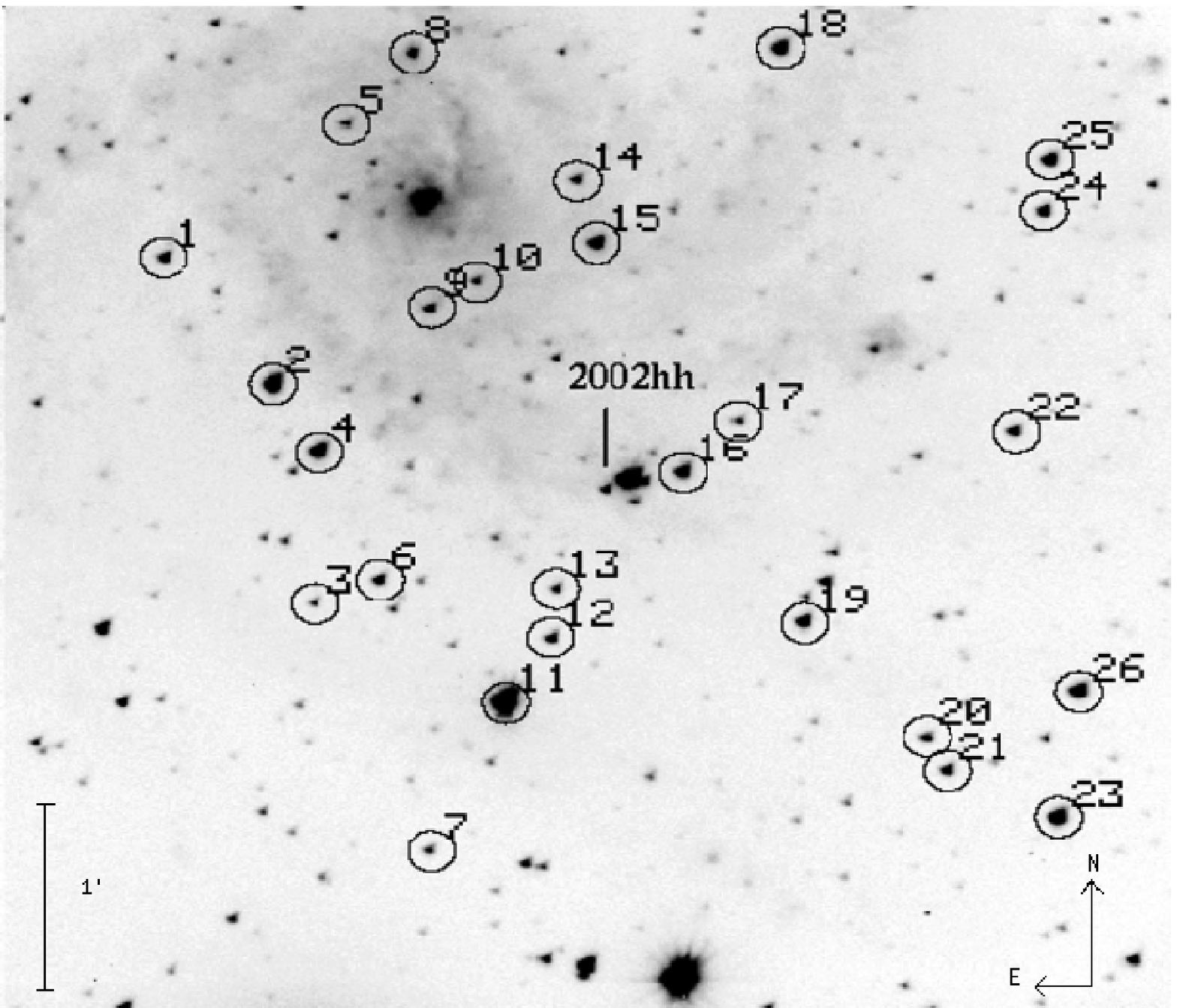}
\caption[] {$R$-band image of the SN~2002hh field taken with KAIT on
2002 December 2 UT, when the SN was about one month
post-explosion. The nucleus of the host galaxy, NGC 6946, is visible
at the top-left corner. Also shown are the local standards used for
the photometric calibration (see Table~1).}
%\label{}
\end{figure}

An important goal of astrophysics is to deepen our understanding of
the progenitors and explosion mechanisms of core-collapse supernovae
(CCSNe).  What kind of stars explode?  What drives the explosion?  Do
CCSNe or their progenitors contribute significant quantities of dust
to the interstellar medium (ISM)?  Observations of CCSNe at late times 
have a unique potential for helping us to answer these questions since by 
these epochs, in the {\it nebular phase} ($t>150$~d), we can view directly
most of the ejecta.  This allows measurements of element/molecule
abundances and distributions such as those achieved for SN~1987A (e.g.,
Meikle et al. 1989, 1993; Varani et al. 1990; Spyromilio, Meikle \&
Allen 1990; Spyromilio et al. 1991).  The drawback is that by the time
the nebular phase is reached the supernova has become
faint. Consequently, only the very nearest events can be effectively
studied in this way.  \\

An excellent opportunity for the study of the
nebular phase came with the discovery of SN~2002hh, one of the
nearest type~II SNe in the past 50~years.  Type~II supernovae make up
the most common subgroup of CCSNe.  SN~2002hh was very highly reddened,
with a $V$-band extinction which reduced its flux by a factor of over 
100. However, the effect of the extinction is much reduced in
the infrared (IR), and indeed it can be argued (see below) that this
wavelength region is at least as important as the optical region for
late-time studies.  \\

IR observations can test the nature of CCSNe in ways not
available in other wavelength regions.  Constraints on the CCSN
progenitor can be obtained with a combination of IR/optical photometry
and IR spectroscopy.  For example, the presence of first overtone CO
emission ($K$ band) in CCSNe is increasingly regarded as ubiquitous,
having been detected in all seven CCSNe which have been observed in
the $K$ band at 3--6 months post-explosion (see Gerardy et al. 2002; 
Meikle et al. 2003, and references therein). 
By modelling the first overtone CO spectrum of the Type~IIn SN~1998S,
Fassia et al. (2001) deduced a CO mass of 10$^{-3}$~M$_{\odot}$, while 
from the bolometric light curve they deduced a $^{56}$Ni mass of 
0.15~M$_{\odot}$.  Combining these two results they inferred 
a core mass exceeding 4~M$_{\odot}$, implying a massive progenitor.  
In addition, the detection of CO is important since it is probably an 
essential precursor to dust condensation in the ejecta.  \\

Infrared spectroscopy during the photospheric phase can probe
ejecta mixing and hence the explosion mechanism. Graham (1988) showed
how the strong He~I 1.083~$\mu$m line at early times could be used to
infer the upward mixing of radioactive iron-group elements in the
ejecta of the peculiar Type~II SN~1987A.  Fassia et al. (1998) and
Fassia \& Meikle (1999) further developed this technique and applied
it, respectively, to the Type~IIP SN~1995V and to SN~1987A.\\

In spite of the high extinction toward SN~2002hh, optical spectra are 
also of value. The prominent, high signal-to-noise ratio (S/N) 
H$\alpha$ and Ca~II lines allow investigation of 
ejecta asymmetry.  The O~I 8446~\AA\ fluorescence line provides 
confirmation of the identification of the O~I 1.1287~$\mu$m line, 
and the cascade pair together may allow us to study the
conditions under which such lines are formed (Spyromilio et al. 1991). \\

A major goal in the study of CCSNe is to provide a significant test of
the proposal that they are, or have been, a significant source of dust
in the universe. While this hypothesis is over 30 years old
(Cernuschi, Marsicano \& Codina, 1967; Hoyle \& Wickramasinghe 1970)
and is still popular (Gehrz, 1989; Tielens 1990; Dwek 1998; Todini \&
Ferrara 2001; Nozawa et al. 2003) {\it direct evidence that SNe are
major dust sources is still very sparse.} Indeed, it is not known if
ordinary Type~II SNe or their progenitors produce large amounts of
dust at all.  \\

Dust condensation in CCSN ejecta can be detected in two
ways. In one method, we can make use of the IR emission from the
hot dust grains.  13 CCSNe (cf. Gerardy et al. 2002, and references 
therein), plus 1 peculiar Type~Ia event (SN~2002ic: Kotak et al. 2004), 
have shown a strong, late-time near-IR (NIR) excess implying the presence of 
hot dust.  However, it is difficult to establish whether this dust was 
newly condensed in the metal-rich SN ejecta or existed previously in the 
progenitor's wind and was detected via an IR echo.  Even the comprehensive 
multi-epoch NIR study of the Type~IIn SN~1998S by Pozzo et al. (2004) could not 
completely resolve this ambiguity.  While their favoured interpretation was 
that at least $10^{-3}$~M$_{\odot}$ of dust formed in the cool dense shell 
at the interface between the ejecta and the circumstellar medium (CSM), 
the alternative IR-echo interpretation was not ruled out.  
Dunne et al. (2003) claimed that their sub-mm observations of 
the SN remnant Cassiopeia~A indicates that $>$1~M$_{\odot}$ of dust formed 
in the ejecta, but this has been recently disproved by Krause et al. (2004) 
as being instead due to interstellar dust in a molecular cloud complex located 
along the line of sight between the Earth and Cas A. \\  

The other way of detecting
newly condensed dust is via its attenuating effect on the red wings of
the broad ejecta lines during the nebular phase.  This technique has
the advantage of being relatively unambiguous in its ability to
demonstrate the presence of new dust, although it may be more
difficult to extract quantitative information about the quantity and
nature of the grains. However, owing to the difficulty of acquiring
high-quality spectra at late times, this method has demonstrated
ejecta dust in just three cases so far: the type~II-pec SN~1987A
(Danziger et al. 1991; Lucy et al. 1991), the type~IIP SN~1999em
(Elmhamdi et al. 2003), and the type~IIn SN~1998S (Pozzo et al. 2004).
Thus, the close proximity of SN~2002hh offers us only the second-ever
opportunity for the study of dust condensation in the most common type
of CCSN. \\

SN~2002hh was discovered (Li 2002) on 2002 October 31.1 (UT dates are used
throughout this paper) during the course of the Lick Observatory Supernova 
Search (LOSS; Li et al. 2000; Filippenko et al. 2001; Filippenko 2005) with
the 0.76-m Katzman Automatic Imaging Telescope (KAIT) at Lick Observatory. 
It was discovered at about 16.5 mag (see Fig.~1) and peaked at $V \approx 15$
mag in mid-November 2002.  SN~2002hh was not detected in a previous KAIT
image taken on October 26.1 at a limiting magnitude of $\sim$19.0. We
shall therefore adopt 2002 October 29 as the explosion epoch, the
uncertainty being about $\pm$2~d.  SN~2002hh occurred in NGC~6946,
an Scd galaxy which has produced 7 other known SNe, five of which 
were classified [SNe~1917A (II), 1948B (IIP), 1968D (II), 1980K
(IIL), and 2004et (IIP)] and two were not [SNe~1939C and 1969P].
SN~2002hh is located at R.A. = 20h34m44s.29, Decl. =
+60$^{\circ}$ 07$'$ 19\farcs0 (2000.0), which is 60\farcs9 west and
114\farcs1 south of the nucleus (Li 2002).  We adopt a host galaxy
distance of $5.9 \pm 0.4$ Mpc (Karachentsev, Sharina \& Huchtmeier
2000).  \\

Spectra taken by Filippenko, Foley \& Swift (2002) on 2002
November 2 revealed broad, low-contrast H$\alpha$ emission and
absorption lines, but with a nearly featureless and very red continuum
(cf. Fig.~6).  Strong, narrow (interstellar) Na~I~D absorption is also
present. Filippenko et al. inferred a very young, highly reddened
Type~II supernova.  The high reddening was confirmed and measured by
Meikle et al. (2002) on Nov. 18.86 from IR images.  Adopting a
phase of 21~days, comparison of the $J-K_s$ colour with the
infrared-template light curves of Mattila \& Meikle (2001) indicated
$E(J-K_s) = 1.0$.  From the reddening law of Cardelli, Clayton \&
Mathis (1989), this yields $A_V \approx 6.1$ mag.  Subtracting the
Galactic extinction (Schlegel, Finkbeiner \& Davis 1998) leads to a
host-galaxy extinction of $A_V \approx 5.0$ mag.  For a distance of 5.9
Mpc, Meikle et al. (2002) derived dereddened absolute IR magnitudes
of $M(J) = -18.3$ and $M(K_s) = -18.5$ (but see our revised values
in Section \ref{extinct}), which are close to the values
given by the templates of Mattila \& Meikle (2001) for normal type~II
SNe. We note that SN~2004et occurred in the same galaxy as SN~2002hh, 
but was not highly reddened (Li et al. 2005). We attribute this to the fact 
that SN~2004et was in a ``clean'' region at the edge of the galaxy while
SN~2002hh is located on a spiral arm.  SN~1917A also occurred in
NGC~6946, just $\sim 22$\farcs0 from SN~2002hh, but it has no measured
extinction.\\

Stockdale et al. (2002) reported detection of radio emission from 
SN 2002hh with the VLA at 17~d (Nov. 15.25 UT) and confirmed the
previously reported optical position. They measured fluxes of
$0.60 \pm 0.10$ mJy at 22.485~GHz and $0.81 \pm 0.08$ mJy at
8.435~GHz.  From the apparently optically thin character of the radio
emission, they suggested that the CSM/ejecta shock interaction was weak at
this epoch, and was evolving unusually rapidly.  Pooley \& Lewin
(2002) reported the detection of X-ray emission from SN~2002hh with
the Chandra X-ray observatory at 27~d, confirming the previously
reported optical and radio positions.  The low X-ray luminosity
detected is consistent with the weak circumstellar interaction
indicated by the radio observations.  Preliminary spectral fits to the
X-ray data indicate a column density of $N_H = 10^{22}$~cm$^{-2}$. 
SN~2002hh was also detected at 1396.75 MHz with the Giant Meterwave Radio
Telescope at 59~d (undetected at 32~d) (Chandra, Ray \& Bhatnagar
2003).  More recently, Beswick et al. (2005) reported VLA measurements
of $0.35 \pm 0.1$ mJy at 4.860~GHz on day~381 and $1.6 \pm 0.2$ mJy at
1.425~GHz on day~899.\\

Barlow et al. (2005) reported the detection of SN~2002hh in archival
{\it Spitzer Space Telescope} images (obtained in the SINGS Legacy Program) 
covering 3.6--24~$\mu$m, in June and November 2004, as well as in an 
11.2~$\mu$m Gemini/Michelle image taken in September 2004. They reported 
a 25\% decline in the total 8~$\mu$m flux between
days~590 and 758, and suggested that most of the mid-IR emission
originates in a massive (10--15~M$_{\odot}$), dusty circumstellar
shell ejected by the progenitor and heated by the supernova
luminosity.  They also proposed that a large fraction of the extinction
is due to this circumstellar dust.  However, they did not rule out the
possibility that a small fraction of the mid-IR emission may have
originated in newly formed dust in the ejecta. \\

In contrast, Meikle et
al. (2005a,b, and in prep.) found a decline of only $\sim$10\% 
in the 8~$\mu$m flux between days~590 and 994.  They concluded that to
produce such a large and slowly declining flux with an IR echo would
require a CSM mass of about 80~M$_{\odot}$ assuming a normal dust/gas
ratio.  They also found that only a small amount of the extinction
should be attributed to this dust.  They concluded that the total
mid-IR emission could {\it not} have arisen from an IR echo within a
dusty CSM ejected by the progenitor. However, the possibility is not
ruled out that the supernova occurred within a dense, dusty ISM and
that this was responsible for a powerful IR echo. Alternatively, most
of the mid-IR flux could simply be due to steady background sources in
the beam. The {\it Spitzer} images show that the region around the supernova
is rich in cool, mid-IR emission structures.  The apparent
disagreement between the two sets of authors is not yet resolved. \\

In this paper we describe our IR and optical photometry and
spectroscopy of SN~2002hh from day 3 to 397.  The paper
is organized as follows. Optical/IR imaging and spectroscopic
observations and data reduction are discussed in Section II; 
we present $VRI$ light curves together with IR
light curves and colour curves. Section III deals with the redshift and
extinction.  In particular, we show that the high extinction toward
the SN is best described with a 2-component extinction model with
different grain sizes.  We also derive the optical/IR luminosity of 
SN 2002hh and the mass of ejected $^{56}$Ni.  Section IV is devoted to the
analysis of the spectroscopic behaviour and the main elements detected
in the optical and IR spectra.  In Section V we compare SN~2002hh with
the well-studied core-collapse SN~1987A. Their optical and IR spectra
at coeval epochs, although similar, show some interesting
differences. The evolution of the CO emission and the unidentified
2.265~$\mu$m feature originally detected in SN~1987A are also
discussed. Finally, the late-time near-IR continuum of SN 2002hh and 
the temporal evolution of the extinction-corrected ($K-L'$)$_0$ colour
are analyzed. Conclusions follow in Section VI.

\begin{table}
\centering
\caption[]{Photometry of the comparison stars for SN~2002hh.} 
\begin{minipage}{\linewidth}
\renewcommand{\thefootnote}{\thempfootnote}
\renewcommand{\tabcolsep}{5.5mm}
\begin{tabular}{rccc} \hline
ID & $V$ & $R$ & $I$  \\ \hline 
 1  & 15.47(1)\footnote{Figures in brackets give the internal error, in units 
of the magnitude's least significant digit.}  & 14.98(1)  & 14.46(1) \\
 2  & 13.57(0)  & 13.14(2)  & 12.79(3) \\
 3  & 16.86(2)  & 16.30(1)  & 15.74(2) \\
 4  & 13.78(1)  & 13.34(1)  & 12.95(3) \\
 5  & 16.97(3)  & 16.45(1)  & 15.94(2) \\
 6  & 15.80(2)  & 15.32(1)  & 14.83(2) \\
 7  & 16.67(2)  & 16.14(1)  & 15.56(2) \\
 8  & 15.94(2)  & 15.28(1)  & 14.64(1) \\
 9  & 16.16(2)  & 15.59(1)  & 15.05(1) \\
10  & 16.75(2)  & 16.18(1)  & 15.60(1) \\
\,\,11\footnote{This local field star was saturated in many frames and therefore was not used for the final photometric calibration. \\} & 12.32(3)  & 11.84(2)  & 11.88(3) \\
12  & 15.82(2)  & 15.37(2)  & 14.88(2) \\
13  & 16.41(2)  & 15.75(1)  & 15.12(1) \\
14  & 16.38(1)  & 15.83(1)  & 15.28(1) \\
15  & 15.06(1)  & 14.17(1)  & 13.36(1) \\
16  & 14.74(2)  & 14.27(2)  & 13.81(3) \\
17  & 16.90(3)  & 16.45(3)  & 15.96(2) \\
18  & 14.81(1)  & 14.14(1)  & 13.48(2) \\
19  & 14.85(2)  & 14.35(2)  & 13.88(2) \\
20  & 16.85(2)  & 16.00(1)  & 15.32(1) \\
21  & 15.48(2)  & 15.00(0)  & 14.53(1) \\
22  & 15.67(2)  & 15.22(1)  & 14.75(0) \\
23  & 13.84(1)  & 13.36(1)  & 12.93(3) \\
24  & 15.63(2)  & 14.93(1)  & 14.36(1) \\
25  & 14.46(1)  & 14.02(1)  & 13.59(1) \\
26  & 14.38(2)  & 13.70(0)  & 13.10(2) \\  \hline
\vspace{-0.8cm}
\end{tabular}  
\end{minipage}
\end{table}

\begin{table}
\centering
\caption[]{$VRI$ photometry of SN~2002hh.} 
\begin{minipage}{\linewidth}
\renewcommand{\thefootnote}{\thempfootnote}
\begin{tabular}{lcccc} \hline
JD--2,450,000 & Epoch\footnote{Days after explosion, assumed to be 2002 October 29 UT (JD 2452577.5).} & $V$ & $R$ & $I$  \\ \hline 
2580.60  & ~~3 &  17.30(7)\footnote{Figures in brackets give the photometric error (see text), in units 
of the magnitude's least significant one or two digits. \\} &   15.64(7) &   14.31(6) \\
2581.62  & ~~4 &  17.28(4)~  &  15.60(3)  &  14.27(3) \\
2583.60  & ~~6 &  17.18(5)~  &  15.53(5)  &  14.19(5) \\
2590.62  & ~13 &  17.29(3)~  &  15.60(3)  &  14.29(6) \\
2594.65  & ~17 &  17.20(6)~  &  15.57(3)  &  14.24(3) \\
2596.61  & ~19 &  17.29(6)~  &  15.60(6)  &  14.26(6) \\
2605.62  & ~28 &  17.32(5)~  &  15.55(3)  &  14.26(4) \\
2610.59  & ~33 &  17.48(4)~  &  15.65(4)  &  14.31(4) \\
2620.62  & ~43 &  17.45(4)~  &  15.67(3)  &  14.23(4) \\
2744.01  & 166 &  19.53(16)  &  17.40(7)  &  16.01(8) \\
2757.01  & 180 &  19.71(13)  &  17.53(6)  &  16.09(7) \\
2765.97  & 188 &  19.52(10)  &  17.56(5)  &  16.18(6) \\
2774.89  & 197 &    --       &  17.45(8)  &  15.96(7) \\
2783.98  & 206 &  19.85(16)  &  17.53(5)  &  16.08(5) \\
2792.96  & 216 &  19.78(13)  &  17.85(6)  &  16.33(6) \\
2801.94  & 224 &  20.16(16)  &  17.89(5)  &  16.52(5) \\
2819.91  & 242 &  20.36(19)  &  18.08(5)  &  16.61(4) \\
2828.88  & 250 &  20.09(21)  &  17.91(6)  &  16.62(5) \\
2837.92  & 260 &  20.74(35)  &  18.17(6)  &  16.85(3) \\
2846.80  & 269 &  20.25(25)  &  18.25(7)  &  17.01(7) \\
2855.81  & 278 &   --        &  18.64(5)  &  16.93(5) \\ \hline
\vspace{-0.8cm}
\end{tabular}  
\end{minipage}
\end{table}

\begin{figure}
\vspace{8.5cm} \includegraphics{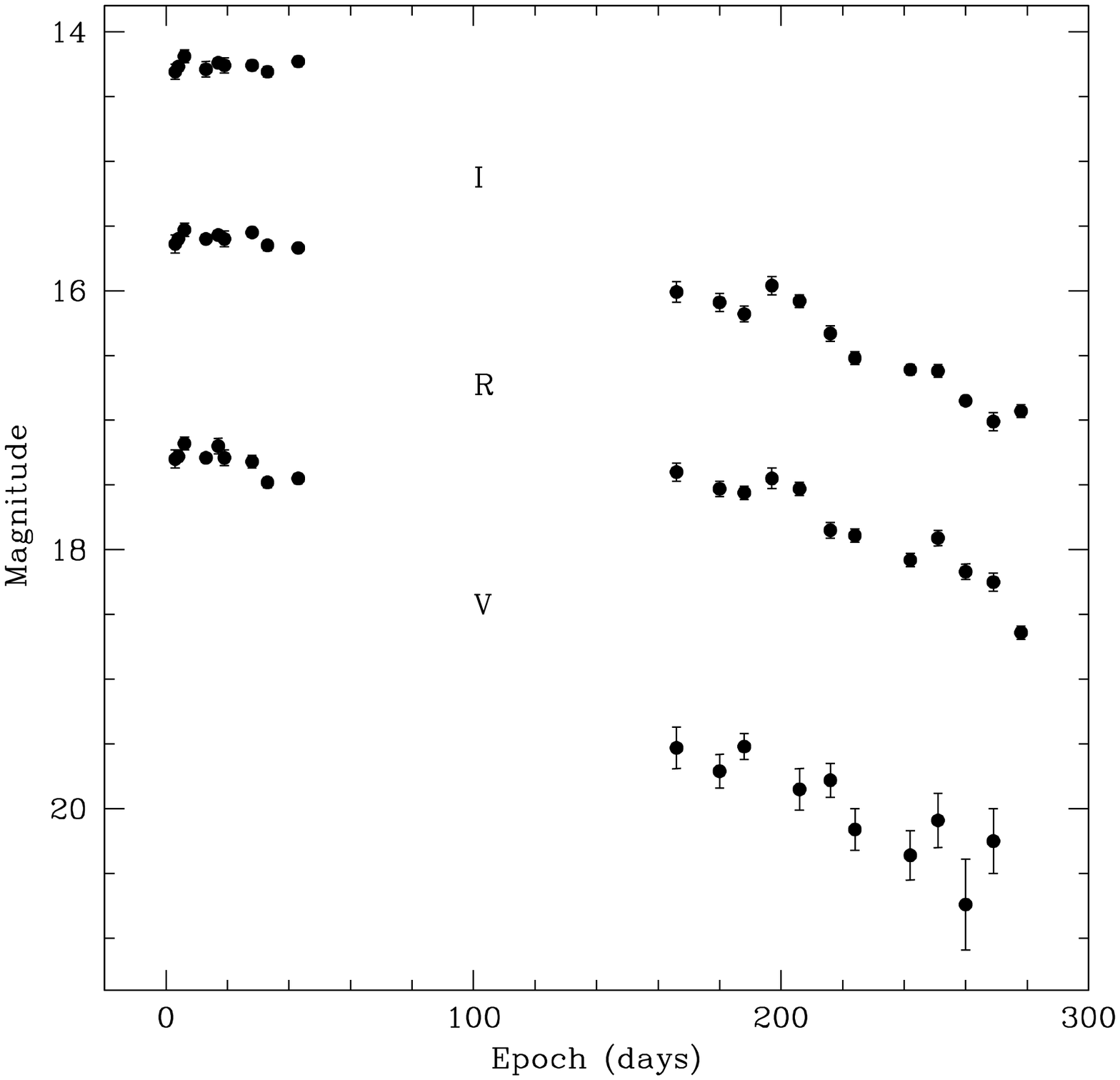}
\caption[]{$VRI$ light curves of SN~2002hh.} 
\end{figure}

\section{Observations}
\subsection{Optical photometry}
We obtained {\it VRI} images of SN~2002hh during the first $\sim 300$
days post-explosion with KAIT. An Apogee AP7 512 $\times$ 512 SITe CCD
camera was used, covering a total field-of-view of $6.7' \times 6.7'$ at
$0.8''$ pixel$^{-1}$. The images were reduced using standard 
{\sc iraf}\footnote{{\sc iraf} (Image Reduction and Analysis Facility) is
distributed by the National Optical Astronomy Observatory (NOAO),
which is operated by the Association of Universities for Research in
Astronomy (AURA), Inc.  under cooperative agreement with the National
Science Foundation.}  routines. \\

Point-spread function (PSF) fitting photometry was performed after galaxy 
subtraction using $VRI$ template images of the SN field
acquired on 2004 Aug 19 after the supernova had faded away.  Local stars 
in the field of the host galaxy (see Fig.~1 and Table~1) were measured on 
three different dates (2002 Nov 5, 2003 May 31, and 2003 Jun 1),
calibrated against 10--15 fields of Landolt standard
stars observed at different airmasses at each night (for a final
root-mean-square [RMS]
solution of about 0.01 mag in all the filters), and finally used for
the photometric calibration of the SN magnitudes. The uncertainty in the
photometry was estimated by adding in quadrature the RMS of the
transformed magnitude of the SN from the observed local standard stars 
and the error in the {\sc iraf} PSF fitting. The photometry may be affected 
by a systematic error (up to 0.02 -- 0.05 mag in $V$ and $I$, 0.07 mag in $R$) 
due to the fact that, although colour terms have been applied to derive 
standard photometry for the KAIT $V$, $R$, and $I$ filters (see Li et al. 2001), 
none of the local standard stars used for the photometric calibration has 
the extreme red colour of SN 2002hh. However, in our use of the photometry to flux
calibrate the spectra, allowance for the colour term effect was achieved
through the use of effective wavelengths (see discussion in section 2.3). \\

Final optical $VRI$ magnitudes and 
associated uncertainties are listed in Table~2, while the corresponding 
light curves are shown in Fig.~2.  The temporal evolution of the 
$VRI$ magnitudes during the first 40~d shows a typical plateau phase, 
which indicates that SN~2002hh is a type~IIP (plateau) supernova, 
the most common kind (see also Fig.~13). 
The late-time decline rates in the $VRI$ bands were roughly 
linear at about 0.007, 0.011 and 0.008 mag day$^{-1}$ (respectively), 
and of the same order as those found in the IR bands (see below).\\

\begin{figure}\vspace{8.5cm} \includegraphics{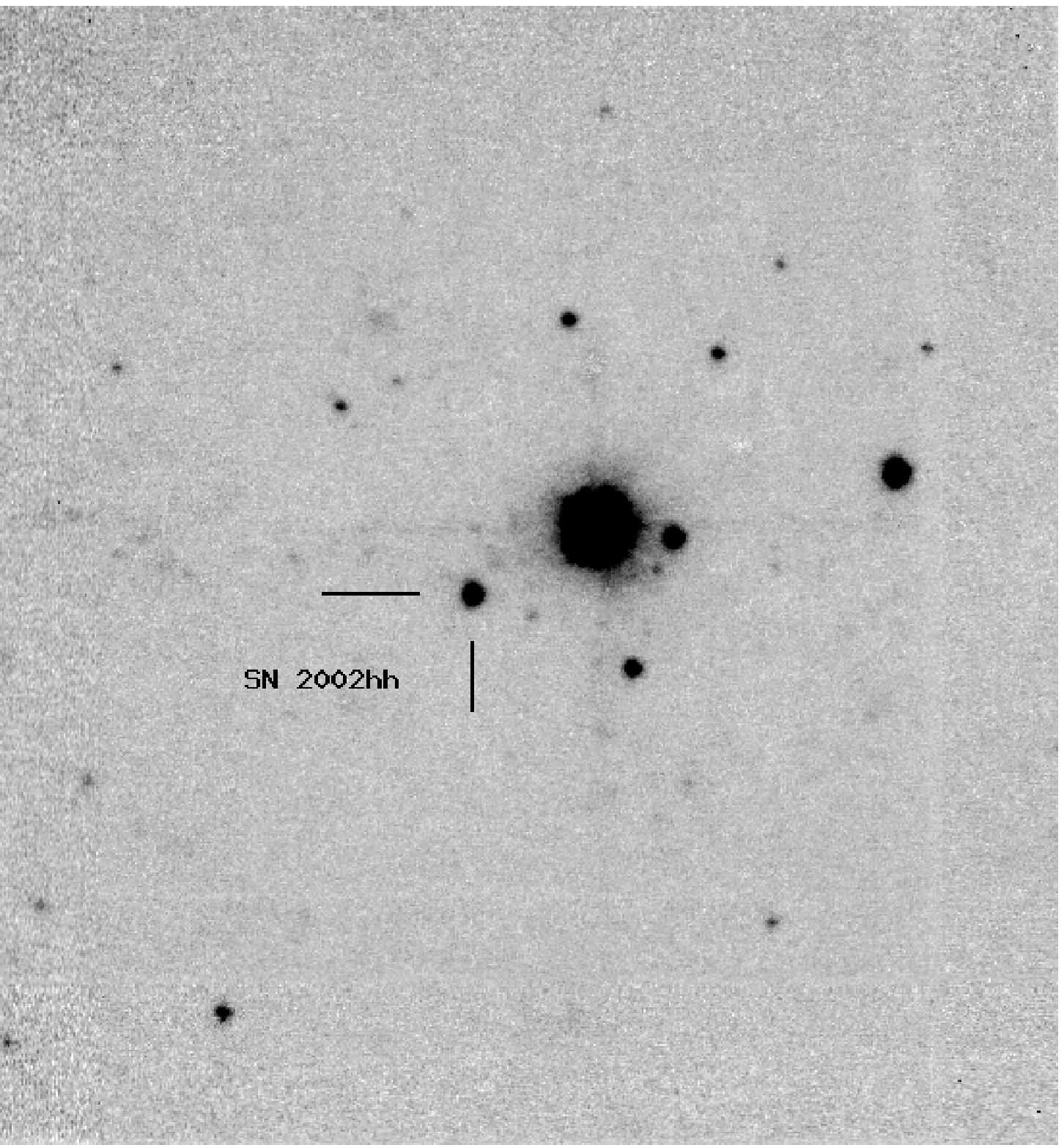}
\caption[]{$K$-band image of SN~2002hh taken with the SpeX imager on 2003 July 22,
when the SN was about nine months post-explosion.} 
\end{figure}

\begin{figure}
\vspace{7.3cm} \includegraphics{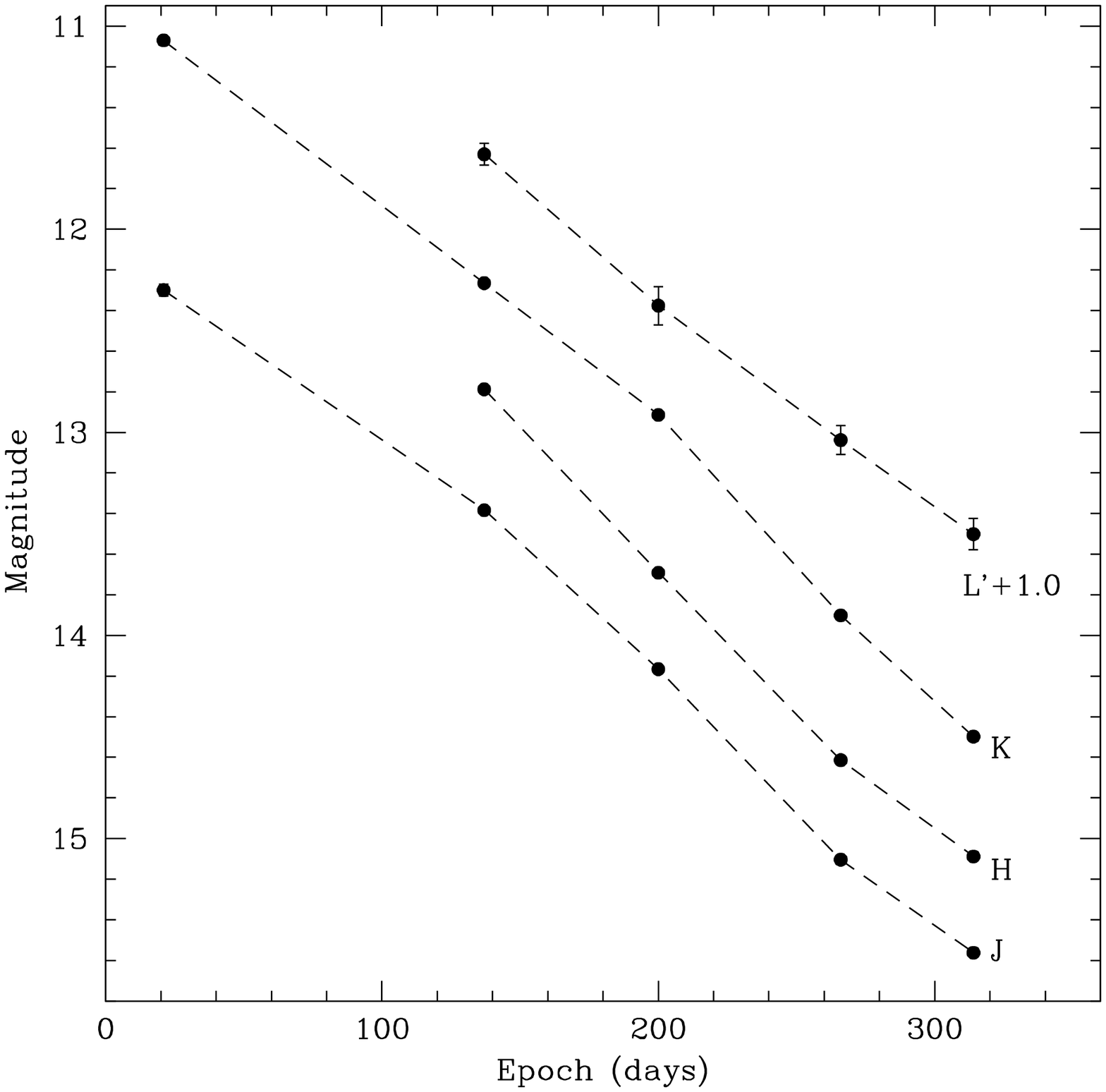}
\caption[]{$JHKL'$ light curves of SN~2002hh. Also plotted are the
the early (+21~d) $J$ and $K$ points. Only the statistical 
errors are shown (see text).} 
%\label{}
\end{figure}

\begin{figure}
\vspace{7.3cm}\includegraphics{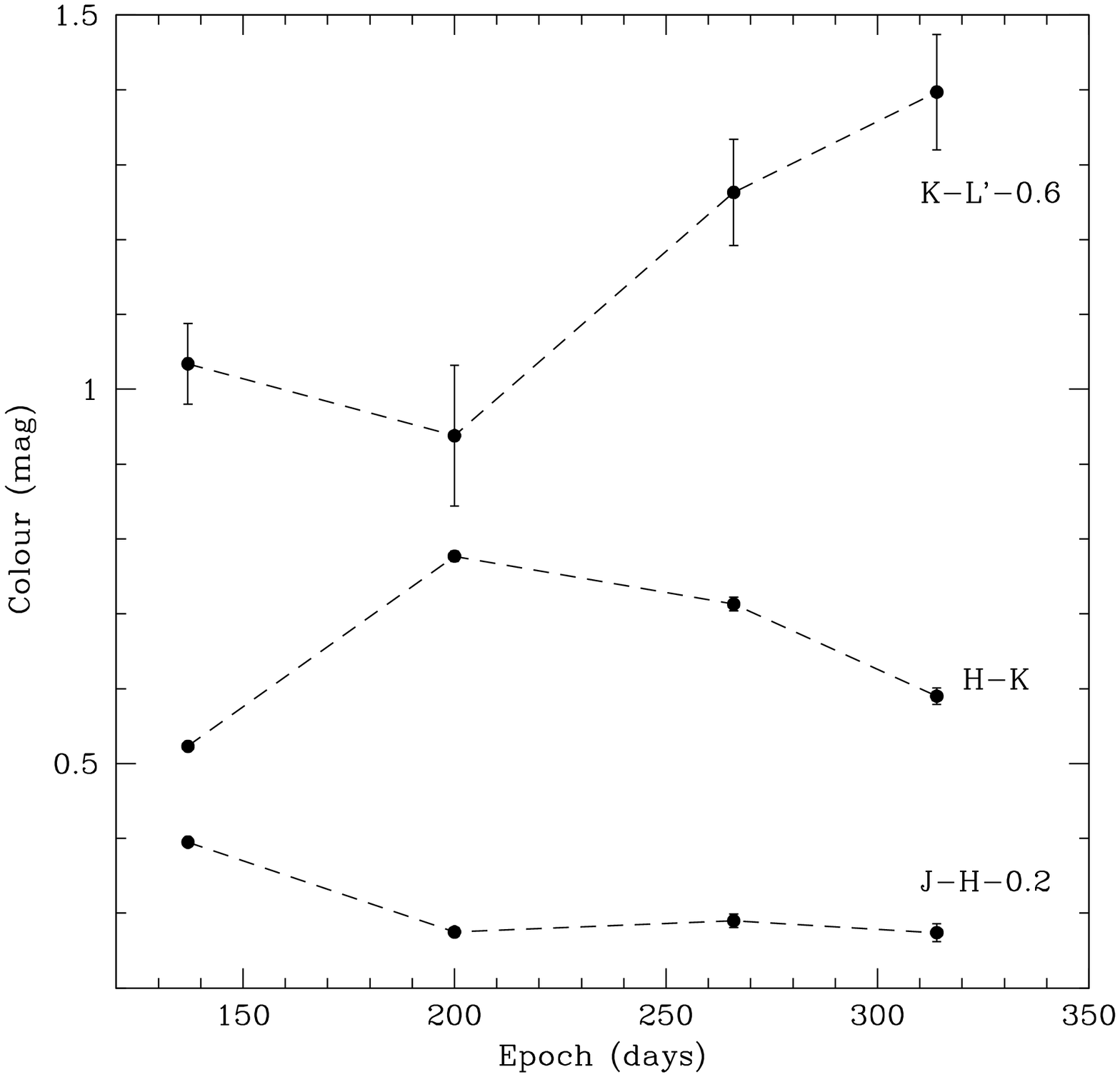}
\caption[]{Post-100~d IR colour evolution of SN~2002hh. Only the statistical 
errors are shown (see text).} 
%\label{}
\end{figure}

\subsection{Near-infrared photometry}

The SpeX imager at IRTF (see Section \ref{irspec}) was used to obtain 
$JHKL'$ images of SN~2002hh at four epochs between days~137 and 314.  
These epochs correspond to four of the five epochs when NIR spectroscopy
was also obtained (see below).  The imager employs an Aladdin-2
$512 \times 512$ pixel InSb array covering a 60\farcs0$\times$60\farcs0 
field-of-view at 0\farcs12 pixel$^{-1}$.  A standard imaging 
procedure was performed, using
5- to 10-point jitter patterns.  The images were reduced using
standard IRAF routines.  The jittered on-source exposures were
median-combined to form sky frames in the $JHK$ bands for both the
supernova and the standard stars. (By median-combine, we mean 
that a median value
was calculated for each pixel over the sequence of on-source images observed in
each band.)  In the $L'$ band, where the sky varies rapidly, sky subtraction 
was performed by subtracting consecutive pairs of on-source offset exposures 
from each other.  In each band, the sky-subtracted frames were then median-combined.
Fig.~3 shows the $K$-band image of SN~2002hh taken with the SpeX imager
on 2003 July 22, when the SN was about nine months post-explosion.\\

\begin{table*}
\centering
\caption{Log of IR photometric observations taken at IRTF with SpeX.}
\begin{minipage}{\linewidth}
\renewcommand{\thefootnote}{\thempfootnote}
\renewcommand{\tabcolsep}{1.5mm}
\begin{tabular}{llccl}
\hline 
JD--2,450,000 & Date (2003) & Epoch\footnote{Days after explosion.} & Photometric standard & Observers \\ \hline
2714.5     & Mar 15 & 137 & FS150 ($JHK$), HD203856 ($L'$) &  R.D. Joseph, J.T. Rayner \\
2777.5     & May 17 & 200\footnote{Mostly cloudy, this night was non-photometric.} & HD201941 ($JHKL'$)\footnote{Due to the non-photometric night, this standard star was not used for calibration; see text for details.} & J.T. Rayner \\
2843.5     & Jul 22 & 266 & FS150 ($JHK$), HD201941 ($L'$) & R.D. Joseph, J.T. Rayner \\
2891.5     & Sep  8 & 314 & FS150 ($JHK$), HD201941 ($L'$) & J.T. Rayner \\ \hline
\vspace{-1cm}
\end{tabular}
\end{minipage}
\end{table*}

\begin{table*}
\centering
\caption[]{IR photometry and colours of SN~2002hh.} 
\begin{minipage}{\linewidth}
\renewcommand{\thefootnote}{\thempfootnote}
\begin{tabular}{lcccccccc} \hline
JD(2450000+) & Epoch\footnote{Days after explosion.}& $J$ & $H$ & $K$ & $L'$ & $J-H$ & $H-K$ & $K-L'$ \\ \hline 
2714.5 & 137 & 13.383(1)\footnote{Figures in brackets give the internal error, in units 
of the magnitude's least significant one or two digits. \\} & 12.788(1)~ & 12.265(1) & 
       10.631(54) & 0.595(1)~~ & 0.523(1)~~ & 1.634(54)\\ 
2777.5 & 200 & 14.166(3)~ & 13.691(4)~ & 12.914(6) & 11.376(94) & 0.475(5) ~ & 
       0.777(7)~~ & 1.538(94) \\ 
2843.5 & 266 & 15.104(6)~ & 14.614(7)~ & 13.901(6) & 12.038(71) & 0.490(9)~~ & 
       0.713(9)~~ & 1.863(71) \\ 
2891.5 & 314 & 15.562(7)~ & 15.088(10) & 14.498(4) & 12.501(77) & 0.474(12) &
       0.590(11) & 1.997(77) \\ \hline
\vspace{-1cm}
\end{tabular}  
\end{minipage}
\end{table*}

Aperture photometry was carried out within the Starlink package {\sc
gaia}\footnote{Graphical Astronomy and Image Analysis Tool, version
2.6-9} (Draper, Gray \& Berry 2002) using an aperture radius of 
18 pixels (equivalent to 2\farcs16) corresponding to 2.5 times the full width 
at half-maximum (FWHM) of the standard star PSF.  The sky background 
was measured using a concentric annular aperture, with inner and outer radii 
respectively of 1.5 and 2.5 times that of the aperture.  The sky variance with 
2~sigma clipping was used to determine the photometric statistical uncertainty.  

Table~3 lists the details of the IR imaging observations, including the 
standard stars used.  On most nights the magnitudes were determined directly 
by comparison with the standards observed on the same nights.  However,
the 200~d epoch was clearly non-photometric and so on this occasion
calibration was carried out via relative photometry of SN 2002hh field
stars.\\

The resulting IR magnitudes and colours of SN~2002hh are listed in
Table~4 and plotted in Figs.~4 and 5.  Associated statistical errors
are shown in parentheses and as error bars.  The late-time decline
rates in the $JHKL'$ bands were roughly linear at, respectively, about
0.012, 0.013, 0.013 and 0.011 mag d$^{-1}$.  The $J-H$ and $H-K$ colours
remained roughly constant at $\sim$0.5 and $\sim$0.7, respectively,
during the 137--314~d period.  The $K-L'$ colour reddened between
days~137 and 314 from about 1.6 to about 2.0 mag.

\subsection{Optical spectroscopy}\label{optspec}
Optical spectra were obtained at 8 epochs spanning 4--397 days
post-explosion.  We used (a) the Kast double spectrograph (Miller \&
Stone 1993) mounted on the Lick Observatory 3.0~m Shane telescope on
Mt. Hamilton, (b) the Echellette Spectrograph and Imager (ESI; Sheinis et
al. 2002) mounted on the 10.0~m Keck~II telescope on Mauna Kea, (c) the
ISIS spectrograph at the 4.2~m William Herschel Telescope (WHT) on
La Palma, (d) the ALFOSC spectrograph at the 2.56~m Nordic Optical 
Telescope (NOT) on La Palma, and (e) the Low Resolution Imaging 
Spectrometer (LRIS; Oke et al. 1995) mounted on the 10.0~m Keck~I 
telescope on Mauna Kea.\\

\begin{table*}
\centering
\caption{Log of optical spectroscopic observations.} 
\begin{minipage}{\linewidth}
\renewcommand{\thefootnote}{\thempfootnote}
\renewcommand{\tabcolsep}{1.4mm}
\begin{tabular}{llllcclll}
\hline 
JD\footnote{JD--2,450,000.} & Date & Epoch\footnote{Days after explosion.} & Telescope/ & Range & Resol. & Slit width/ & Spectroscopic & Observers \\ 
 &  &     & Instrument &  ($\mu$m) & ($\lambda/\Delta\lambda$)   & pos. angle  & standard\footnote{For Kast and LRIS observations, we give the standard stars used 
(respectively) for the blue and red part of the combined spectrum.}  & \\ \hline
2581.8   & 2002 Nov  2 & ~~~4 & Lick/Kast\footnote{Lick observations used a D550 dichroic.} & 0.31--1.04 & 600 & 2\farcs0/166$^{\circ}$ & BD$+28^{\circ}4211$, & 
           A. Filippenko, R. Foley, B. Swift \\
 &  &  &  &  &  & & BD$+17^{\circ}4708$ & \\
2585.9   & 2002 Nov  6 & ~~~8 & Keck/ESI & 0.39--1.04 & 4000 & 1\farcs0/164$^{\circ}$ & BD$+28^{\circ}4211$ & A. Filippenko, S. Jha, R. Chornock  \\
2621.8   & 2002 Dec 12 & ~44 & Lick/Kast & 0.31--1.04 &  550 & 2\farcs0/270$^{\circ}$ & BD$+28^{\circ}4211$, & R. Foley, M. Papenkova \\
 &  &  &  &  &  & &  HD 19445 & \\
2738.8   & 2003 Apr  8 & 162 & Lick/Kast & 0.32--1.04 &  500 & 2\farcs0/79$^{\circ}$ & Feige 34, & R. Foley, M. Papenkova, D. Weisz \\
 &  &  &  &  &  & &  HD 84937 & \\
2804.7   & 2003 Jun 14 & 227 & WHT/ISIS  & 0.47--0.78 & 1800 & 1\farcs5/176$^{\circ}$ & BD$+17^{\circ}4798$ & C. Benn\\
2828.4   & 2003 Jul  7 & 250\footnote{ High temperature and lots of dust reported, 
         very high sky brightness.} & 
         NOT/ALFOSC & 0.50--0.96 &  900 & 1\farcs3/181$^{\circ}$ & BD$+28^{\circ}4211$ & J. Sollerman et al. \\
2839.7   & 2003 Jul 19 & 262 & WHT/ISIS  & 0.64--1.03 & 2050 & 1\farcs0/139$^{\circ}$  & HD 340611 & I. Soechting \\
2973.9   & 2003 Nov 29 & 397 & Keck/LRIS\footnote{LRIS observations used a D560 dichroic.} & 0.31--0.94 & 1050 & 1\farcs0/127$^{\circ}$ & BD$+28^{\circ}4211$, & A. Filippenko  \\ 
 &  &  &  &  &  & &  BD$+17^{\circ}4708$ & \\ \hline
\vspace{-1cm}
\end{tabular}
\end{minipage}
\end{table*}

\begin{table*}
\centering
\caption{Log of IR spectroscopic observations.} 
\begin{minipage}{\linewidth}
\renewcommand{\thefootnote}{\thempfootnote}
\renewcommand{\tabcolsep}{1.4mm}
\begin{tabular}{llllcccll}
\hline 
JD\footnote{JD (245 0000+).} & Date & Epoch\footnote{Days after explosion.} & Telescope/ & Range & Resol. & Slit width & Spectroscopic & 
Observers \\ 
 &  &     & Instrument &  ($\mu$m) & ($\lambda/\Delta\lambda$)  & (arcsec)   & standard & \\ \hline
2714.5   & 2003 Mar 15 & 137 & IRTF/SpeX & 0.81--2.42 & 1200 & 0.5 & HD 199217 & R.D. Joseph, J.T. Rayner \\ 
2777.5   & 2003 May 17 & 200 & IRTF/SpeX & 0.80--2.42 & 2000 & 0.3 & HD 205314  & J.T. Rayner\\ 
   &        &     &           & 2.84--4.20 & 1500 & 0.5 & \quad \quad ''   & \\
2843.5   & 2003 Jul 22 & 266 & IRTF/SpeX & 0.80--2.42 & 2000 & 0.3  & HD 194354 & R.D. Joseph, 
         J.T.  Rayner \\ 
   &        &     &           & 2.86--4.14 & 1500 & 0.5  & \quad \quad '' & \\
2891.5   & 2003 Sep 8  & 314 & IRTF/SpeX & 0.80--2.42 & 2000 & 0.3  &  HD 194354 & J.T. Rayner \\ 
2958.5   & 2003 Nov 14 & 381\footnote{Bad seeing: thick cirrus most of the night, 
         $A_K$ extinction ranging from $\sim 1$ mag (when spectrum taken) to opaque.} & 
         IRTF/SpeX & 0.93--2.42 & 750 & 0.8 & HD 194354 &  R.D. Joseph, J.T. Rayner \\ \hline
\vspace{-1cm}
\end{tabular}
\end{minipage}
\end{table*}

With Lick/Kast we used the 452/3306 grism and the 300/7500
grating, together with $180 \times 1200$ and $210 \times 1200$ pixel 
CCD arrays, each having a
0\farcs798 pixel$^{-1}$ scale.  With Keck/ESI we used the 175 lines mm$^{-1}$
grating and $32.3^\circ$ blaze grisms, with a $2048 \times 4096$ CCD
array having a 0\farcs12--0\farcs17 pixel$^{-1}$ scale.  With WHT/ISIS we
used the R316R grating and the EEV4290 MARCONI2 detector with a 0\farcs20
pixel$^{-1}$ scale.  With NOT/ALFOSC we used Grism 5
(0.5--1.025~$\mu$m) and the Loral $2048 \times 2048$ pixel array (CCD7, now
retired) with a 0\farcs19 pixel$^{-1}$ scale.  With Keck/LRIS we used the
400/3400 grism and the 400/8500 grating equipped with the
$4620 \times 4096$ and $1000 \times 2245$ pixel CCD arrays, respectively, 
having scales of 0\farcs211 and 0\farcs135 pixel$^{-1}$. For the
Kast, ESI, ISIS and LRIS spectra the position angle of the slit was
generally aligned along the parallactic angle to reduce differential
light losses (Filippenko 1982). For the NOT spectrum, the slit was
positioned at PA 181$^{\circ}$ in order to minimise contamination from the
bright nearby star (2MASS J20344320+6007234) lying about 9\farcs0 WNW from
the supernova.  Table~5 gives details of the optical spectroscopic
observations. Standard stars used are also listed, together with 
slit widths and position angles.  \\

\begin{figure*}
\vspace{10.62cm} 
\includegraphics{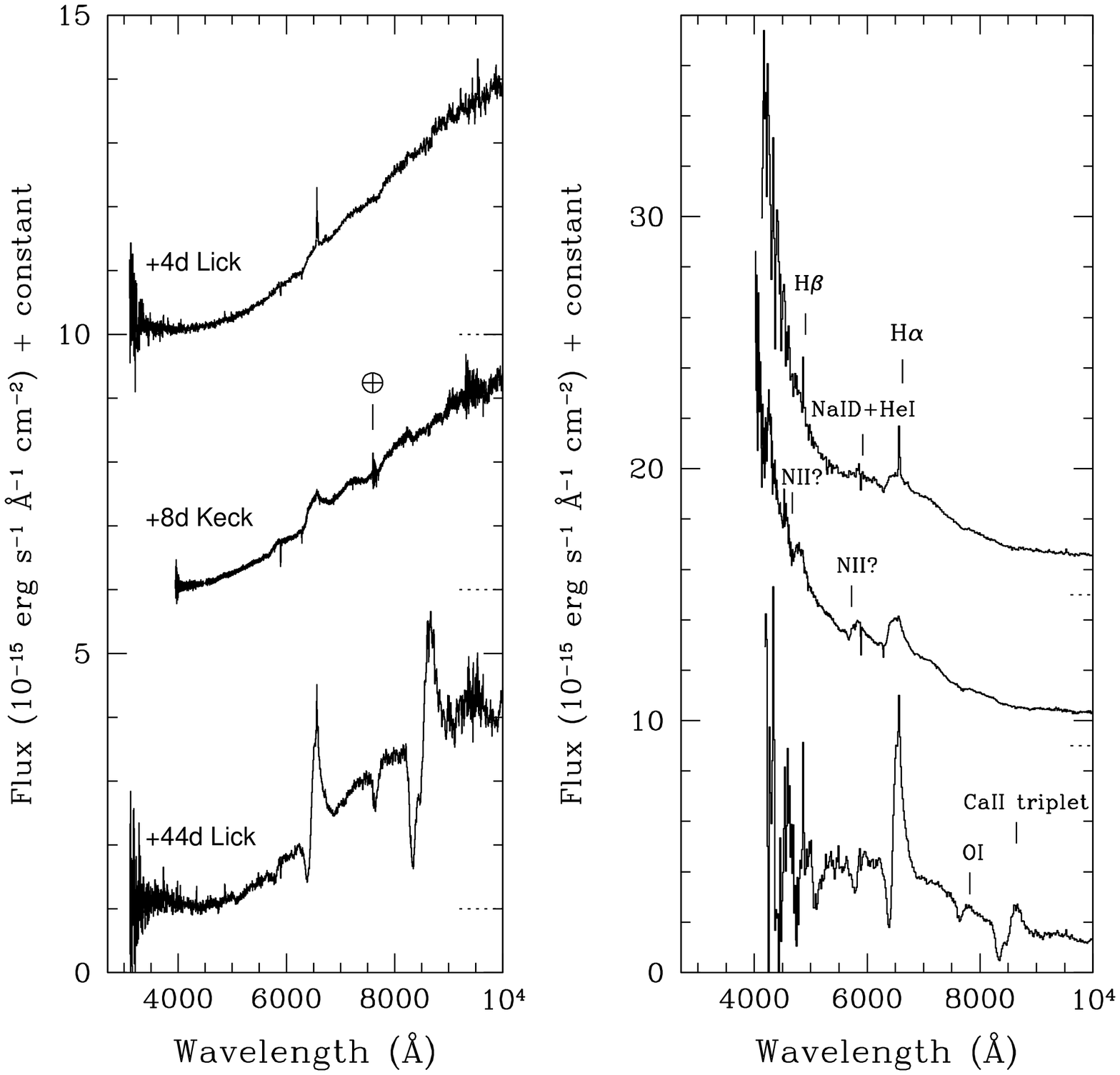}
\caption[] {Evolution of the optical spectra of SN~2002hh during the
photospheric phase.  The left and right panels show the spectra
(respectively) before and after dereddening (see Section 3.2).  The
spectra have been displaced vertically for clarity (the zero-flux
levels are indicated by the dashed lines).  No correction has been
made for redshift. The narrow emission spikes in the H$\alpha$ (6563
\AA) and H$\beta$ (4861 \AA) line profiles are discussed in Section
\ref{extinct}. Line identifications are discussed in Section \ref{optid}. 
The dereddened spectra have been truncated to remove very low S/N regions 
toward the blue. Residuals from removal of telluric lines are marked.}
\vspace{-0.5cm}
\end{figure*}

\begin{figure*}
\vspace{10.62cm}
\includegraphics{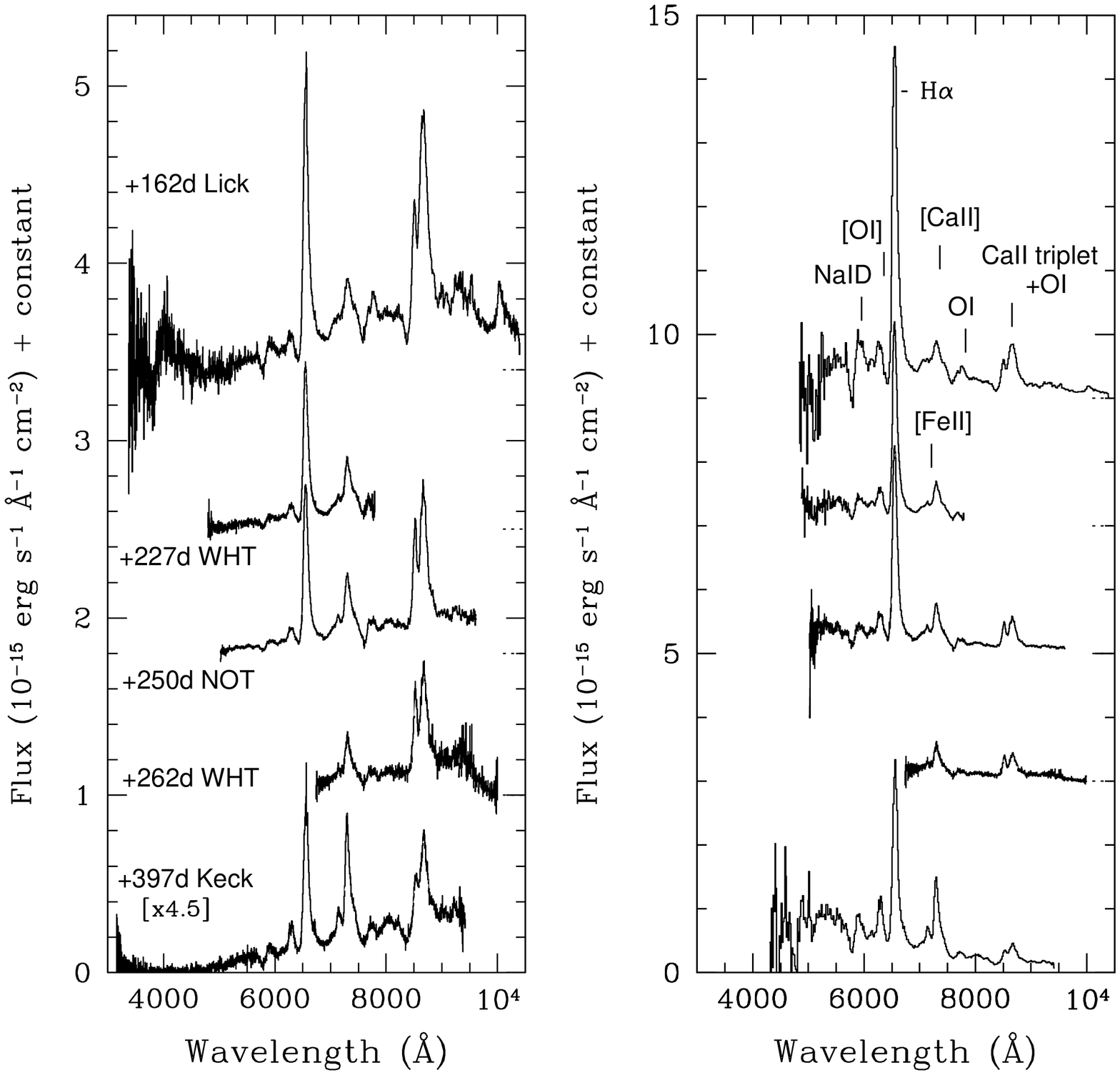}
\caption[] {As in Fig.~6, but showing the evolution of the optical spectra 
of SN~2002hh during the nebular phase.}
%\label{}
\end{figure*}

The spectra from Kast, ESI and LRIS were reduced using standard
techniques as described by Li et al. (2001) and references
therein. Flatfields for the red CCD were taken at the position of the
object to reduce NIR fringing effects. The spectra were corrected
for atmospheric extinction and telluric bands (Bessell 1999; Matheson
et al. 2000a), and then flux-calibrated using standard stars observed
at similar airmass on the same night as the SN (see Table~5).
Approximate spectral resolutions were derived from the widths of 
night-sky lines.
The wavelength calibration is accurate to about $\pm$0.2~\AA\, for
Kast, from about $\pm$0.02~\AA\, to $\pm$0.06~\AA\, (depending on the
order) for ESI, and about $\pm$0.08~\AA\, for LRIS.  The ISIS
spectra were reduced using standard procedures within FIGARO
(Shortridge 1991).  The wavelength calibration is accurate to about
$\pm$1~\AA.  The ALFOSC spectrum was reduced using standard procedures
within {\sc iraf}; the wavelength calibration is accurate to
$\pm$3~\AA. \\

\begin{figure*}
\vspace{22.5cm} \includegraphics{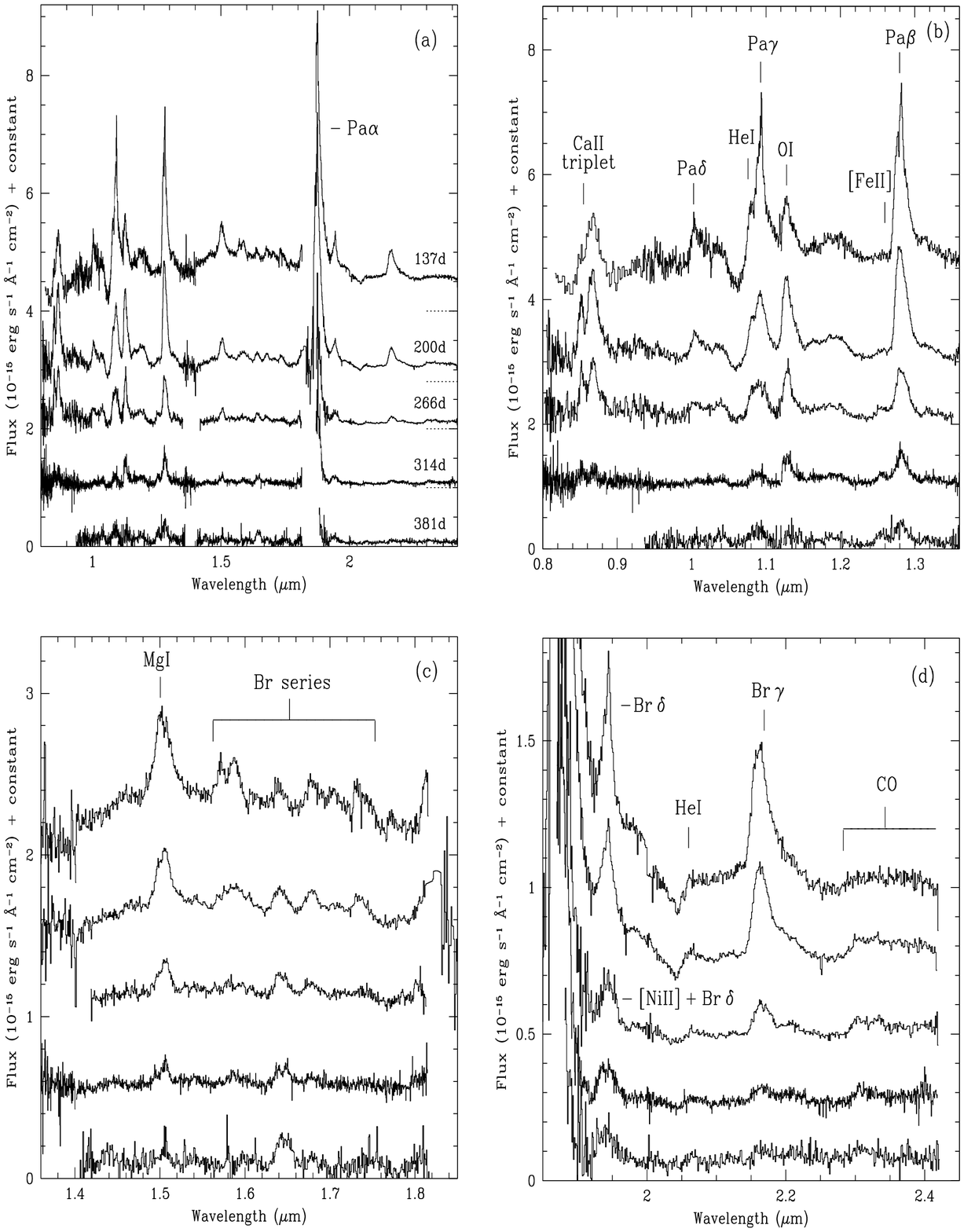}
\caption[] {Panel~(a): Evolution of the $IJHK$-band IR spectra of
SN~2002hh taken at IRTF during the period 137 to 381 days
post-explosion.  The spectra have been displaced vertically for
clarity (the zero-flux levels are indicated by the dotted lines).  No
correction has been applied for redshift or reddening. Panels~(b-d):
Detailed subsections of Panel~(a) showing, respectively, spectral
ranges 0.8--1.36~$\mu$m, 1.36--1.85~$\mu$m and 1.85--2.45~$\mu$m.  The
detailed evolution of the Pa\,$\alpha$ 1.875~$\mu$m line is shown
separately in Fig.~9.}
%\label{}
\end{figure*}

Final fluxing of all except the 397~d spectrum was achieved using
near-contemporary $VRI$ photometry taken with KAIT at Lick
Observatory (see Table~2).  $VRI$ transmission functions were formed
taking into account the atmospheric transmission function for Lick
Observatory, the KAIT imaging filters, and the CCD quantum efficiency.
These were then applied to the SN~2002hh spectra and to a model
spectrum of Vega.  The resulting optical spectra were integrated and
the total fluxes compared with those of Vega for each $VRI$ band to
derive spectra-based $VRI$ magnitudes.  These were then compared with
contemporary $VRI$ photometric magnitudes derived by interpolation of
the light curves.  Corrections ranged from a factor of $\sim$0.6 to
$\sim$2. \\

 It was found that the variation
with wavelength could be reasonably described with linear functions.
Linear fits were therefore made to the set of $VRI$ correction factors
for each epoch. For each waveband point the effective wavelength was
first determined from $\int \lambda F_{\lambda} \,d\lambda/\int
F_{\lambda} \,d\lambda$.  The resulting linear function was then used
to scale and tilt the corresponding supernova spectrum.  The blue
limit of the ALFOSC spectrum (250~d) is 5020~\AA\ and consequently
does not encompass the full width of the $V$-band transmission
function.  Therefore only the $R$ and $I$ transmission functions were
used to derive the flux correction function for this epoch.  The
wavelength coverage of the ISIS spectra is such that on days +227 and
+262 only the $V$~band and $I$~band (respectively) are fully spanned by
the spectra. Therefore, in these cases only single correction factors
were applied (0.97 for the +227~d spectrum, and 0.7 for the 
+262~d spectrum). We judge the accuracy of the
final fluxing of the flux-corrected optical spectra to be 
$\sim\pm10$\%.  For the 397~d spectrum, no contemporary photometry was
available and so no flux correction was attempted.  The fluxing error
in this case could be as much as 30\%.  \\

The optical spectra are
shown in Figs.~6 (photospheric phase) and 7 (nebular phase).  Each
figure comprises 2 panels. The left and right panels show the spectra
respectively before and after dereddening. The dereddening is
discussed in subsection~3.2.  Optical line identification is discussed
in Section \ref{optid}.\\

\subsection{IR spectroscopy}\label{irspec}
IR spectra were obtained using the 3.0~m NASA Infrared Telescope
Facility (IRTF) on Mauna Kea (Hawaii), equipped with SpeX, a
medium-resolution 0.8--5.5~$\mu$m cryogenic spectrograph and imager
(Rayner et al. 2003). We used both the 0.8--2.4~$\mu$m cross-dispersed
(SXD) and the 1.9--4.1~$\mu$m cross-dispersed (LXD) grating modes which
provide resolution of up to $R = \lambda/\Delta\lambda = 2500$.  
The spectrograph contains an Aladdin-3 $1024 \times 1024$ pixel 
InSb array, with a 0\farcs15 pixel$^{-1}$ scale.  $JHK$-band 
spectra were acquired at five epochs,
spanning 137--381 days post-explosion.  In addition, $I$-band spectra
were obtained at 4 epochs (137--314~d), and $L'$-band spectra were
obtained at 200~d and 266~d.  See Table~6 for the log of IR
spectroscopic observations.\\

\begin{figure}
\vspace{7.5cm} \includegraphics{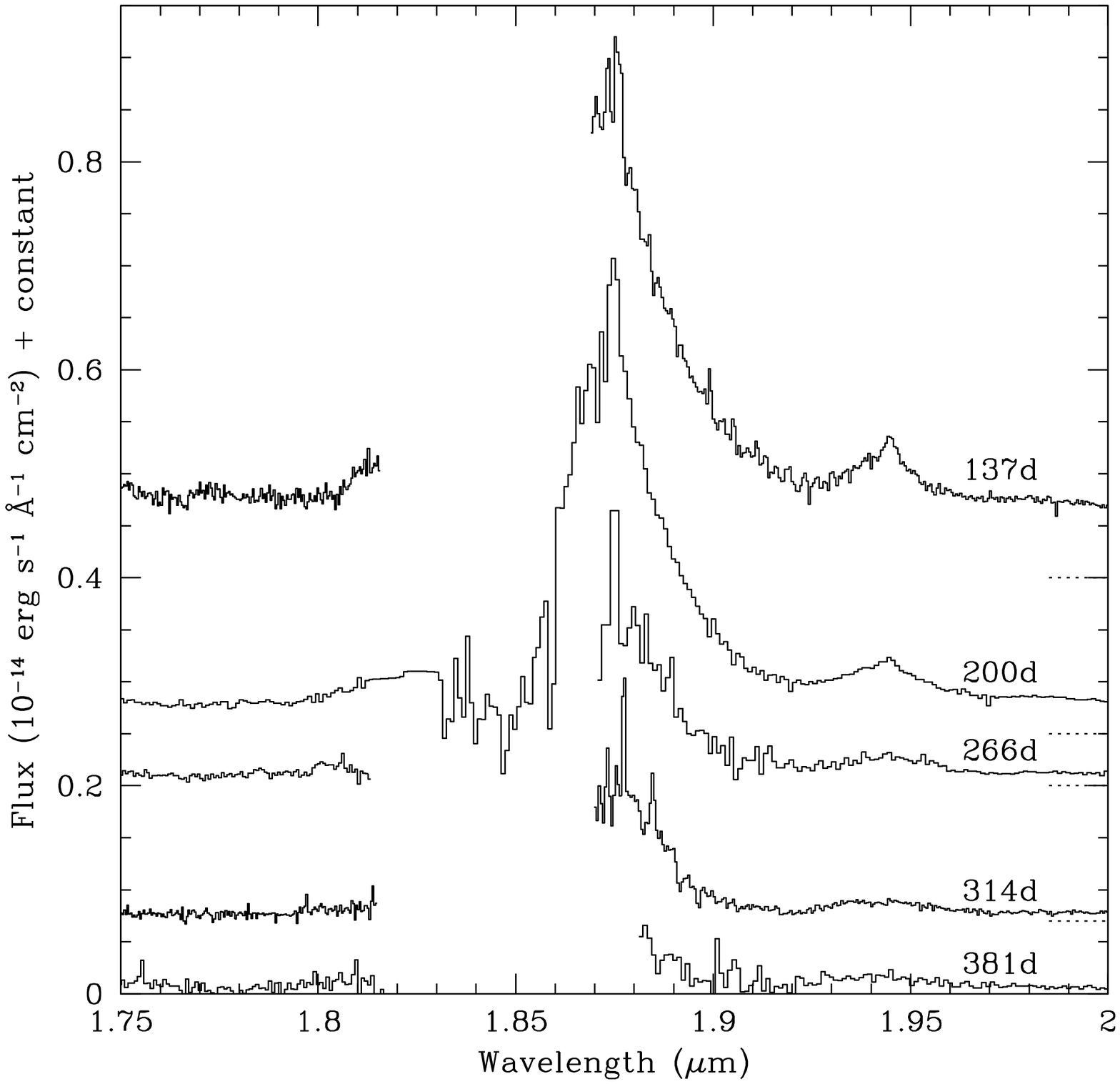}
\caption[] {Evolution of the Pa\,$\alpha$ line from the IR spectra 
shown in Fig.~8. The spectra have been displaced vertically for clarity 
(the zero-flux levels are indicated by the dotted lines).
The spectra have not been corrected for redshift or reddening.} 
%\label{}
\end{figure}

\begin{figure}
\vspace{7.3cm} \includegraphics{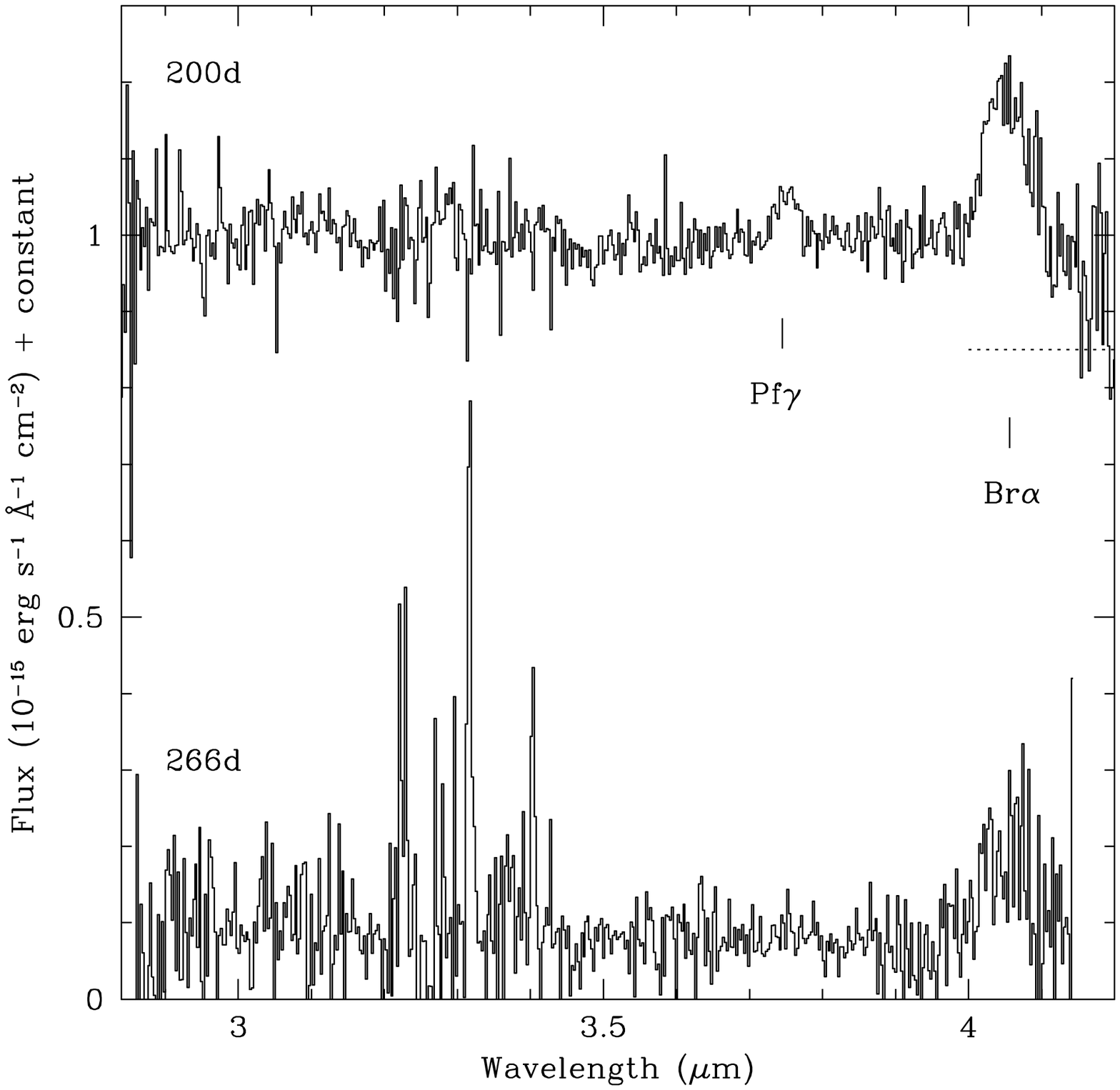}
\caption[] {$L$-band IR spectra of SN~2002hh taken, respectively, 
at 200~d and 266~d post-explosion.  The 200~d spectrum has been
displaced vertically for clarity (the zero-flux level is indicated by
the dotted line).  No correction has been applied for redshift or
reddening.}
%\label{}
\end{figure}

The spectra were reduced using {\sc FIGARO} (Shortridge 1991).  Most of the
sky emission was removed by subtraction of successive pairs of
exposures taken at two nod positions.  The pairs were then coadded.
The spectral orders were individually straightened, and residual sky
emission lines removed.  The spectra were optimally extracted (Horne
1986) and the effects of cosmic rays and bad pixels removed.
Wavelength calibration was by means of Argon arc spectra and is
accurate to better than $\pm$0.0010~$\mu$m. The telluric standards are
listed in Table~6.  \\

Final fluxing of the spectra was by means of the contemporary IR
photometry (see Table~4). Transmission functions were formed to represent the
combined transmission of the Mauna Kea atmosphere and the SpeX imaging
filters.  A model atmosphere (IRTRANS) for a water vapour column of
1.6~mm and an airmass of 1.5 was used to provide the atmospheric
transmission.  The filters are the Mauna Kea filter set.  Each filter
function was multiplied by the atmospheric transmission function to
provide net transmission functions in each band. These were then
applied to the SN~2002hh spectra and to a model spectrum of Vega. The
resulting spectra were integrated and the total fluxes compared with
those of Vega for each band to derive spectra-based magnitudes.  These
were then compared with contemporary photometric magnitudes.  For the
SXD spectra the dispersion over all epochs in the differences between
the spectroscopic and photometric values was about 0.3 mag.
This is a respectable value, given the narrowness of the slit and the
differences in conditions between the measurements of the supernova
and standards. \\ 

It was also found that at any given epoch a systematic
difference existed between the spectroscopic and
photometrically derived magnitudes. We therefore applied corrections
to the SXD spectra to match them to the contemporary photometry.  
The corrections applied were as follows: day~137 ($\times$1.43),
day~200 ($\times$0.8), day~266 ($\times$1.03), 
and day~314 ($\times$1.11).  In
the $L$~band the required corrections were day~200 ($\times$0.51) and
day~266 ($\times$0.71).  We judge the accuracy of the final fluxing to
be $\sim\pm10$\% for the $IJHK$~band spectra and $\sim\pm20$\%
for the $L'$~band spectra. For the 381~d spectrum, no
contemporary photometry was available and so no flux correction was
attempted; the fluxing error in this case could be as much as 50\%.
The $IJHK$-band spectra are shown in Fig.~8, while the detailed
evolution of the Pa\,$\alpha$ 1.875~$\mu$m line is shown separately in
Fig.~9.  The $L'$-band spectra for epochs 200~d and 266~d are shown in
Fig.~10.  IR line identification is discussed in Section \ref{optid}.\\

The NIR spectroscopic coverage is among the most complete ever
achieved for a supernova at late times.  Indeed, SN~2002hh is the
first supernova beyond the Local Group for which $L$-band spectra
have been acquired. The temporal coverage of the Pa\,$\alpha$ line
(Fig.~9) is also unsurpassed for a supernova at late times.  The
strong atmospheric absorption in this region means that this line has
seldom been observed in Type~II SNe.  \\

\section{Redshift and extinction}\label{extinct}
\subsection{Redshift}

Strong, narrow Na~I~D 5890,~5896 \AA\, and K~I 7699~\AA\ absorption
was detected in the early-time spectra. In Fig.~11 we show this
spectral region at high resolution on day~8 obtained with Keck/ESI.  
This reveals two sets of Na~I~D lines corresponding to velocities of 
$-20\pm10$~km s$^{-1}$ and $+110\pm10$~km s$^{-1}$.  
The K~I line has components at $-15\pm15$~km
s$^{-1}$ and $+113\pm10$~km s$^{-1}$.  The Galactic coordinates of
SN~2002hh are {\it l}~$=95.68^{\circ}$ and {\it b}~$=+11.67^{\circ}$.  
We suggest that the blueshifted component is due to absorption within 
the Milky Way, perhaps within the Perseus Arm. It is unlikely that the 
+110/113~km s$^{-1}$ absorptions are local, given the kinematic distribution 
of the Milky Way (Martos et al. 2004) and the Galactic longitude of
SN~2002hh. \\

Narrow emission lines of H$\alpha$, H$\beta$ (4861 \AA), [N~II]
(6548, 6584 \AA) and possibly [S~II]
(6717, 6731 \AA) are present in some of the spectra, 
showing redshift velocities consistent with the derived redshift of
$+110$~km s$^{-1}$.  The narrow lines in the
6300--6800~\AA\ region from 4~d to 397~d are shown in detail in Fig.~14
in the next Section. During the earlier phase, narrow lines are present 
on days 4 and 44 at comparable intensity, fade by 162~d, and have vanished by
227~d. However, the lines are also completely absent at 8~d. A
somewhat broader slit was used on the days when the narrow lines were
apparent (see Table~5) and so we suspect that the cause is incomplete
background subtraction.  H~II regions 420 and 421 (Bonnarel, Boulesteix
\& Marcelin 1986) lie within 3\farcs0 of the supernova, and may be
responsible for the narrow emission.  By 397~d, narrow lines have
re-appeared. They are unresolved at a resolution of 285 km s$^{-1}$ and are
again centred at about $+110\pm10$~km s$^{-1}$. They are more than a
factor of 10 weaker than at 44~d.  Inspection of the 2D frames again
suggests that the origin of the narrow lines is due to incomplete
background subtraction. While a relatively narrow 1\farcs0 slit was
used at 397~d, the much weaker ejecta emission by this epoch is
probably the reason that residual narrow lines from the background
have become apparent.  That the narrow emission lines show redshift
velocities consistent with the $\sim$+110 km s$^{-1}$ derived from the
Na~I~D and K~I absorptions may indicate a physical association
between the H~II regions and the line-of-sight absorbing material.\\

We adopt $+110\pm10$~km s$^{-1}$ as the redshift of the SN
centre-of-mass, although neither the narrow absorption lines nor
narrow emission lines necessarily have any physical connection with
the supernova. This value is larger than the redshift of the host
galaxy ($+48\pm10$~km s$^{-1}$); the difference is caused
by the rotation of NGC 6946 (see Walsh et al. 2002).  \\

\begin{figure}
\vspace{7.3cm} \includegraphics{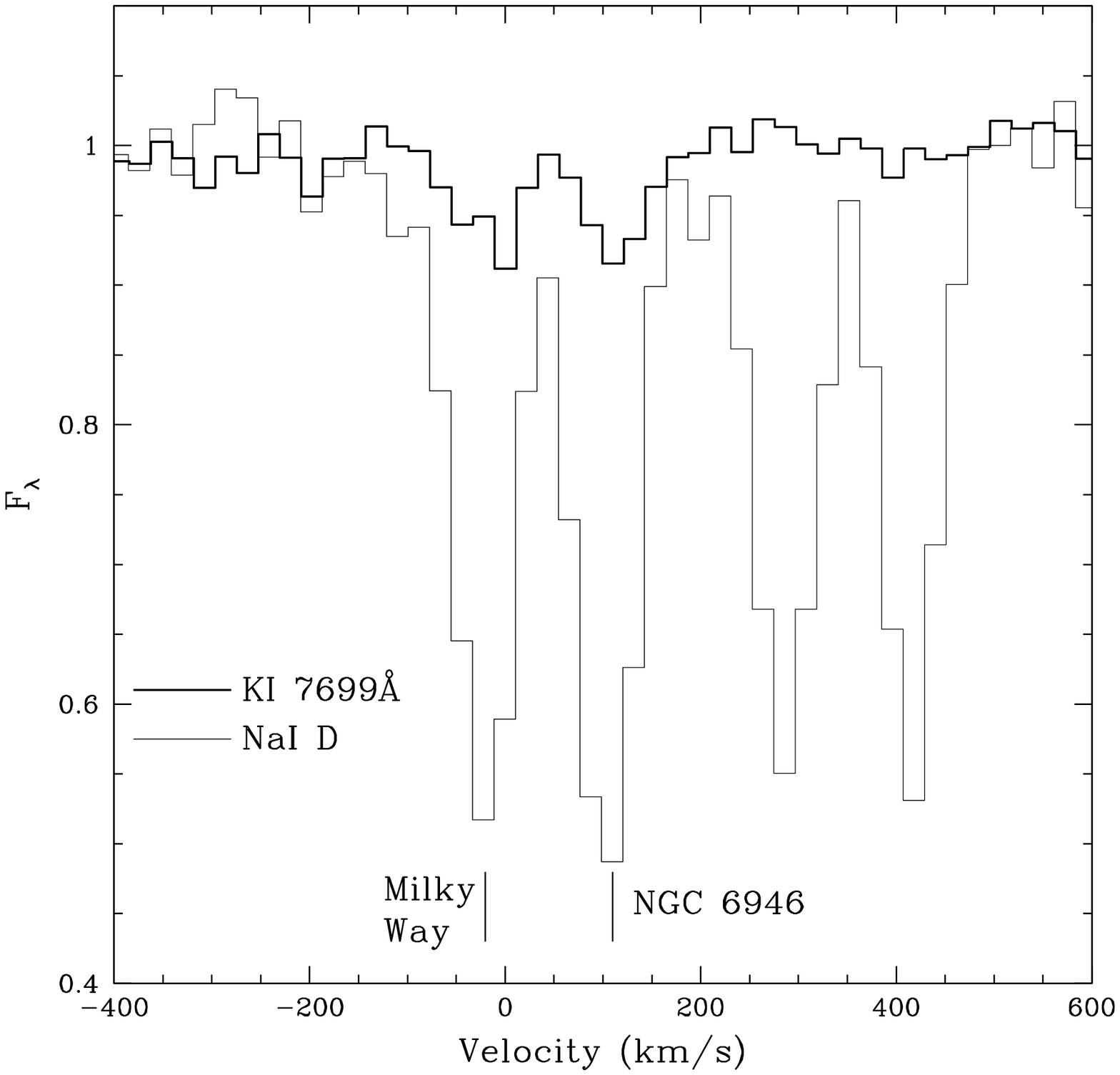}
\caption[] {Interstellar absorption lines toward SN~2002hh at
+8~days. The thick (upper) line is K~I 7699~\AA\ and the thin (lower)
line is the Na~I~D doublet. The spectra are plotted in velocity space
relative to the rest frames of the K~I and Na~I~D1 lines,
respectively. The Na~I~D2 line can also be seen at the right. 
Each line
has two components produced by absorption in, respectively, the Milky
Way and NGC~6946.  The lines are unresolved at a resolution of
75~km s$^{-1}$.}
%\label{}
\end{figure}

\subsection{Extinction}
As mentioned in the Introduction, the extinction to SN~2002hh is very high.  
To estimate and correct for the extinction, we compared the spectra and
light curves of SN~2002hh with those of SN~1999em. The justification
for this was that the light curves of both supernovae indicate that
both events were normal Type~IIP. We therefore assumed that the two
events were identical, and endeavoured to match their coeval optical
spectra (epoch $\sim$+7~days) by adjusting the amount of dereddening
for SN~2002hh.  We adopted the Cardelli et al. (1989) extinction law, 
initially with $R_V=3.1$.  For the SN~1999em extinction we took $A_V=0.31$ mag 
(Baron et al. 2000) and for the explosion epoch we adopted JD2451475.4 
(1999 Oct. 23.9), which is the weighted mean of the estimates of Baron 
et al. (2000), Hamuy et al. (2001), Leonard et al. (2002a) and Elmhamdi 
et al. (2003).  The SN~1999em spectrum was also scaled by a factor of 
4 to allow 
for the difference in distance (cf. Leonard et al. 2003).  However, we found
that no value of $A_V$ would provide a satisfactory match between the
spectra.  We therefore allowed both $A_V$ and $R_V$ to vary. From
this, we found that satisfactory matches could be obtained with
$A_V$(SN~2002hh)$=5.2\pm0.2$ mag and $R_V=1.9\pm0.2$. \\

\begin{figure}
\vspace{7.3cm} \includegraphics{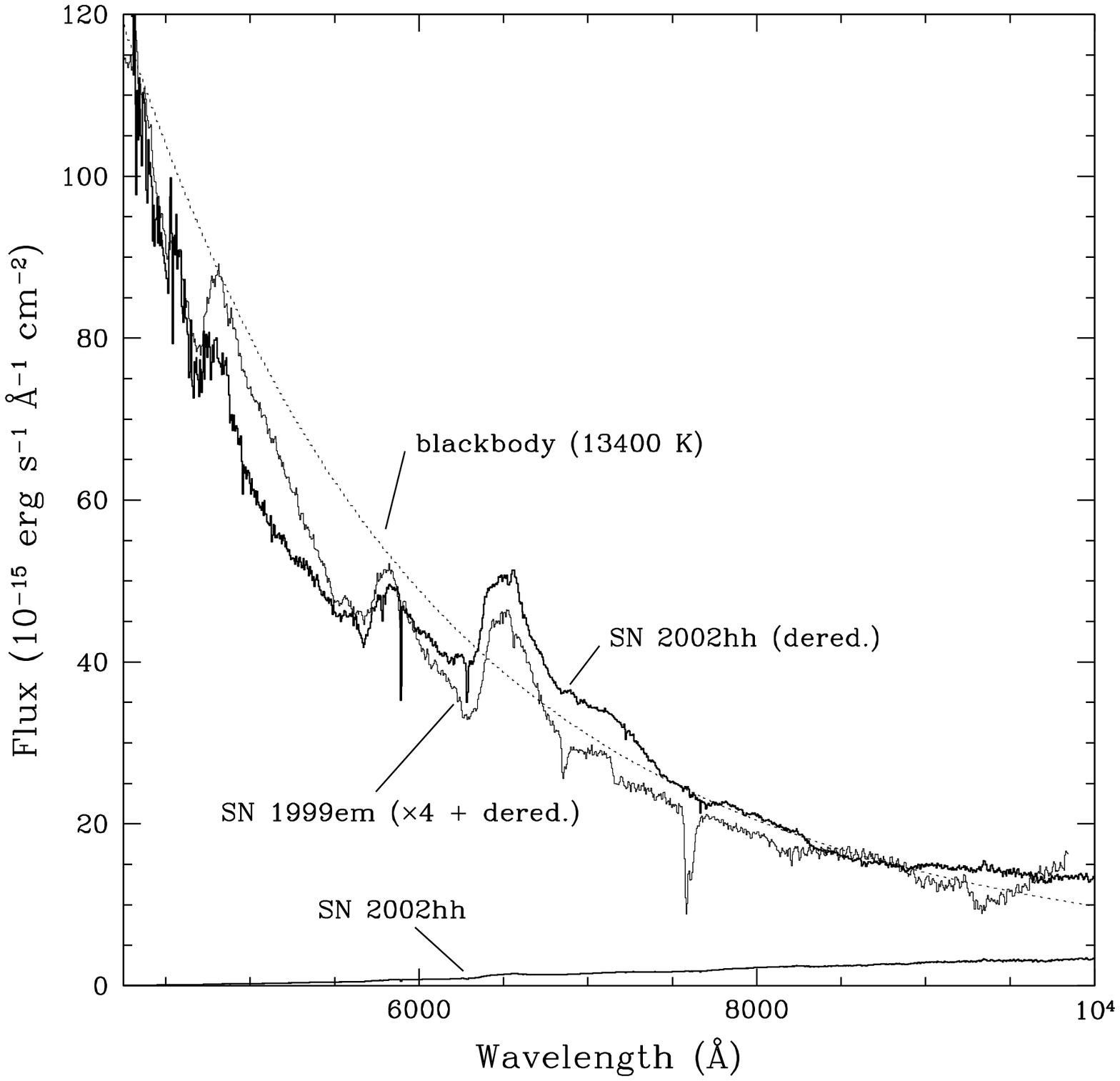}
\caption[] {Comparison of coeval spectra of SN~2002hh and SN~1999em
taken at about +7~days post-explosion.  The SN~2002hh spectrum has
been dereddened according to the 2-component extinction model {\it
viz}. $A_V=3.3$ mag, $R_V=3.1$ plus $A_V=1.7$ mag, $R_V=1.1$
(see text for details). The SN~1999em spectrum has been dereddened
with $A_V=0.31$ mag, $R_V=3.1$, and scaled by $\times$4 to correct for
the difference in distance between the two SNe.  The extinction law of
Cardelli et al. (1989) was used in both cases. Also shown is a
blackbody curve for a temperature of 13,400~K. The faint red spectrum
near the bottom is the original un-dereddened spectrum of SN~2002hh
plotted on the same scale.  This illustrates the very large extinction
correction that has been applied.  In spite of this, the dereddened
spectra of the two SNe appear remarkably similar.  }
%\label{}
\end{figure}

\begin{figure}
\vspace{7.8cm} \includegraphics{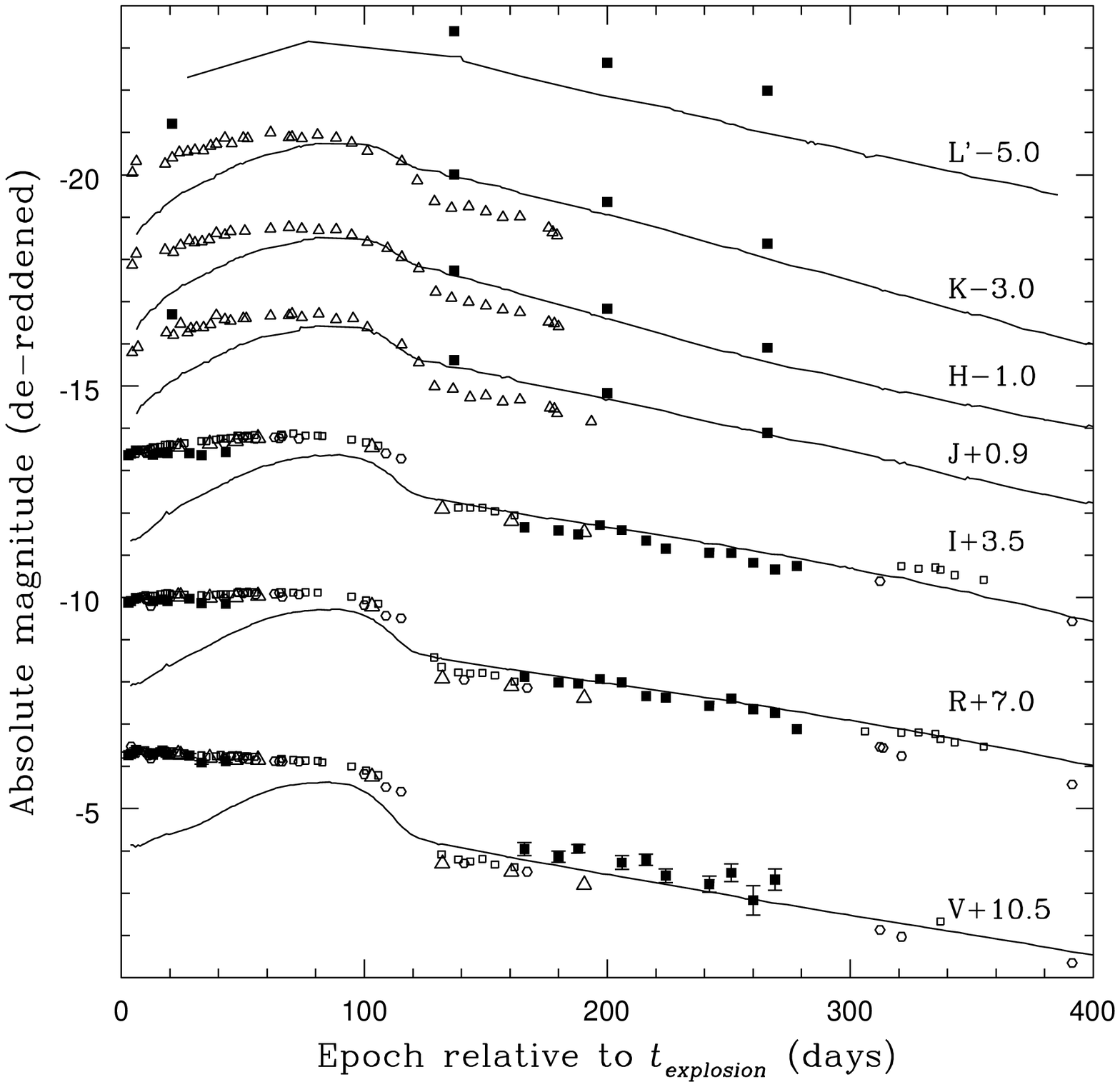}
\caption[] {Dereddened absolute magnitude optical and NIR light curves
of SN~2002hh (filled squares) compared with those of the
Type~IIpec SN~1987A (lines) and the Type~IIP SN~1999em (open squares --
Leonard et al. 2002a; triangles -- Hamuy et al. 2001, see text for details; 
circles -- Elmhamdi et al. 2003). For clarity, the light curves have been displaced
vertically by the amounts indicated at the RHS. See text for further
details. Note that the early SN~2002hh IR points are J and K at +21~d
(see also text).}
%\label{}
\end{figure}

As a check on the above extinction estimate, we examined the narrow
Na~I~D and K~I interstellar absorption lines (Fig.~11). At such a high
extinction (see above), it is recognised (e.g., Munari \& Zwitter 1997)
that the equivalent width (EW) of Na~I~D is of little value since the
lines are too close to saturation. That this is the case here is
confirmed by the EW ratio of the Na~I 5890~\AA\ to 5896~\AA\ lines,
which is 1.2 for the Milky Way and 1.3 for NGC~6946 (toward
SN~2002hh) --- that is, both are close to the total saturation ratio of 1.1
and far from the unsaturated value of 2.0.  However, the K~I line is
expected to remain unsaturated for much higher extinctions (Munari \&
Zwitter 1997). We obtain K~I line EW values of 0.131 $\pm 0.021$~\AA\
and 0.146 $\pm 0.026$~\AA\ for the Milky Way and NGC~6946,
respectively.  From the EW vs. $E(B-V)$ curve of Munari \& Zwitter (1997),
we deduce corresponding $E(B-V)$ values of 0.50 and 0.57 mag.  The
former is reasonably consistent with the $E(B-V) = 0.34$ mag value from
the far-IR maps of Schlegel et al. (1998). For $R_V \approx 3$, the latter
is consistent with the $A_V = 1.2^{+1.3}_{-1.7}$ mag which Barlow et
al. (2005) derive from the extinction profiles of Holwerda et
al. (2005). However, Munari \& Zwitter caution that the derived value 
of $E(B-V)$
may be an overestimate if the absorption line is formed by unresolved
multiple components. Alternatively, the presence of a dense
``dust pocket'' might not be traced by the absorption lines, leading
to an underestimate of $E(B-V)$. \\

 Notwithstanding these comments, the
total $E(B-V) = 1.07$ mag corresponds to only $A_V = 3.3 \pm 0.3$ mag
($R_V=3.1$). Taken at face value, this suggests that the remaining
$A_V=1.9\pm0.2$ mag of the total 5.2 mag extinction was unrevealed by
the K~I lines. This additional extinction may be associated with the
line-of-sight H~II regions and so did not show the correlation with
the K~I absorption EW which occurs in H~I regions. \\

To investigate this
further, we dereddened the day~+8 SN~2002hh spectrum using a
2-component extinction where the first component, due to the combined
effect of the ISMs of the Milky Way and NGC~6946, was assigned
$A_V=3.3$ mag and $R_V=3.1$.  A second component was then added, with
$A_V$ and $R_V$ as free parameters. A satisfactory match was obtained
with $A_V=1.7\pm0.2$ mag and $R_V=1.1$ for the second component.  The
dereddened spectra of SNe~1999em and 2002hh are shown in Fig.~12.
Also shown is the spectrum of a blackbody corresponding to $T = 13,400$~K
and an expansion velocity of 6,800~km s$^{-1}$, with the expansion beginning
7~days earlier.  While the value obtained for $R_V$ is small, such
values have been invoked in the past in multi-component extinction
models (see Wang et al. 2004 and references therein).  It implies that the
``dust pocket'' grain size is small compared with the average value
for the local ISM.  Also, as already indicated, the EW-derived ISM
extinction may be an overestimate.  A smaller ISM extinction would
allow a larger ``dust pocket'' extinction and hence a larger value for
$R_V$, with the upper limit being the $R_V=1.9$ obtained in the
single-component case.  \\

Wang (2005) has suggested that the anomalous reddening seen toward a number 
of Type~Ia supernovae could be produced by {\it normal-sized} grains within 
a dusty circumstellar cloud. In this ``light-echo'' scenario, not only is 
light scattered out of the beam, but also we see light that has been scattered 
into our line-of-sight.  Thus, in addition to the scattering cross-section
we also need to consider the (wavelength-dependent) albedo of the
grains. Wang finds that a dust cloud of inner radius $10^{16}$cm would
produce a significant reduction in $R_V$. However, we suggest that
survival of grains at such a small distance from most supernovae of
any type is rather unlikely. In Section 5.4 below, we find that the
evaporated dust-free cavity of SN~2002hh has a radius of
$\sim10^{17}$cm ($8750$~A.U.).  Dust at this large distance would have
only a modest effect on $R_V$ (see Fig. 4 of Wang 2005).  We therefore
remain persuaded that the most plausible explanation for the 
small-value $R_V$ component is due to a line-of-sight dust pocket where the
grain size is atypically small. \\

As noted in the Introduction, Pooley \& Lewin (2002) derived a 
column density of $N_H = 10^{22}$~cm$^{-2}$ from their X-ray data. Adopting
$A_V=0.53\times10^{-21}$[$N$(HI)+$2N$(H$_2$)] (Bohlin, Savage \& Drake
1978) we obtain $A_V = 5.3$ mag, in good agreement with the values derived
from spectral matching and the K~I interstellar lines.\\

In Fig.~13 we show the dereddened absolute magnitude optical and IR
light curves of SN~2002hh and SN~1999em, as well as those of the
peculiar Type~II SN~1987A.  To the data for SN~2002hh we have added
the +21~d $J$ and $K$ points of Meikle et al. (2002), which, according to
our 2-component extinction model, correspond to dereddened absolute magnitudes 
of $M(J)= -17.6$ and $M(K)= -18.2$.  The SN~1999em optical magnitudes are from 
Hamuy et al. (2001), Leonard et al. (2002a) and Elmhamdi et al. (2003). 
The IR magnitudes of SN~1999em are instead from Fig.~1 of Hamuy et al. (2001): 
no tabulated data were presented there, nor were these subsequently published.
The optical magnitudes for SN~1987A are from Hamuy et 
al. (1988) and Suntzeff et al. (1988), $JHK$ magnitudes
from Bouchet et al. (1989) and Bouchet \& Danziger (1993), and $L'$
magnitudes from Catchpole et al. (1988) and Whitelock et al. (1988).
Dereddening of the SN~1999em and SN~2002hh light curves was by means
of, respectively, the single-component and 2-component extinction laws
described above.  The adopted distances are 5.9~Mpc for 
SN~2002hh (see above) and 11.7~Mpc for SN~1999em 
(Leonard et al. 2003).  For SN~1987A we
used a distance of 51.4~kpc and an $A_V=0.6$ mag (Romaniello et
al. 2000, 2002).  Its magnitudes were dereddened using the Cardelli et
al. law with $R_V=3.1$.\\

Since the $A_V$ and $R_V$ values used for dereddening the SNe~1999em
and 2002hh light curves were obtained by matching their day~+7 optical
spectra, we checked that a match was also obtained between the $VRI$
light curves of the two supernovae at around +7~d. In the $R$ and $I$
bands a good match (difference $<$0.1 mag) was indeed found.  However,
in the $V$ band, the SN~2002hh magnitude was about 0.3 mag fainter
than that of SN~1999em, in spite of the good match obtained between
the spectra in this wavelength range. We attribute this discrepancy to
differences in the true effective wavelengths (used in the light curve
dereddening) of the $V$ band between the two SNe due to the large
difference in extinction together with the intrinsically steep
spectral slope. We find that by shifting the effective wavelength of
the $V$ band for SN~2002hh to the blue by 200~\AA\ a good match is
obtained at +8~d. The same effective $VRI$ wavelengths are used for
dereddening the optical light curves at all epochs. This may explain,
at least partly, why the light-curve match between SNe~1999em and
2002hh becomes poorer at later epochs. \\

The optical light curves of SNe~1999em and 2002hh are generally
well-matched in both the photospheric and nebular phases. The
early-time difference with respect to SN~1987A is not surprising given
its rather different nature. Nevertheless, the radioactive tails of
all three supernovae match reasonably well in the optical. In the NIR,
the early-time difference between SN~1987A and SN~2002hh is similar to
that seen in the optical.  However, in contrast to the optical region,
by 266~d the $HK$ luminosities of SN~2002hh slightly exceed those of
SN~1987A. Moreover the excesses increased by 314~d.  In addition, the
$L'$ luminosity of SN~2002hh is significantly brighter than that of
SN~1987A from as early as 137~d, and this excess also shows some sign
of increasing with time. This ``IR excess'' is discussed below.  \\

The NIR light curves of SNe~1999em and 2002hh, while consistent with
having similar shapes, display what appears to be an approximately
constant relative offset of 0.5--0.7 mag spanning 10 to
200~days, with SN~1999em being the fainter of the two.  Given the
good optical match between SNe~1999em and 2002hh at early times, and
between all 3 supernovae during the radioactive tail, this large NIR
mismatch between SN~1999em and 2002hh is puzzling. The occurrence of
this mismatch both at early times and in the $J$ band probably rules out
emission from warm dust in SN~2002hh (see below) as the source of this
difference. We have tried to match the light curves by leaving both 
$R_V$ and $A_V$ as free parameters, but no satisfactory match 
was found for the $VRIJHK$ photometry, not even with the two-component 
dust model: while the optical light curves agree, the IR light curves
are always offset by the amount specified above. We therefore suspect 
that a spurious systematic offset is present in the SN~1999em NIR data, 
possibly related to the IR calibrations. \\

The above analysis confirms that the extinction toward SN~2002hh is
high and that the bulk of it occurs within NGC~6946. There is some
indication that the extinction within NGC~6946 comprises two
components.  One of these has an exceptionally small $R_V$ and is not
traced by the K~I absorption lines. It may be due to a localised dust
cloud associated with line-of-sight H~II regions.  \\

\subsection{Optical/IR luminosity and the $^{56}$Ni mass}\label{oirlum}

As explained above, we derived the total extinction by assuming that
SN~1999em and SN~2002hh were alike in the optical region at about
1~week post-explosion.  The fact that the subsequently dereddened
$VRI$ absolute light curves also agree to as late as $\sim$270~d
tends to support the validity of this assumption.  We also find good
agreement with the late-time $VRI$ tails of SN~1987A.  From these
matches we conclude that the mass of $^{56}$Ni ejected by SN~2002hh
was similar, about $0.08 \pm 0.01$~M$_{\odot}$. \\

A more direct approach to estimating the mass of $^{56}$Ni ejected 
is by determining the total late-time luminosity during the radioactive tail.  
In the earlier part of the nebular phase, Type~II SNe become
largely optically thin to UV/optical/IR (UVOIR) radiation, 
but remain optically thick
to the high-energy gamma rays from the radioactive decay.  Thus, in
principle, the late-time UVOIR light curve can allow us to estimate
the amount of $^{56}$Ni produced in the SN explosion.  The problem in
attempting such an analysis for SN~2002hh is the enormous extinction
correction required. One possible concern is that we have no useful
data shortward of the $V$ band which cuts off at about 4300~\AA.
However, given the late epochs considered, this may only result in a
small underestimate.  In SN~1987A at 203~d the luminosity peaked in the
$I$ band, while less than 2\% of the total was emitted shortward of
4300~\AA\ (Catchpole et al. 1988, Fig.~4). Thus, if the similarity of
SN~2002hh to SN~1987A is maintained at these shorter wavelengths, we
can safely ignore the lack of $UB$-band data.  Nevertheless, even in
the $I$ band we estimate an extinction reducing the flux by a factor of
$8 \pm 1$. \\

We have estimated the $^{56}$Ni mass in SN~2002hh using our late-time
$VRIJHKL'$ photometric data obtained at 200~d and 266~d to derive
optical/IR (OIR) luminosities. The optical and IR magnitudes (see Tables~2
and 4) were converted into fluxes at the effective wavelengths of the
filters using the flux calibrations of Bessell (1979) and those of
Bersanelli, Bouchet \& Falomo (1991), respectively.  Adopting the
two-component extinction model ($A_{V1}$=$3.3\pm0.3$ mag, $R_{V1}=3.1$;
$A_{V2}$=$1.7\pm0.2$ mag, $R_{V2}=1.1$; see above), the
fluxes were then dereddened according to the Cardelli et al. (1989)
extinction law.  The total fluxes at the two epochs were then derived
by fitting spline curves (Press et al. 1992) to the dereddened
$VRIJHKL'$ fluxes and integrating over the wavelength range
4288--44993 \AA\, using a trapezoidal rule (the range is fixed by the
limits of the filter functions used in the photometric observations)
(see Table~7, Col.~2).  Finally, the OIR luminosities were derived
from the total fluxes by adopting a distance of $5.9 \pm 0.4$ Mpc to
the host galaxy (see Table~7, Col.~3).  The reported uncertainties take into
account the internal errors from the spline fit plus the uncertainties
in both the distance and extinction values.  As discussed below, a
fraction of the IR luminosity may be due to an IR echo.  At 
200~d and 266~d the echo model described below suggests 
that around 10\%
of the late-time luminosity could have been due to an echo. \\

\begin{table}
\centering
\caption[]{Nebular OIR flux and luminosity of SN 2002hh.} 
\begin{minipage}{\linewidth}
\renewcommand{\thefootnote}{\thempfootnote}
\begin{tabular}{ccc} \hline
Epoch & $F_{\rm OIR}$\footnote{\hspace{-1mm}$^{,\,\,b}$ Error given in
brackets; see text.}  & $L_{\rm OIR}^b$  \\ 
(days) & (10$^{-10}$ erg s$^{-1}$ cm$^{-2}$) & (10$^{41}$ erg s$^{-1}$) \\ \hline 
200 & 0.442(0.053) & 1.84(0.25) \\
266 & 0.215(0.034) & 0.89(0.16) \\ \hline
\vspace{-0.85cm}
\end{tabular}  
\end{minipage}
\end{table}

The mass of $^{56}$Ni ejected from SN~2002hh is obtained from
$M(^{56}Ni)=M(^{56}Ni)_{87A}\times L_{02hh}/L_{87A}$ M$_{\odot}$,
where $M(^{56}Ni)_{87A}$ is the mass of $^{56}$Ni produced by
SN~1987A, $L_{02hh}$ is the OIR luminosity of SN~2002hh and
$L_{87A}$ is the coeval UVOIR luminosity of SN~1987A.  We adopt
$M(^{56}Ni)_{87A}=0.073\pm0.012$ M$_{\odot}$, which is the weighted
mean of values given by Arnett \& Fu (1989) and by Bouchet, Danziger
\& Lucy (1991), and includes uncertainties in distance and
extinction.  Thus, for SN~2002hh we obtain an ejected $^{56}$Ni mass
of $0.08\pm0.02$~M$_{\odot}$ (where the given error includes the
uncertainty in the distance and extinction of SN~2002hh plus the
uncertainty in the $^{56}$Ni mass of SN~1987A).  Reducing this by 10\%
to allow for the IR echo contribution, we obtain
$0.07\pm0.02$~M$_{\odot}$.  To within the errors, this is the same
value as obtained above from the similarity of the $VRI$ light curves.
The large uncertainty is primarily due to the high extinction
correction.\\

Similar coeval luminosities for the [Fe~II]~1.257~$\mu$m line are
obtained in SNe~1987A and 2002hh although the SN~2002hh values are of
low S/N (see Fig.~21 in Section~5).  Assuming a similar ionisation
degree and temperature for the two SNe and that most of the iron
results from the decay of $^{56}$Ni, this provides additional support
for our conclusion that similar masses of $^{56}$Ni were ejected by
the two SNe.

\section{Spectroscopic Behaviour}\label{optid}
\subsection {Overview}
\subsubsection{Photospheric phase}

During the photospheric phase, the 4~d and 8~d dereddened optical
spectra (Fig.~6, right) are characterized by a blue continuum with few
discrete features. Broad, low-contrast P-Cygni-like H$\alpha$ is
visible at 4~d, together with H$\beta$ 4861 \AA, He~I 5876~\AA, and
strong, narrow (interstellar) Na~I~D absorption (see Section 3.1).
Narrow emission lines of H$\alpha$ and H$\beta$ are also present and
are probably due to nearby H~II regions (see Section 3.1). By 44~d,
following the typical spectral evolution of Type~IIP SNe, the broad
H$\alpha$ and H$\beta$ P-Cygni emission had become much more prominent
while the blue continuum had faded significantly.  In addition, broad
P-Cygni lines of O~I 7773~\AA\, and the Ca~II triplet had appeared. \\

In the 8~d spectrum, features occur at about 4600 \AA\ and 5500 \AA.
Similar features are also visible in near-coeval spectra of SN~1987A,
SN 1999em and SN 1999gi (see Elmhamdi et al. 2003; Leonard et
al. 2002b).  For SN~1999em, Baron et al. (2000) suggest N~II $\sim
5679$ \AA\, (multiplet 3) as the origin of the second of these.
Recently, Dessart \& Hillier (2005) have presented a quantitative
spectroscopic analysis of the photospheric phase of Type~IIP SNe: from
synthetic fits to early-time spectra they confirm the Baron et
al. identification.  They conclude that the 4600 \AA\ and 5500
\AA\ features in SNe~1987A, 1999em and 1999gi are due, respectively,
to N~II $\sim 4623$ \AA\, (multiplet 5) and N~II $\sim 5679$ \AA\,
(multiplet 3).  We therefore propose that the 4600 \AA\ and 5500 \AA\
features in the 8~d spectrum of SN~2002hh are also due to the N~II
multiplets.  As pointed out by Dessart \& Hillier, this indicates an
overabundance of nitrogen in the outer layers of the progenitor star
prior to core collapse.

\subsubsection{Nebular phase}
In our investigation of the nebular-phase spectra (Figs.~7--10), we
were guided by the spectroscopic studies of SN~1987A performed by
Meikle et al. (1989; 1993) and Spyromilio et al. (1991). The optical
and NIR spectra are dominated by broad emission lines of neutral or
singly ionised species. In Tables~8 and 9 we list, for the optical and
NIR regions respectively, the identifications of the stronger
nebular-phase lines (uncorrected for redshift) together with their
intensities.  The evolution of the velocity widths for the most
prominent single emission lines is described in Table~10. In the
optical region, a weak flat continuum is present up to at least 262~d
(see Fig.~7, right). However, by the time the SN is recovered on day
397~d, a strong blue continuum has developed. A strong, persistent IR
continuum was also present in the nebular phase and this is discussed
in the next section. \\

\begin{table*}
\centering
\caption[]{Optical emission features observed on days 162 to 397 in SN~2002hh.} 
\begin{minipage}{\linewidth}
\renewcommand{\thefootnote}{\thempfootnote}
\renewcommand{\tabcolsep}{3mm}
\begin{tabular}{cccc} \hline
ID & Epoch (days) & $\lambda_{\it peak}$ (\AA) & Intensity (10$^{-13}$ erg s$^{-1}$ cm$^{-2}$) \\ \hline
[0~I] 6300, 6364 \AA\, & 162   & 6267 & 0.16 \\
                      & 227   & 6298  & 0.09 \\
                      & 250   & 6304  & 0.08 \\
                      & 262   & -- & -- \\
                      & 397   & 6300 & 0.03 \\ \\ 

H$\alpha$ 6563 \AA\,\footnote{The H$\alpha$ 6563 \AA\, intensity
excludes, when present, interstellar nebular emission (see the 162~d
and 397~d spectra).}
                      & 162   & 6543 & 1.50 \\
                      & 227   & 6545 & 0.86 \\
                      & 250   & 6546 & 0.73 \\
                      & 262   & -- & -- \\
                      & 397   & 6554 & 0.18 \\ \\ 

[Ca~II] 7291, 7323 \AA\,& 162  & 7303 & 1.00 \\
                       & 227  & 7298 & 0.43 \\
                       & 250  & 7303 & 0.54 \\
                       & 262  & 7305 & 0.39 \\
                       & 397  & 7298 & 0.17 \\ \\ 

O~I 7771~\AA\ + K~I 7676~\AA\,& 162  & 7763 & 0.82 \\
                             & 227  & 7678 & 0.18 \\
                             & 250  & 7770 & -- \\
                             & 262  & 7685 & 0.16 \\
                             & 397  & 7718 & 0.05 \\ \\ 

O~I 8446 \AA\,+ Ca~II 8498 \AA\,
%(\footnote{This feature is blended with the Ca~II components at 8542, 8662 \AA.}) 
 & 162 & 8505 & 0.77 \\
 & 227 & -- & -- \\
 & 250 & 8517 & 0.63 \\
 & 262 & 8520 & 0.50 \\
 & 397 & 8529 & 0.11 \\ \\

Ca~II 8542 \AA\,+ Ca~II 8662 \AA\,
 & 162 & 8676 & 2.44 \\
 & 227 & --   & -- \\
 & 250 & 8662 & 1.32 \\
 & 262 & 8672 & 0.99 \\
 & 397 & 8677 & 0.24 \\ \hline
\vspace{-1cm}
\end{tabular}  
\end{minipage}
\end{table*}

\begin{table*}
\centering
\caption[]{IR emission features observed on days 137 to 381 in SN~2002hh.} 
\begin{minipage}{\linewidth}
\renewcommand{\thefootnote}{\thempfootnote}
\renewcommand{\tabcolsep}{3mm}
\begin{tabular}{cccc} \hline

ID & Epoch (days) & $\lambda_{\it peak}$ ($\mu$m) & Intensity (10$^{-13}$ erg s$^{-1}$ cm$^{-2}$) \\  \hline
Ca II triplet 0.85-0.87~$\mu$m & 137 & 0.871 & 1.49 \\
                               & 200 & 0.867 & 3.74 \\ 
                               & 266 & 0.867 & 2.21\\
                               & 314 & -- & -- \\
                               & 381 & -- & -- \\ \\ 

Pa\,$\delta$ 1.005~$\mu$m + ? (\footnote{This feature is blended with
an unidentified feature around 1.03~$\mu$m.  The given intensity is
for the whole double-peaked feature at peak emission.}) & 137 & 1.003 & 0.92 \\
                         & 200 & 1.003 & 1.20 \\
                         & 266 &  --   &  -- \\
                         & 314 &  --   &  -- \\
                         & 381 &  --   &  -- \\ \\ 
               
Pa\,$\gamma$ 1.094~$\mu$m + He~I 1.083~$\mu$m & 137 & 1.093 & 2.57 \\
                         & 200 & 1.092 & 2.29 \\
                         & 266 & 1.090 & 1.51 \\
                         & 314 & 1.095 & 0.51 \\
                         & 381 & 1.094 & 0.52 \\  \\

O~I 1.129~$\mu$m & 137 & 1.128 & 1.23 \\
                 & 200 & 1.127 & 1.67 \\
                 & 266 & 1.130 & 0.88 \\
                 & 314 & 1.128 & 0.66 \\
                 & 381 &  --   &  -- \\ \\ 

Pa\,$\beta$ 1.282~$\mu$m & 137 & 1.281 & 3.60 \\
                 & 200 & 1.279 & 2.66 \\
                 & 266 & 1.279 & 1.40 \\
                 & 314 & 1.281 & 0.73 \\
                 & 381 & 1.281 & 0.62 \\ \\

Mg~I 1.503~$\mu$m & 137 & 1.501 & 0.87 \\
                  & 200 & 1.505 & 0.68 \\
                  & 266 & 1.507 & 0.36 \\
                  & 314 & 1.506 & 0.16 \\
                  & 381 & 1.505 & 0.08 \\ \\

Pa\,$\alpha$ 1.875~$\mu$m & 137 & 1.876 & -- \\
                         & 200 & 1.875 & 7.75 \\
                         & 266 & 1.875 & 3.70 \\
                         & 314 & 1.878 & 1.10 \\
                         & 381 &  --   &  -- \\ \\

Br\,$\delta$ 1.945~$\mu$m & 137 & 1.944 & 0.81 \\
                         & 200 & 1.944 & 0.76 \\
                         & 266 & 1.938 & 0.50 \\
                         & 314 & 1.945 & 0.34 \\
                         & 381 & 1.945 & 0.40 \\ \\

Br\,$\gamma$ 2.166~$\mu$m  & 137 & 2.162 & 1.08 \\
                          & 200 & 2.162 & 0.86 \\
                          & 266 & 2.162 & 0.32 \\
                          & 314 & 2.164 & 0.20 \\
                          & 381 &  --   &  --  \\ \\

Pf\,$\gamma$ 3.740~$\mu$m  & 200 & 3.750 & 0.29 \\ \\

Br\,$\alpha$ 4.051~$\mu$m  & 200 & 4.056 & 1.37 \\ 
\hline
\vspace{-0.8cm}
\end{tabular}  
\end{minipage}
\end{table*}

\subsubsection{Individual Elements}
{\it Hydrogen and Helium}\\

\noindent The late-time spectra (see Figs.~7, 8) are dominated by broad
H$\alpha$ emission together with lines of the Paschen and Brackett
series.  H$\beta$ is undetected.  Given a typical intrinsic
H$\beta$/H$\alpha$ intensity ratio plus the high extinction, this is
unsurprising.  Pf\,$\gamma$ can be seen seen in the day~200
2.85--4.15~$\mu$m spectrum (see Fig.~10).  Strong He~I~1.083~$\mu$m
and 2.058~$\mu$m are also present.  Both He~I features show
P-Cygni-like troughs which persist right up to to the last epoch.  As
the supernova evolved, the H~lines gradually weakened with respect to
other lines.  The H/He line profiles and their evolution are discussed
later. \\

\noindent{\it Oxygen}\\

\noindent We identify the strong 1.13~$\mu$m feature with the O~I
1.129~$\mu$m Bowen resonance fluorescence line.  Ly\,$\beta$ photons
pump the 1025 \AA\, triplet and the subsequent cascade feeds the
O~I~8446~\AA\ and 1.129~$\mu$m lines.  An alternative identification
is [S~I]~1.131~$\mu$m (the other [S~I] line at 1.082~$\mu$m would be
blended with He~I).  However, the O~I 1.129~$\mu$m identification is
confirmed by the presence of the strong, narrow feature at
$\sim$8500~\AA\, (see Fig.~7) which we identify with the O~I~8446~\AA\
line, blended with the Ca~II IR triplet.  The O~I 1.129~$\mu$m line
increased in prominence during the 137--314~d period with respect to
Pa\,$\gamma$ (see Table~9 for details).  In addition, its width
narrowed up to day~266. Its presence is important since, for the Bowen
mechanism to operate, there must have been intimate mixing of hydrogen
and oxygen. Other oxygen lines visible in our spectra include
[O~I]~6300, 6364 \AA\, and O~I~7771~\AA\, (probably blended with
K~I~7676~\AA), although the different components are not resolved.
The P-Cygni absorption may be affected by the uncorrected telluric
absorption at $\sim 7600$ \AA. \\

\noindent{\it Sodium}\\ 

\noindent By 162~d, the broad He~I 5876~\AA\ P-Cygni line had given way to a
P-Cygni feature due to Na~I~D. The presence of sodium emission is
confirmed by Na~I 2.206~$\mu$m emission, initially seen as a red wing to the 
Br$\gamma$ line.  In addition, the red wing of the O~I 1.13~$\mu$m feature 
is probably due to the Na~I~1.138~$\mu$m doublet. \\

\noindent{\it Magnesium}\\

\noindent We identify the strong 1.5~$\mu$m feature with
Mg~I~1.503~$\mu$m.  This dominated the $H$-window to day~266, but faded
quite rapidly thereafter. Its optical transition at 8807~\AA\, 
probably contributes to the red wing of the Ca~II triplet. \\

\noindent{\it Silicon}\\ 

\noindent At all five epochs a broad feature in the 1.16--1.22~$\mu$m
range is present.  The same feature occurred in SN~1987A and was
attributed to a blend of K~I(multiplet~6), Mg~I(6) and Si~I(4) by
Meikle et al. (1989).  However, Fassia, Meikle \& Spyromilio (2002)
noted that, given this multiple origin, the constancy of the shape
between days 112 and 1822 is remarkable and may argue against a blend
of different species. The virtual absence of [Si~I] 1.6068~$\mu$m
emission in SN~2002hh suggests a significantly lower silicon abundance
than was seen in SN~1987A (see section~5.2).  \\

\noindent{\it Calcium}\\ 

\noindent The Ca~II IR triplet is prominent in both the optical and IR
spectra, probably blended with O~I~8446~\AA, and possibly with
[C~I]~8727~\AA\ and Mg~I~8807~\AA. We also identify the [Ca~II]
7291, 7323 \AA\ feature.  [Fe II]~7388, 7452~\AA\, possibly may be 
contributing to its red wing.  The lack of prominent emission at
1.99~$\mu$m tends to rule out a significant contribution from
Ca~I~1.94, 1.98~$\mu$m. \\

\noindent{\it Iron, Cobalt, Nickel}\\ 

\noindent We identify [Fe~II]~1.257~$\mu$m and possibly
[Fe~II]~1.644~$\mu$m in the NIR plus a weaker feature in the optical
region due to [Fe~II]~7155 \AA.  [Fe~II]~1.257~$\mu$m became
increasingly prominent with time and by 381~d was quite significant
despite the low S/N of the spectrum (partly due to atmospheric
conditions; see note in Table~6). In the $1.4-1.8~\mu$m window, as the
Brackett lines faded, a prominent feature at $\sim$1.643~$\mu$m
emerged, probably due to [Fe~II]~1.644~$\mu$m. While
[Si~I]~1.645~$\mu$m may also have been contributing, there is little
sign of the other component of this doublet at 1.607~$\mu$m which
should be 40\% of the 1.645~$\mu$m intensity. We conclude that
[Si~I]~1.645~$\mu$m made at most a minor contribution. In
Section~\ref{oirlum} above, we have used the [Fe~II]~1.257~$\mu$m
line to estimate the mass of ejected $^{56}$Ni.\\

There is little sign of [Co~II]~1.547~$\mu$m emission.
In SN~1987A during the period 255--377~days the intensity ratio of the
features at 1.54~$\mu$m (a blend of [Fe~II] 1.533~$\mu$m and
[Co~II]~1.547~$\mu$m) and 1.64~$\mu$m was typically 0.25. Thus, given
the relatively low S/N of the 1.64~$\mu$m feature in SN~2002hh during
this period, it is likely that the 1.54~$\mu$m feature is buried in
the noise.\\

As Br\,$\gamma$ and Br\,$\delta$ faded it is clear that by 266~sd
another species was contributing to the feature at 1.94~$\mu$m.  By
the final epoch, this feature is the most prominent in this window.
In SN~1987A Meikle et al. (1989, 1993) attributed this to a blend of
Br\,$\delta$, Ca~I~1.94~$\mu$m and [Ni~II]~1.94~$\mu$m (plus a small
contribution from [Fe~II]~1.95~$\mu$m). They found that at 377~d the
[Ni~II] line contributed 30\% of the blend, increasing to 60\% by
574~d. However, any Ca~I~1.94~$\mu$m contribution should be
accompanied by a comparably strong Ca~I~1.98~$\mu$m feature.  The lack
of a strong emission feature at 1.98~$\mu$m in SN~2002hh tends to rule
out a significant contribution of Ca~I to the 1.94~$\mu$m blend during
the 266--381~d period, although it may have made a more significant
contribution at earlier epochs.  We conclude that by 266~d the
1.94~$\mu$m feature is a blend of [Ni~II]~1.94~$\mu$m and
Br\,$\delta$.\\

\begin{figure}
\vspace{7.3cm} \includegraphics{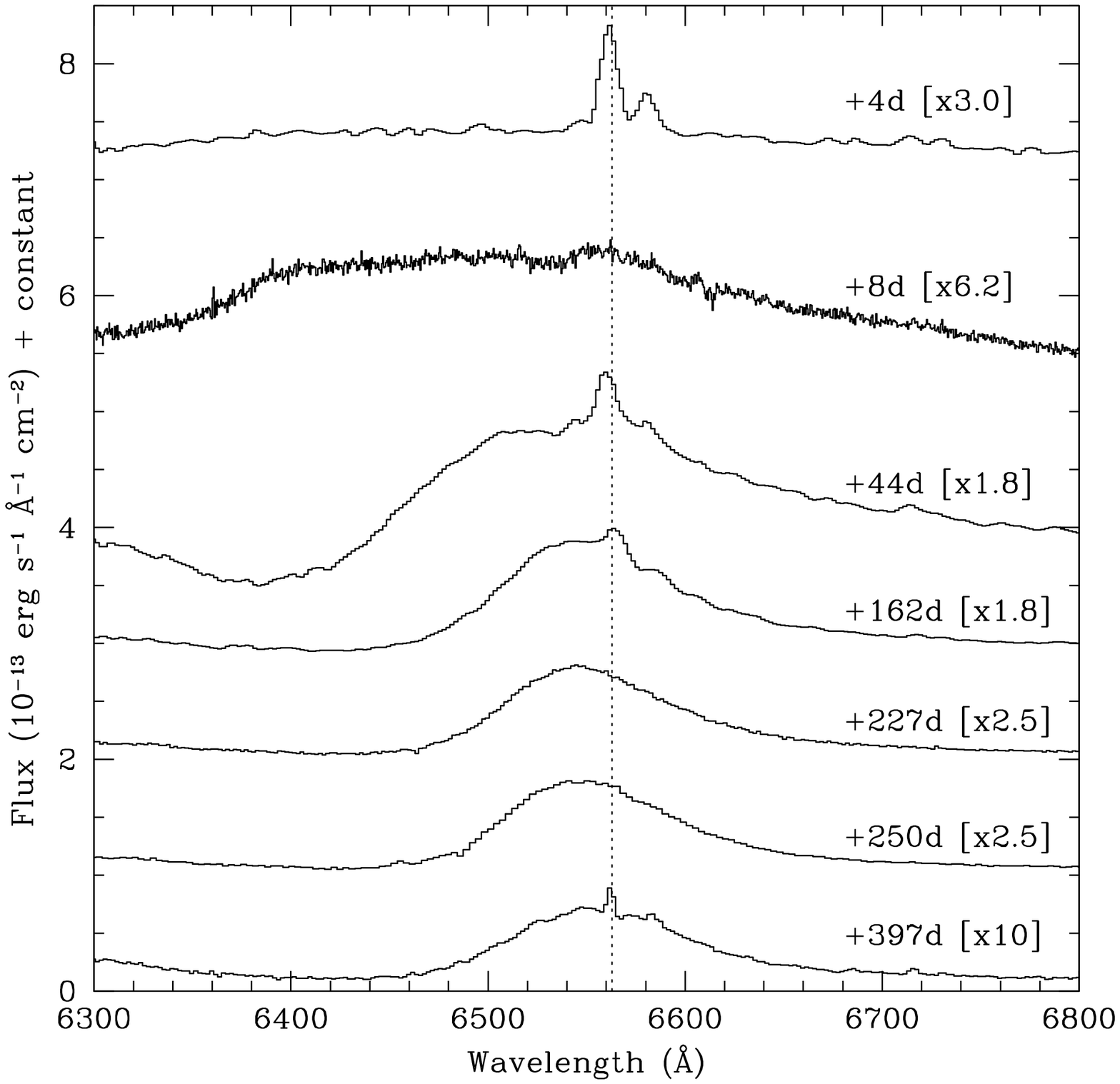}
\caption[] {Evolution of the H$\alpha$ emission-line profile during
the 4-397~d period. The spectra have been dereddened and
redshift-corrected, and have been displaced vertically for
clarity. The vertical dotted line corresponds to the adopted redshift
of 110 km s$^{-1}$ for the SN centre of mass (see text). Narrow emission lines
of H$\alpha$, [N~II] $\lambda\lambda$ 6548, 6584 and [S~II]
$\lambda\lambda$ 6717, 6731 from the H~II regions are also visible
(see Section \ref{extinct} for details).}
%\label{}
\end{figure}

\begin{figure}
\vspace{7.3cm} \includegraphics{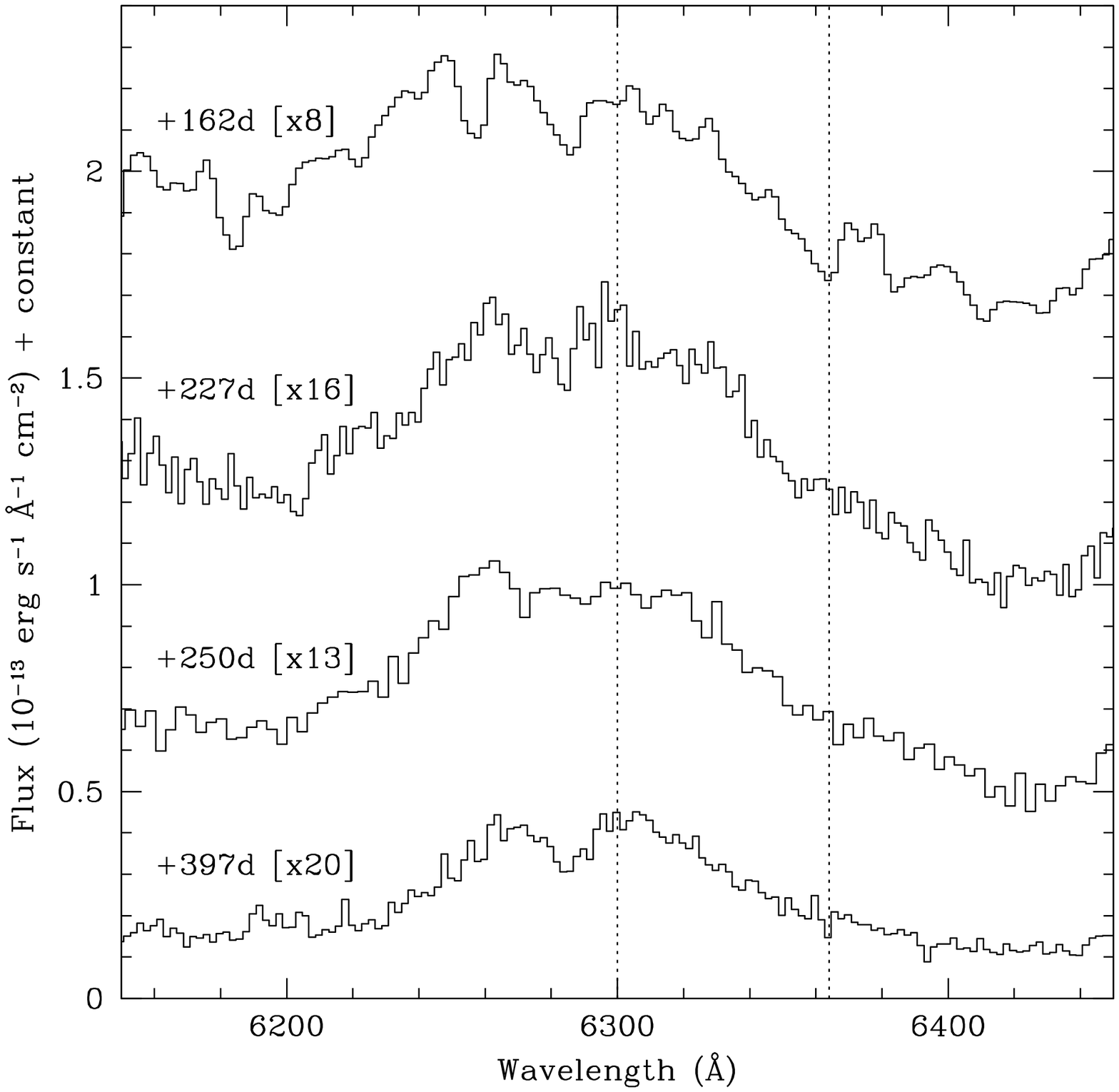}
\caption[] {As in Fig.~14, but showing the evolution of the O~I 
$\lambda\lambda$6300, 6364
emission-line profile during the nebular phase.}
%\label{}
\end{figure}

\noindent{\it Carbon Monoxide}\\ 

\noindent CO first-overtone emission is detected by 137~d.  
As time passes the longer-wavelength components fade faster as the
cooling ejecta depopulates the higher vibrational levels
(Spyromilio et al. 1989).  The presence of CO emission is
important as it may presage the condensation of dust in the ejecta
(see Introduction). The CO temporal evolution in the period 137-381~d 
is discussed in detail in Section \ref{CO}.\\
 
\begin{figure}
\vspace{7.3cm} \includegraphics{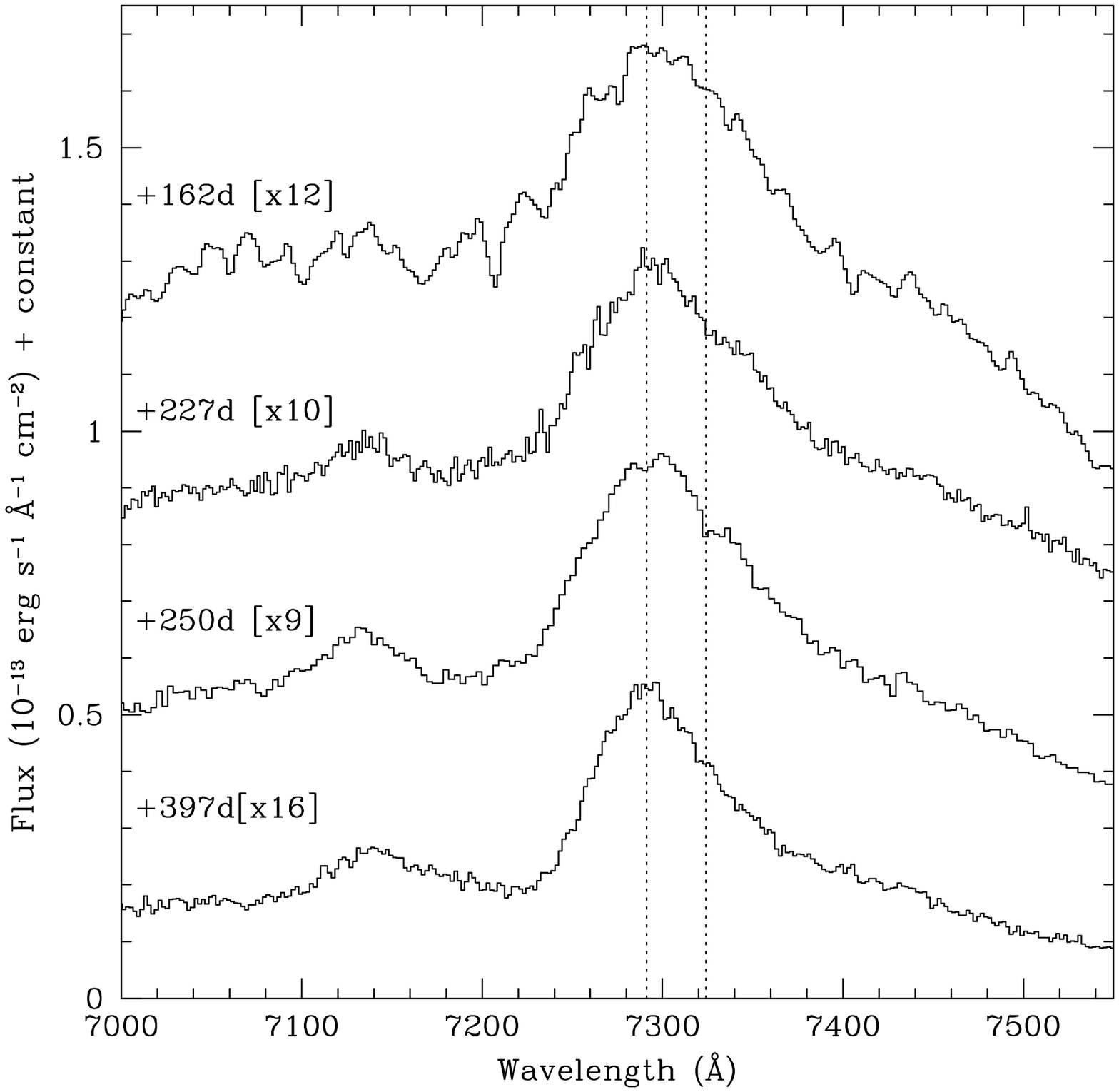}
\caption[] {As in Fig.~14, but showing the evolution of the [Ca~II] 
$\lambda\lambda$7291, 7323 emission-line profile during the nebular phase.}
%\label{}
\end{figure}

\begin{figure}
\vspace{7.3cm} \includegraphics{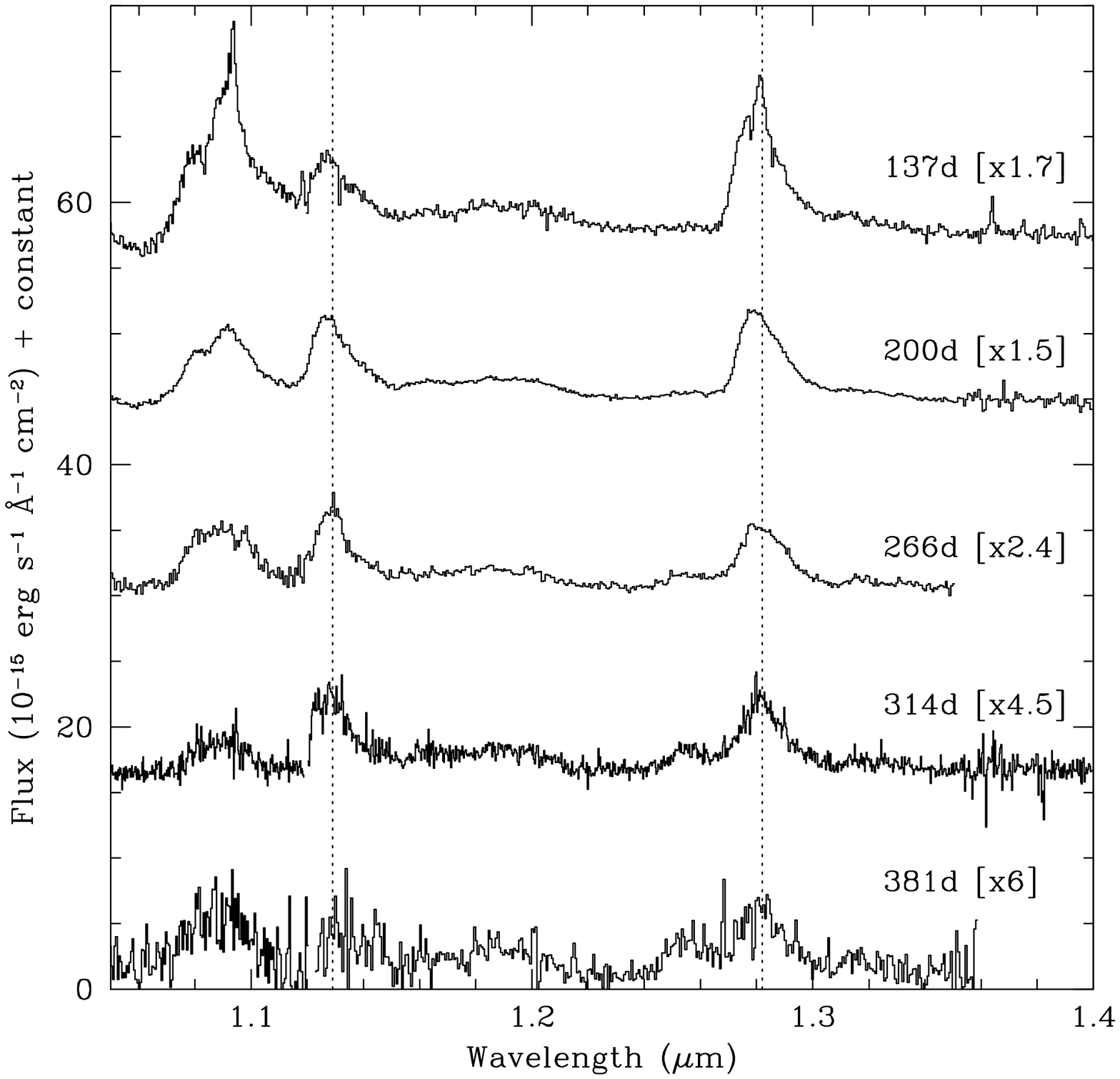}
\caption[] {Evolution of the O~I (1.129~$\mu$m)  and 
the Pa\,$\beta$ (1.282~$\mu$m) 
emission lines in SN~2002hh (days 137 to 381). All the spectra
have been dereddened and shifted to the rest frame.}
%\label{}
\end{figure}

\begin{figure}
\vspace{7.3cm} \includegraphics{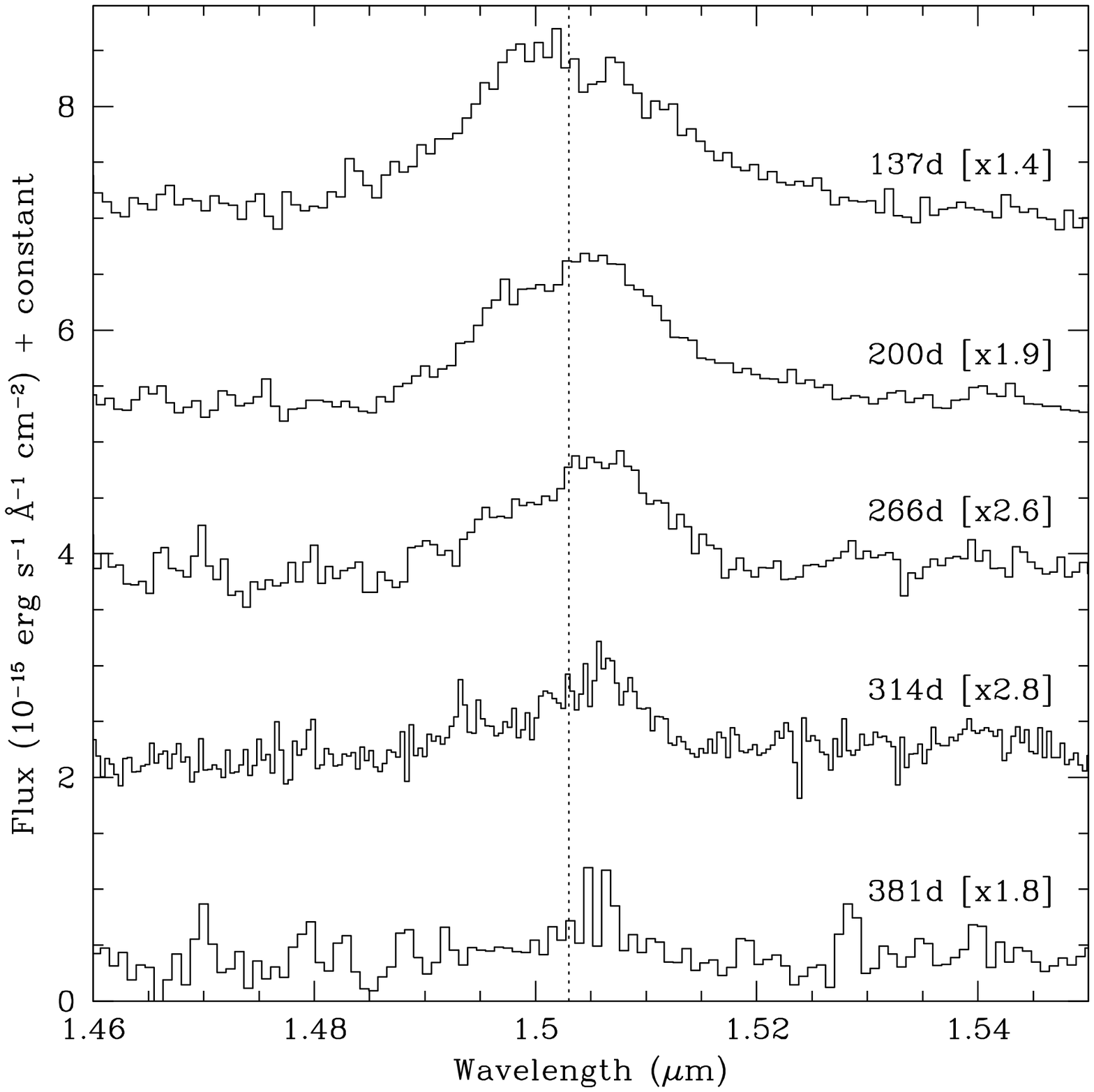}
\caption[] {As in Fig.~17 but showing the evolution of the Mg~I~1.503~$\mu$m emission line 
in SN~2002hh.}
%\label{}
\end{figure}

\begin{figure}
\vspace{7.3cm} \includegraphics{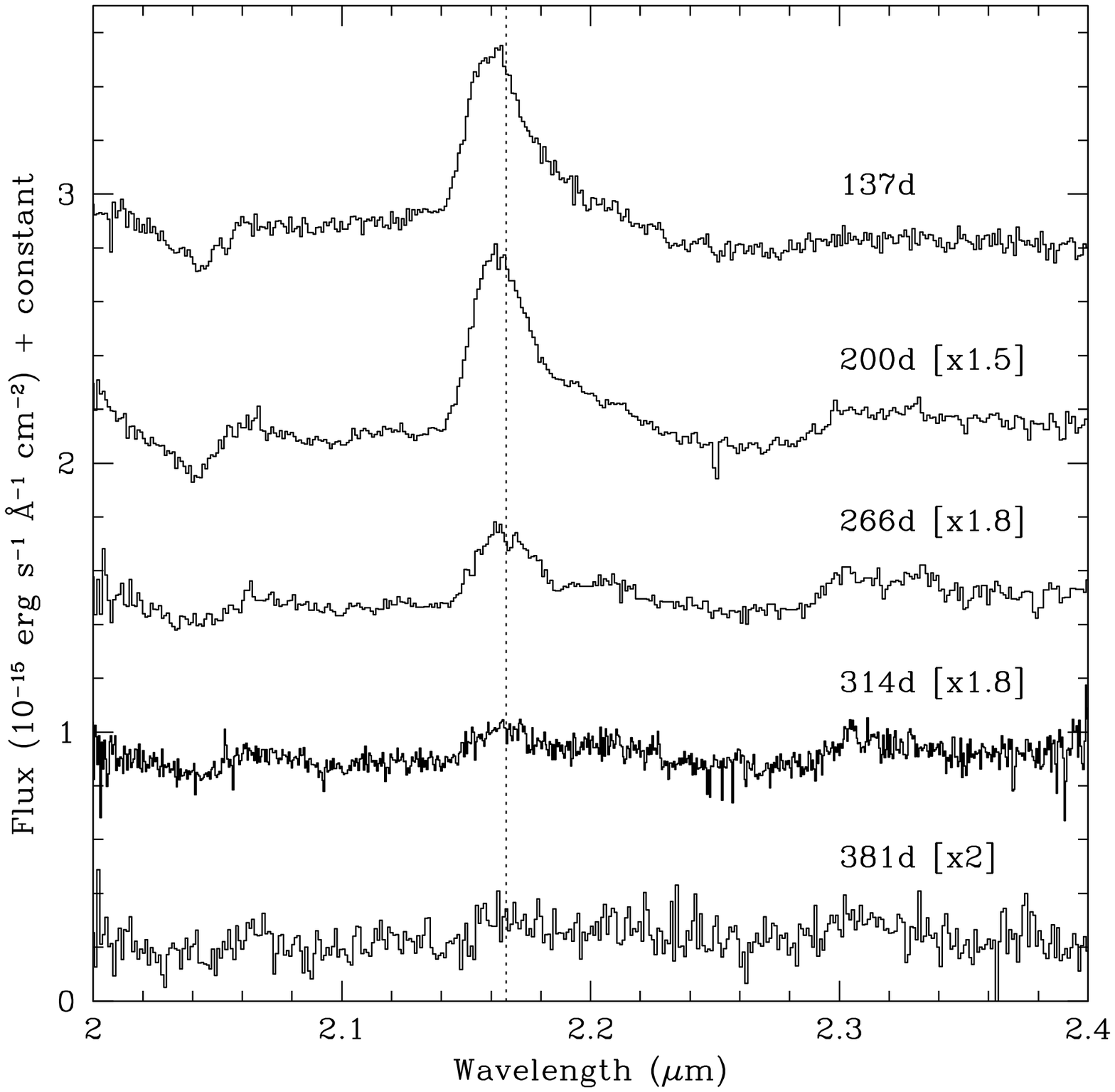}
\caption[] {As in Fig.~17 but showing the evolution of the Br\,$\gamma$ 2.166~$\mu$m 
emission line and the CO emission in SN~2002hh.}
%\label{}
\end{figure}

\subsubsection{Nebular line-profile evolution}\label{nebevol}
In Figs.~14 to 19 we show the evolution of the most prominent,
isolated lines.  In all these plots, the rest-frame wavelengths of the
lines are indicated by vertical dotted lines.  As the supernova
evolved, the H~lines gradually weakened with respect to other
lines. Due to difficulty in disentangling contributions from other
species, we were able to measure the velocity widths (FWHM) and shifts
of just H$\alpha$ in the optical region, and of O~I, Pa$\beta$,
Mg~I and Br$\gamma$ in the IR region.  The line widths and shifts are
listed in Table~10. \\

\begin{table}
\centering
\caption[]{Emission-line velocity widths (FWHM) and shifts.} 
\begin{minipage}{\linewidth}
\renewcommand{\thefootnote}{\thempfootnote}
\renewcommand{\tabcolsep}{1mm}
\begin{tabular}{lllll} \hline
\multicolumn{5}{c}{Optical line velocity widths, [shifts] \footnote{Associated uncertainties are $\pm 180$ km s$^{-1}$ for the velocity
widths, and $\pm 30$ and $\pm 100$ km s$^{-1}$ for the optical and IR line shifts, respectively. }(km s$^{-1}$)} \\ \hline 
\multicolumn{5}{c}{Epoch (days)} \\ 
Feature & 162  & 227 & 250 & 397 \\ \hline
H$\alpha$ & 4530, [--915] & 4475, [--805] & 4475, [--750] & 4630, [--430] \\
& & & & \\ \hline 
\multicolumn{5}{c}{Infrared line velocity widths, [shifts] (km s$^{-1}$)} \\ \hline 
\multicolumn{5}{c}{Epoch (days)} \\ 
Feature & 137 & 200 & 266 & 314 \\ \hline 
O~I  & 3565, [--530] & 3295, [--530] & 2750, [+50] & 2470, [--400] \\
Pa$\beta$\footnote{Pa$\alpha$ is blueshifted by about --250 km s$^{-1}$.} & 3830, [--235] & 4115, [--700] & 4285, [--700] & 3865, [--235] \\
Mg~I  & 3890, [--300] & 3530, [+400] & 2820, [+800] & 1940, [+600] \\
Br$\gamma$ & 4350, [--555] & 3840, [--555] & 3915, [--555] & -- \\ \hline  
\vspace{-0.85cm}
\end{tabular}  
\end{minipage}
\end{table}

During the nebular phase, line widths of around 3800--4600~km s$^{-1}$ (FWHM)
are seen in the hydrogen lines.  There is weak evidence of a modest
decline in the widths of the other lines with time.  During the earlier part of the
nebular phase both the Balmer and Paschen lines tend to be asymmetric
with enhanced red wings.  The profiles of [O~I] 6300, 6364~\AA\ and 
Mg~I~1.503~$\mu$m 
(see Figs.~15, 18) exhibit pronounced irregularities during the earlier 
part of the nebular phase.  This may indicate clumping in the SN ejecta as
was inferred for SN 1985F (Filippenko \& Sargent 1989), SN~1987A, and 
SN 1993J (Matheson et al. 2000b, and references therein).  
In contrast, the lack of well defined small-scale
structures in [Ca~II]~7291, 7323~\AA\ (see Fig.~16) suggests that calcium is more
uniformly distributed in the SN ejecta.  The [Ca~II] feature does
exhibit a persistent asymmetry with an extended red wing. However, a
3/2 intensity ratio is expected for the 7291,7323~\AA\ lines of the
doublet (see Spyromilio, Stathakis \& Meurer 1993), which may explain
most of the extended red wing. In addition, as inferred for SN~1987A
(see Spyromilio et al. 1991), both [Fe II]~7388,7452~\AA\, and [Ni
II]~7379,7413 \AA\, may be contributing to the red wing. \\

The H$\alpha$ peak exhibits a blueshift (with respect to 
the adopted +110 km s$^{-1}$
centre-of-mass redshift) of $-915\pm30$ km s$^{-1}$ at 162~d, declining
to $-430\pm30$~km s$^{-1}$ at 397~d (see Table~10 and Fig.~14). (An
even larger blueshift, $\sim -1700$ km s$^{-1}$, is seen at the
photospheric epoch of 44~d.).  The Pa$\beta$ line peak exhibits a
blueshift of $\sim -200\pm100$~km s$^{-1}$ at 137~d, but by 200~d this
has increased to $-700\pm100$~km s$^{-1}$. A similar shift is seen at
266~d, after which the shift reverts to around $-200\pm100$~km
s$^{-1}$ again.  Br$\gamma$ shows an unchanging blueshift of about
$-550\pm100$~km s$^{-1}$ from 137~d to 266~d (see Fig.~19). The O~I 1.129~$\mu$m
(Fig.~17) emission line shows an almost constant blueshift of $\sim
-500\pm100$~km s$^{-1}$, except at 266~d when it shows a redshift 
of about $+50\pm100$~km s$^{-1}$. In contrast, Mg~I~1.503~$\mu$m (Fig.~18) 
shows an early blueshift of $\sim -300\pm100$~km s$^{-1}$, but from
200~d to 314~d it is redshifted by $\sim$+600~km s$^{-1}$. \\

The pattern of emission-line blueshifts is probably indicative of
variations in relative abundancies and conditions across the ejecta.
However, apart from Pa$\beta$, no lines show evidence of an {\it
increasing} blueshift with time nor progressive attenuation of the red
wing, suggesting that, up to 381~d, there was no significant dust
condensation in the ejecta. \\

\section{Spectroscopic comparison of SN~2002hh and SN~1987A}
We now compare near-coeval late-time optical and NIR spectra of
SN~2002hh and SN~1987A during the period of 4.5 months to $\sim$1~year
(see Figs.~20--24). The excellent late-time spectroscopic coverage of
SN~2002hh means that it is the first CCSN for which such a comparison
is possible. This comparison can be justified at the outset since, as
we have shown, the nebular-phase light curves of the two supernovae are
quite similar and were powered by a similar amount of $^{56}$Ni.  Only
at early times do the light curves differ significantly, primarily due
to the different physical sizes of their progenitor stars (see Woosley
1988).  In all the comparison figures the spectra have been corrected
for redshift and extinction.  The recession velocities are $+283$ and
$+110$ km s$^{-1}$ for SN~1987A and SN~2002hh, respectively.  The
dereddening made use of the extinction law of Cardelli et al. (1989).
For SN~1987A we adopted $A_V=0.6$ mag and $R_V=3.1$. For SN~2002hh, we
used the two-component extinction model ($A_{V1}=3.3$ mag ,
$R_{V1}=3.1$; $A_{V2}=1.7$ mag , $R_{V2}=1.1$; see Section
\ref{extinct}).  In addition, the SN~1987A spectra have been scaled to
the distance of SN~2002hh (5.9~Mpc).

\subsection{Optical spectra}
In Figure~20 we compare optical spectra of SN~2002hh at epochs of
162~d, 250~d and 397~d with near-coeval spectra of SN~1987A.  The
SN~1987A spectra are from Spyromilio et al. (1991) and Stathakis
(1996).  The overall spectral form and evolution of the two SNe are
rather similar, especially at 162~d and 250~d. At all three epochs the
H$\alpha$ luminosity is similar between the two SNe.  The Na~I~D
P-Cygni feature is also similar in the two events. The most striking
difference between the two SNe is that both [Ca~II]~7291, 7323~\AA\
and the Ca~II IR triplet are much stronger in SN~1987A throughout the
162--397~d period. Moreover the strong Ca~I~1.99~$\mu$m feature in
SN~1987A at 192~d is virtually absent in the SN~2002hh 200~d spectrum
(see following section). We infer a significantly lower calcium
abundance in SN~2002hh. \\

\begin{figure*}
\vspace{22cm} \includegraphics{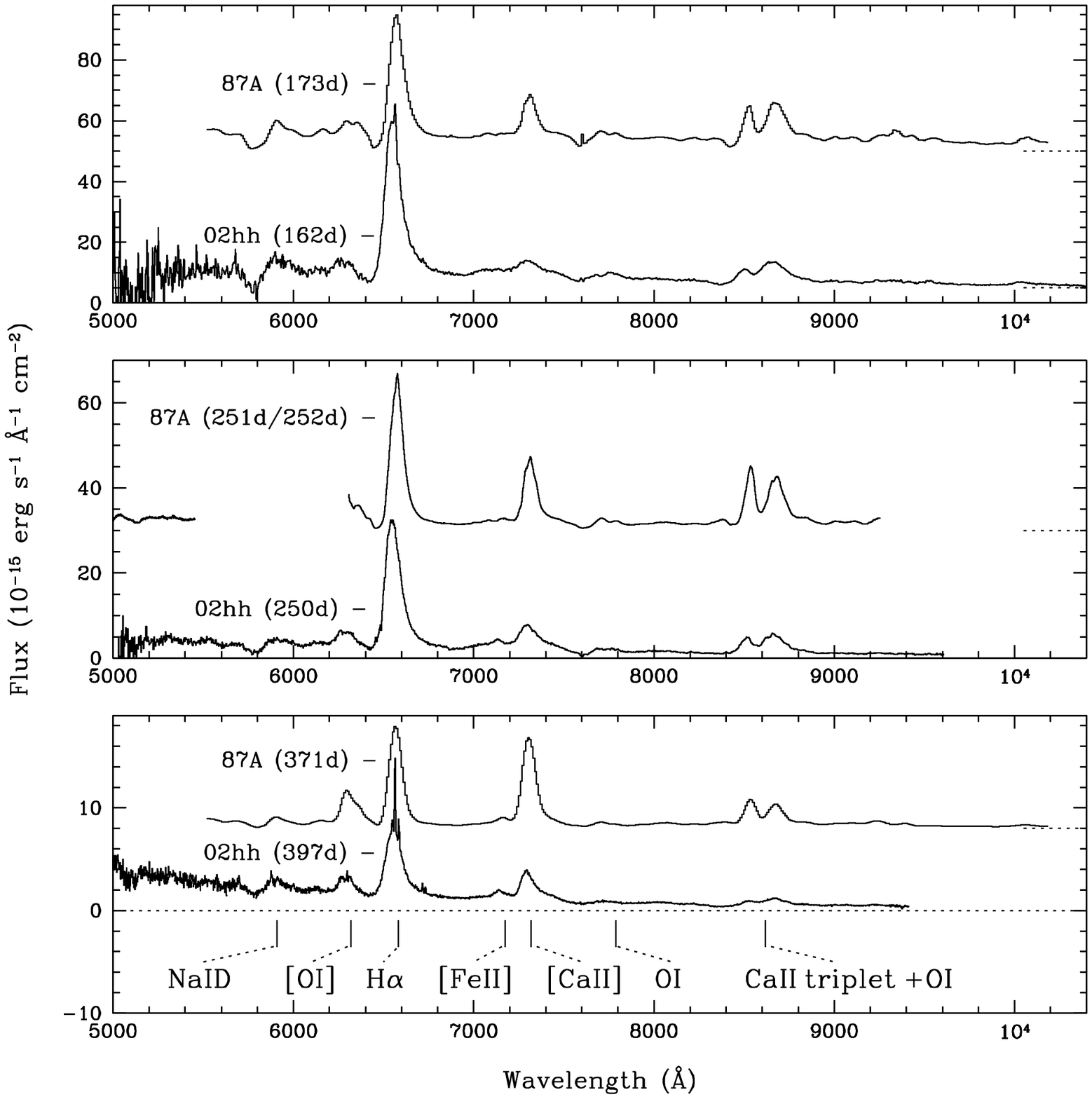}
\caption[] {Comparison of the optical spectra of SN~2002hh with
near-coeval spectra of SN~1987A. The spectra have been corrected
for redshift and extinction.  The SN~1987A spectra have been scaled to
the distance of SN~2002hh (5.9~Mpc) and displaced vertically for
clarity. At the bottom we show the line identifications for SN~1987A
as given in Spyromilio et al. (1991).}
%\label{}
\end{figure*}

\begin{figure*}
\vspace{22cm} \includegraphics{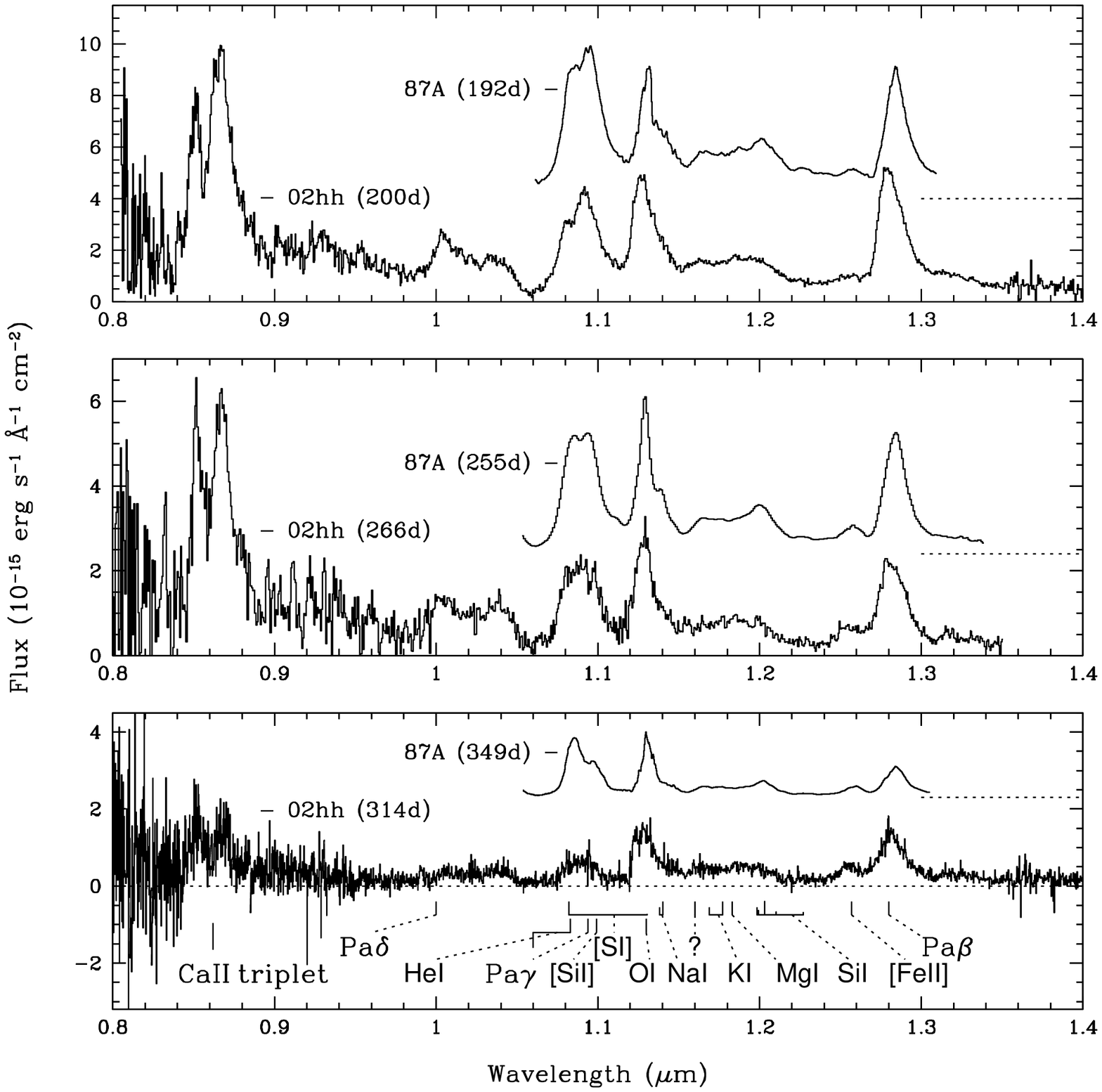}
\caption[] {Comparison of the NIR spectra of SN~2002hh with
near-coeval spectra of SN~1987A in the 0.8--1.4 $\mu$m
region (which includes the $J$ band).  
The spectra have been corrected for redshift and extinction.
The SN~1987A spectra have been scaled to the distance of SN~2002hh
(5.9~Mpc) and displaced vertically for clarity. At the bottom we show 
the line identifications for SN~1987A as given in Meikle et al. (1993).}
\end{figure*}

\begin{figure*}
\vspace{22.5cm} \includegraphics{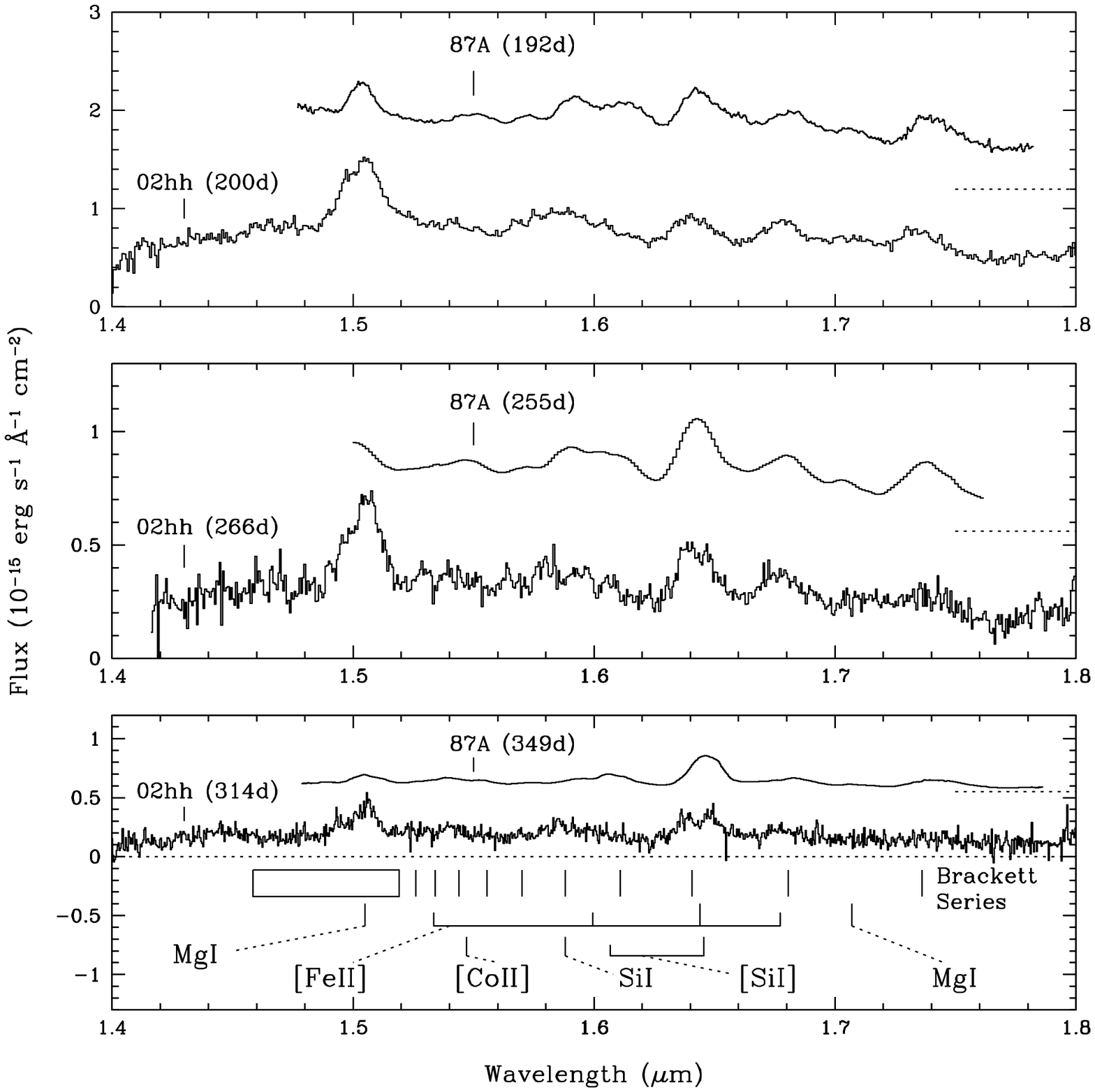}
\caption[] {As for Fig.~21, but for the $H$ band (1.4--1.8 $\mu$m).}  
\end{figure*}

\begin{figure*}
\vspace{22.5cm} \includegraphics{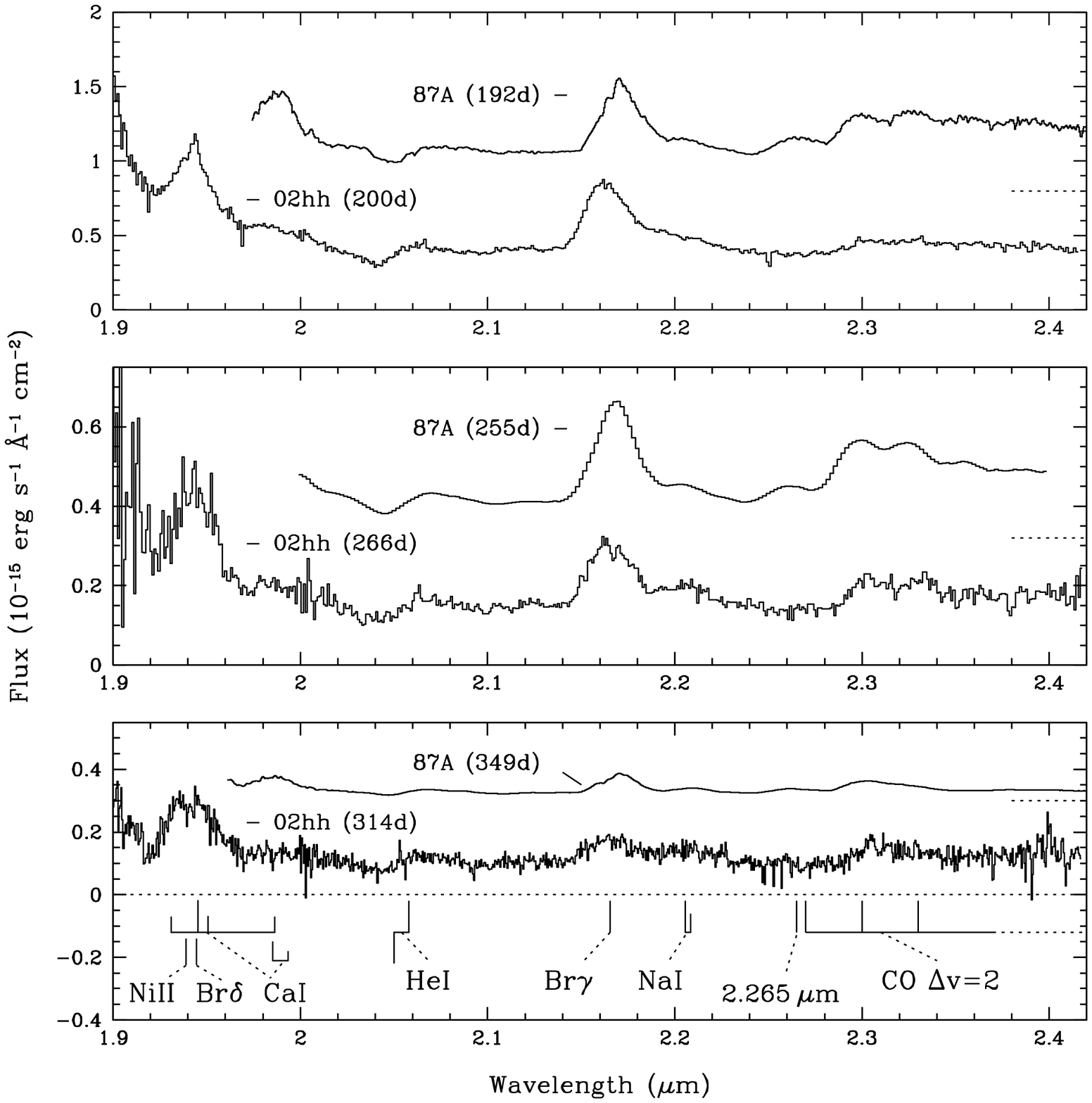}
\caption[] {As for Fig.~21, but for the $K$ band (1.9--2.4 $\mu$m).}  
\end{figure*}

\begin{figure*}
\vspace{22.5cm} \includegraphics{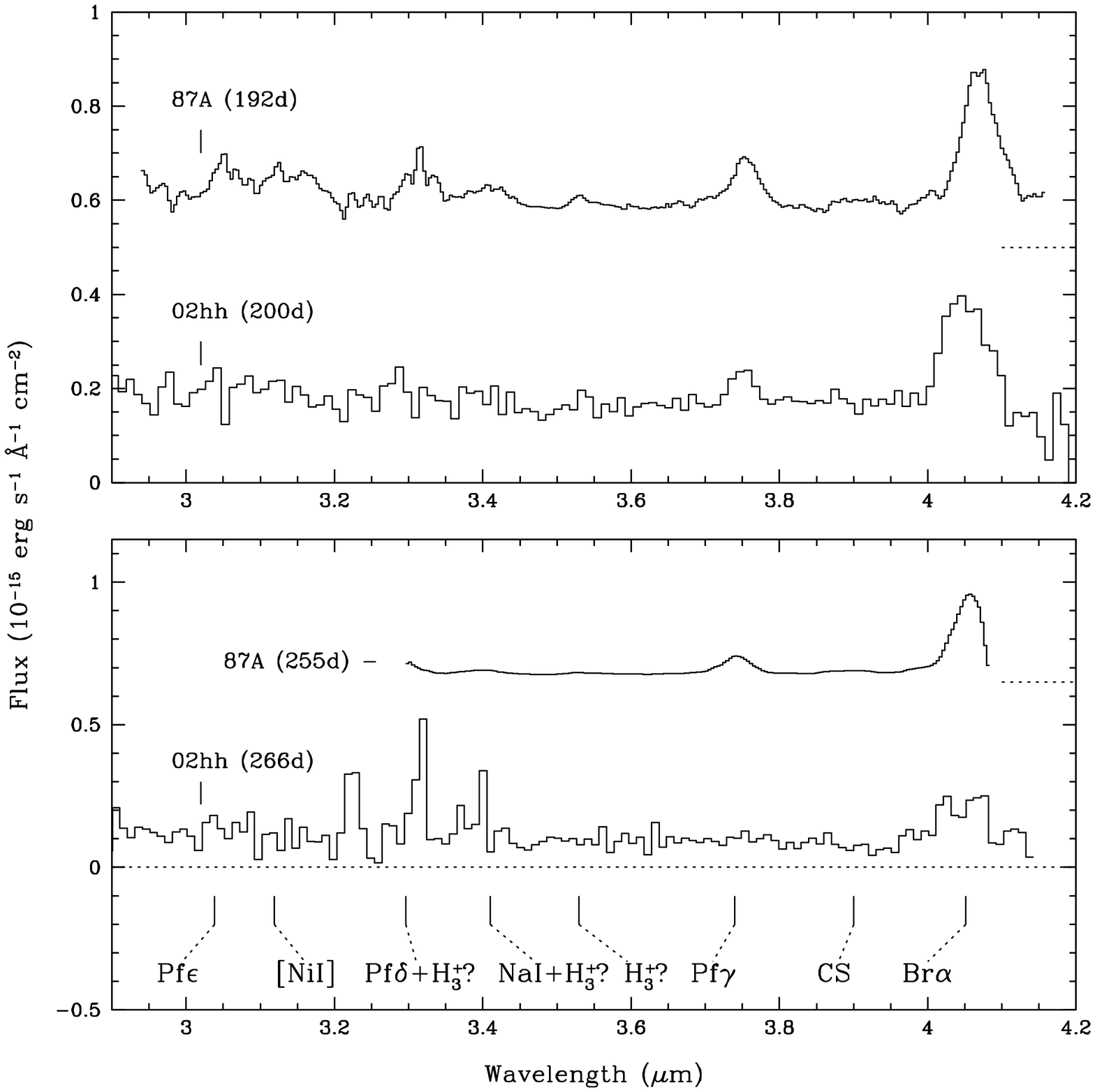}
\caption[] {As for Fig.~21, but for the $L$ band (2.9--4.2 $\mu$m).}  
\end{figure*}

At 162~d and 397~d the SN~2002hh optical continuum is distinctly bluer
than that of SN~1987A, while at 250~d they are about the same.  At 162~d the
flux levels toward the blue are similar, but by 1~$\mu$m the SN~1987A
continuum flux exceeds that of SN~2002hh by more than a factor of
2. Comparison of the overlap of the red end of the 162~d
optical spectrum with the blue end of the 137~d and 200~d infrared
spectra of SN~2002hh suggests that some of the SN 2002hh/SN 1987A red
discrepancy ($\sim\times$1.35) is due to error in the linear flux
correction function (see Section 2.3) for the 162~d SN~2002hh optical
spectrum. At least some of the remaining discrepancy may be due to
error in the SN~1987A fluxing. Unfortunately, we are unable to apply an
overlap test here since there is a 0.12~$\mu$m gap between the
SN~1987A optical and IR spectra. \\ 

In the 397~d optical spectrum of
SN~2002hh, the continuum flux toward the red agrees well with that of
SN~1987A at 371~d.  However, as we move to the blue we see a growing
excess in the SN~2002hh continuum relative to that of SN~1987A.  We
were unable to check the spectral fluxing of SN~2002hh against
photometry for the 397~d optical or 381~d IR spectra, but nevertheless
we find fair flux agreement in the small optical/IR overlap region
around 0.94~$\mu$m.  This gives us confidence that the fluxing toward
the red end of the SN~2002hh optical spectrum is good. However, it is
possible that at least some of the blue discrepancy could be due to
fluxing error which increases toward shorter wavelengths (see Section 2.3).
Another possibility is the presence of uncorrected background
contamination, as Milne \& Wells (2003) concluded was the case for the
Type~Ia SN~1989B.  A third explanation is that in SN~2002hh a small
fraction of the peak luminosity is being scattered by circumstellar or
interstellar dust. Thus, as the supernova flux fades between 162~d and
397~d, the relative contribution of the scattered light increases.
Dust-echo enhancement of late-time optical spectra has been reported
for the Type~Ia SNe~1991T and 1998bu (Patat 2005 and references
therein).  Spectral modelling of the blue excess may distinguish
between these possibilities for SN~2002hh. However, we are much less
persuaded that the blue continuum is due to ejecta-CSM interaction
since no signs of this are seen in the contemporary spectral lines.

\subsection{Near-IR spectra}
In Figures~21 to 24, we compare $\sim$0.8--4.2~$\mu$m spectra of
SN~2002hh at 200~d, 266~d and 314~d with near-coeval spectra of
SN~1987A taken from Meikle et al. (1989).  At the bottom of each
figure we show the line identifications for SN~1987A as given in
Meikle et al. (1993).  The fluxing for SN~1987A has been revised to
match the IR photometry of Bouchet \& Danziger (1993), Catchpole et
al. (1987, 1988) and Whitelock et al. (1988).  In particular, the 
$L'$-band spectra of SN~1987A at 192~d and 255~d have been scaled by
factors of 1.31 and 1.46, respectively.  In addition, the SN~1987A fluxes
have been scaled to 5.9~Mpc, the distance of SN~2002hh.  All the
SNe~1987A and 2002hh spectra have been dereddened as described
earlier.  \\
  
As in the optical region, the spectral form and evolution of the two
SNe are very similar.  The similarity of the form and fluxes of the
spectra in the $J$ and $H$ bands is particularly impressive especially
when one considers that the only corrections have been to the
extinction and the difference in distance.  \\

\noindent {\it J-band spectra}\\ 

\noindent The Pa$\gamma$~1.094~$\mu$m\,-\,He~I~1.083~$\mu$m blend is
stronger in SN~1987A, especially by 314~d, and yet the He~I absorption
remains similar in the two SNe (although this could simply be due
to saturation).  The Pa$\beta$~1.282~$\mu$m and O~I~1.129~$\mu$m
emission peaks are at about the same level in both events throughout
the 200--314~d period.  The [Fe~II]~1.257~$\mu$m is about the same in
both SNe up to 266~d, but at 314~d it is about twice as strong in
SN~2002hh compared with SN~1987A.  The Si~I~1.20~$\mu$m emission is somewhat
stronger in SN~1987A, but this difference fades with time.  Some of
the Pa$\gamma$~1.094~$\mu$m\,-\,He~I~1.083~$\mu$m blend difference may 
be due to the additional presence of [Si~I]~1.099~$\mu$m at a stronger intensity 
in SN~1987A due to a higher abundance of silicon in this event.  However,
it seems unlikely that this can provide the whole explanation for the
difference. \\

\noindent {\it H-band spectra}\\ 

\noindent 
The overall appearance of the $H$-band spectra is similar between the
two SNe.  The most striking and persistent difference is the factor of
2.5 greater luminosity of Mg~I~1.503~$\mu$m in SN~2002hh.
Redward of the Mg~I line some other differences are seen between the
two events.  SN~1987A shows a persistent emission feature at
1.61~$\mu$m which is virtually absent in SN~2002hh. Meikle et
al. (1989) ascribed the 1.61~$\mu$m emission to [Si~I]
1.6068~$\mu$m. The weakness of the corresponding emission in SN~2002hh
suggests a lower silicon abundance in this SN.  Interestingly, the
1.64~$\mu$m feature is comparable to, or slightly stronger than in
SN~1987A during the 200-314~d period. Meikle et al. (1989) attribute
this feature to a blend of [Fe~II]~1.644~$\mu$m, [Si~I]~1.645~$\mu$m
and Br12. However, the weakness of the [Si~I] 1.6068~$\mu$m component
in SN~2002hh suggests that the 1.64~$\mu$m feature is predominantly
due to iron and hydrogen in this SN. \\

\noindent {\it K-band spectra}\\ 

\noindent The two SNe show very similar features from 200~d to 314~d,
although no comparison can be made shortward of about 1.97~$\mu$m
where the SN~1987A data terminate.  The peak-base intensity of
Br$\gamma$~2.166~$\mu$m is similar at 200~d, at 266~d it is stronger
in SN~1987A, but by 314~d it is stronger in SN~2002hh.  The
He~I~2.058~$\mu$m feature is of comparable strength at $\sim$200~d in
both SNe, but is stronger in SN~2002hh thereafter.  The Na~I
2.206~$\mu$m line, which shows up in both SNe on the red wing of the
Br$\gamma$ line at $\sim$200~d, is increasingly prominent in SN~2002hh
at later epochs.  The CO first overtone is about a factor of 3
stronger in SN~1987A at 192~d than in SN~2002hh. However, the SN~1987A
CO intensity fades more rapidly than in SN~2002hh so that by 314~d the
situation is reversed.  Indeed, by 381~d the SN~2002hh CO emission
appears to be about twice as strong as in SN~1987A (see below).
There is little sign of the unidentified 2.265~$\mu$m feature in
SN~2002hh (see below). In addition, the strong Ca~I~1.99~$\mu$m
feature in SN~1987A at 192~d is virtually absent in the SN~2002hh
200~d spectrum.  A striking difference between the two SNe is that
SN~2002hh shows a much stronger underlying continuum. At 200~d, the
excess continuum amounts to $\sim1.4\times10^{-16}$~erg s$^{-1}$
cm$^{-2}$ \AA$^{-1}$, falling to about half this value at 266~d and
maintaining this level until 314~d.  The (SN~2002hh):(SN~1987A)
continuum ratio rises from 1.6 to 3.6 between 200~d and 314~d.  The
continuum is discussed below. \\

\begin{figure}
\vspace{7.2cm} \includegraphics{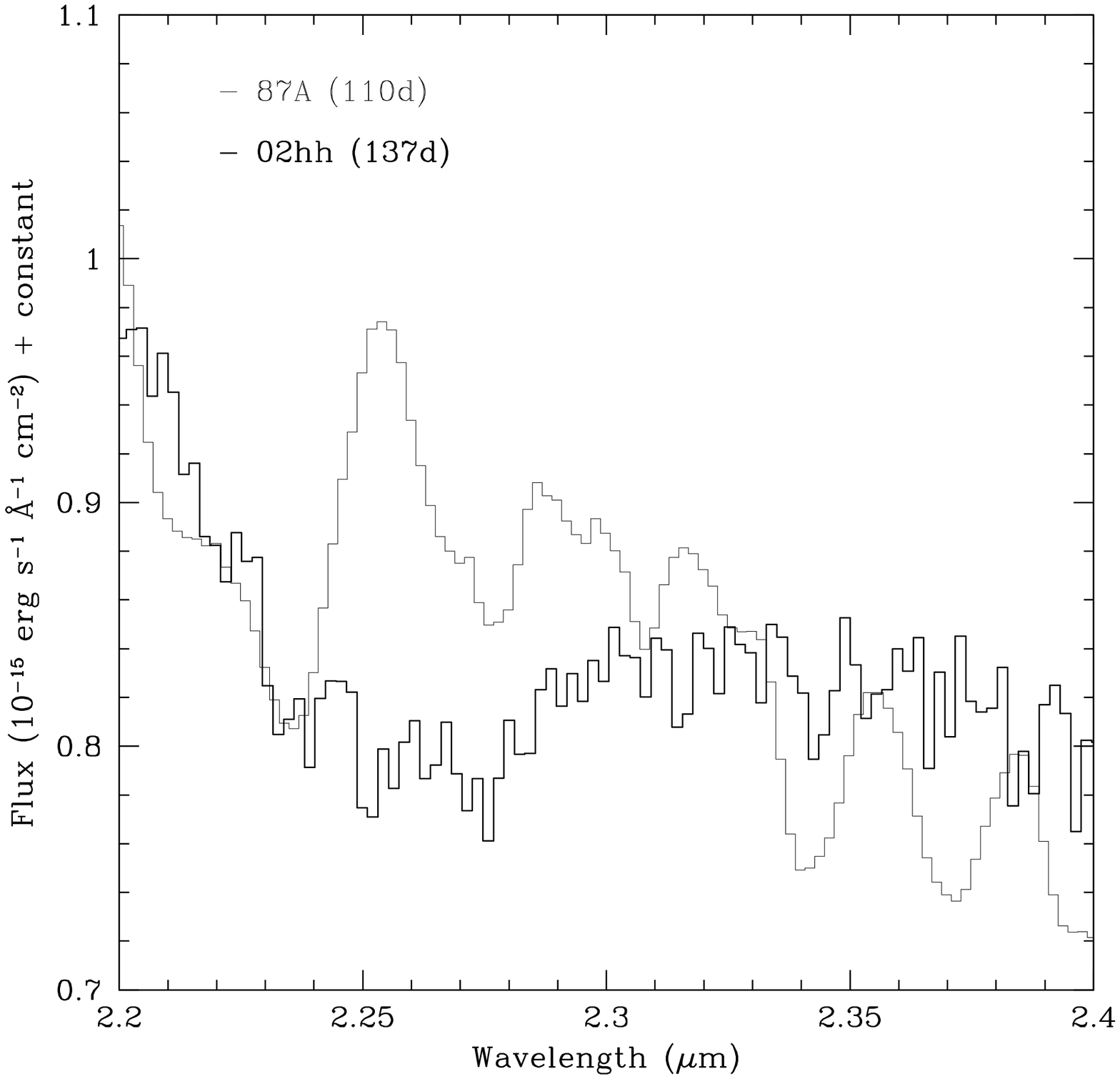}
\caption[] {Comparison of the first-overtone CO profile of SN~2002hh
at 137~d with that of SN~1987A at 110~d.  The SN~1987A spectrum has been
scaled to the distance of SN~2002hh, and has been displaced vertically
so that the underlying continuum matches that of SN~2002hh (see text).
The spectra have been corrected for redshift and extinction.}
%\label{}
\end{figure}

\begin{figure}
\vspace{7.2cm} \includegraphics{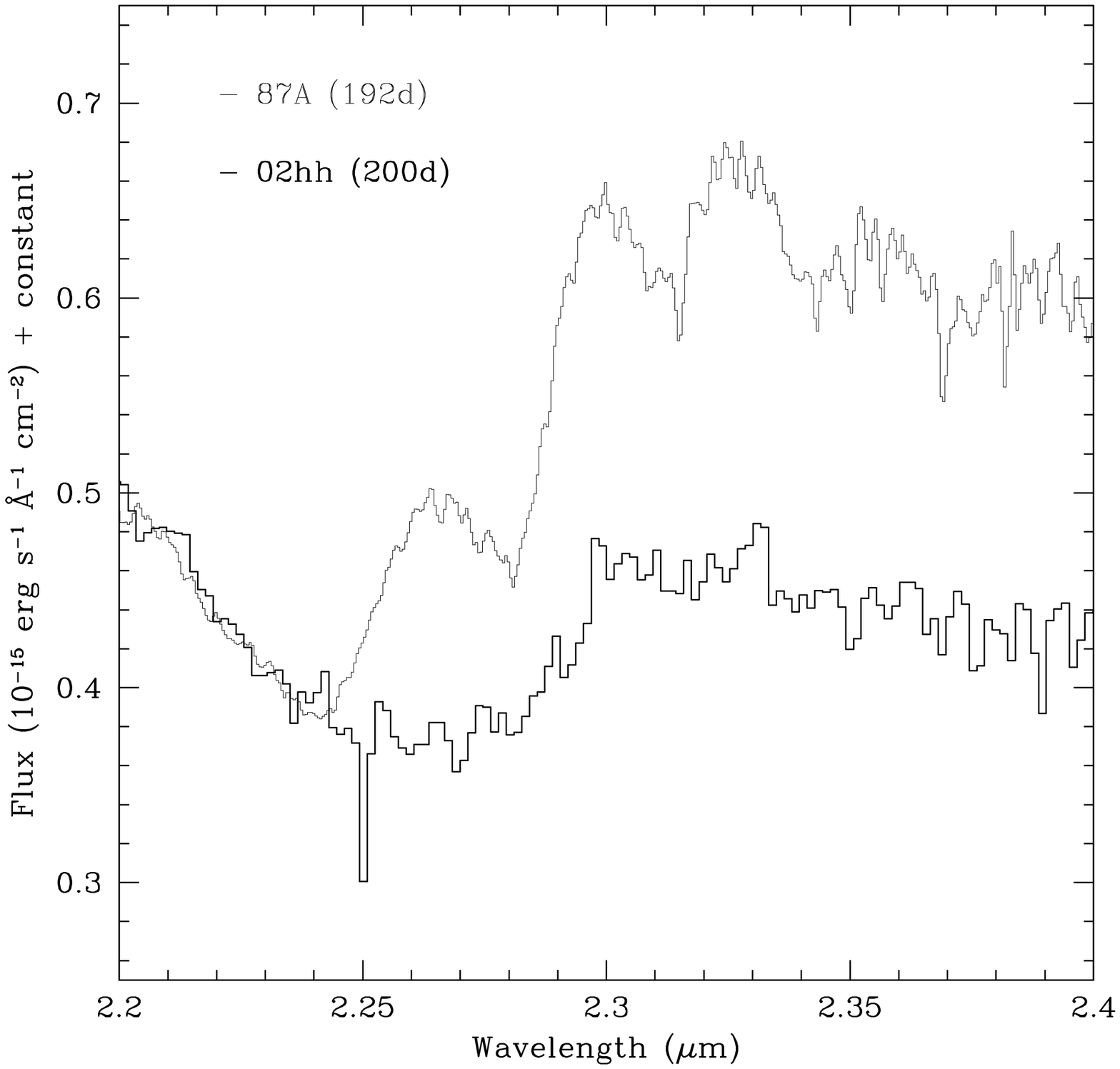}
\caption[] {As for Fig.~25 but for SN~2002hh at 200~d and SN~1987A at 192~d.}
%\label{}
\end{figure}

\begin{figure}
\vspace{7.2cm} \includegraphics{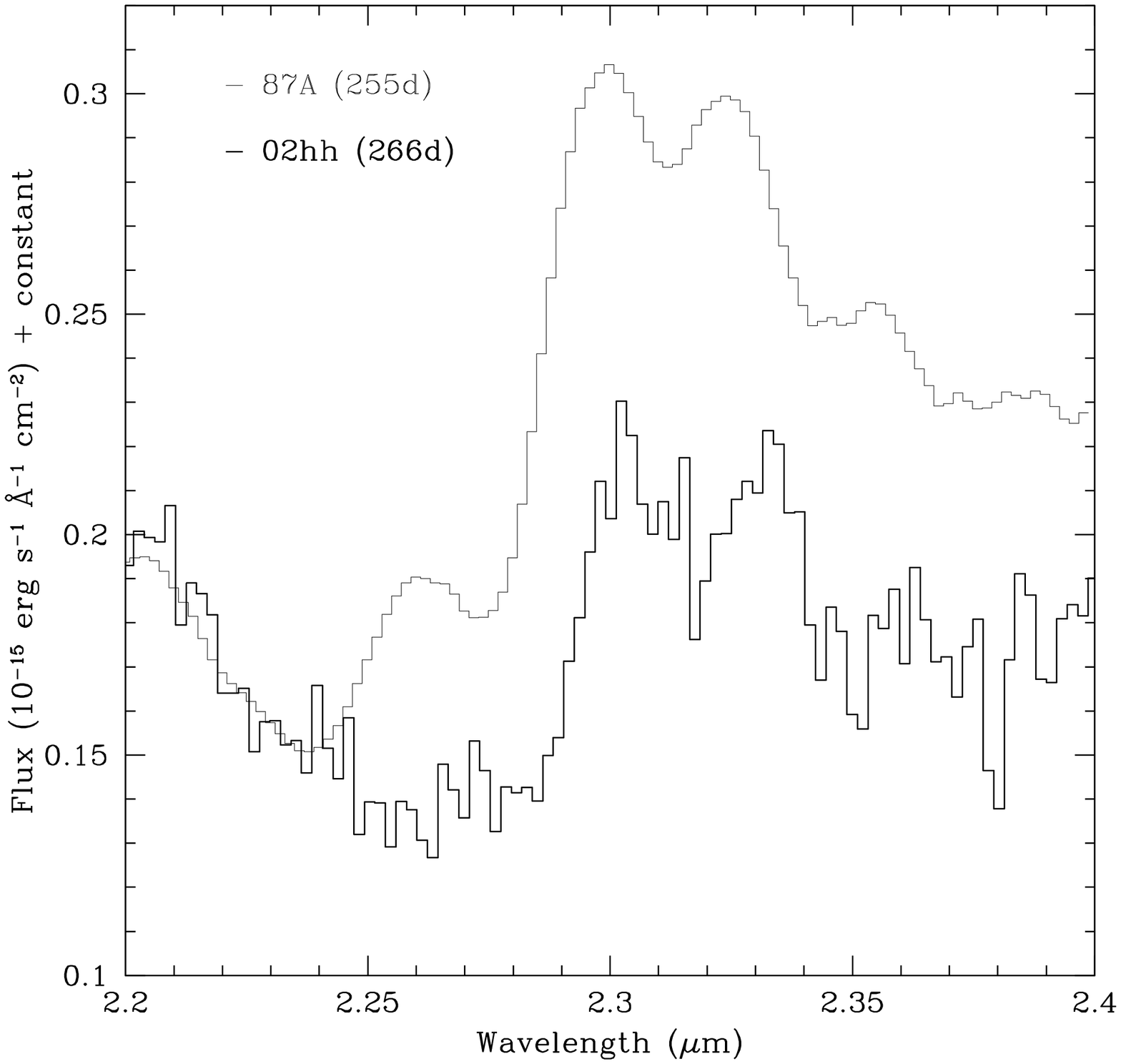}
\caption[] {As for Fig.~25 but for SN~2002hh at 266~d and SN~1987A at
255~d.}
%\label{}
\end{figure}

\begin{figure}
\vspace{7.2cm} \includegraphics{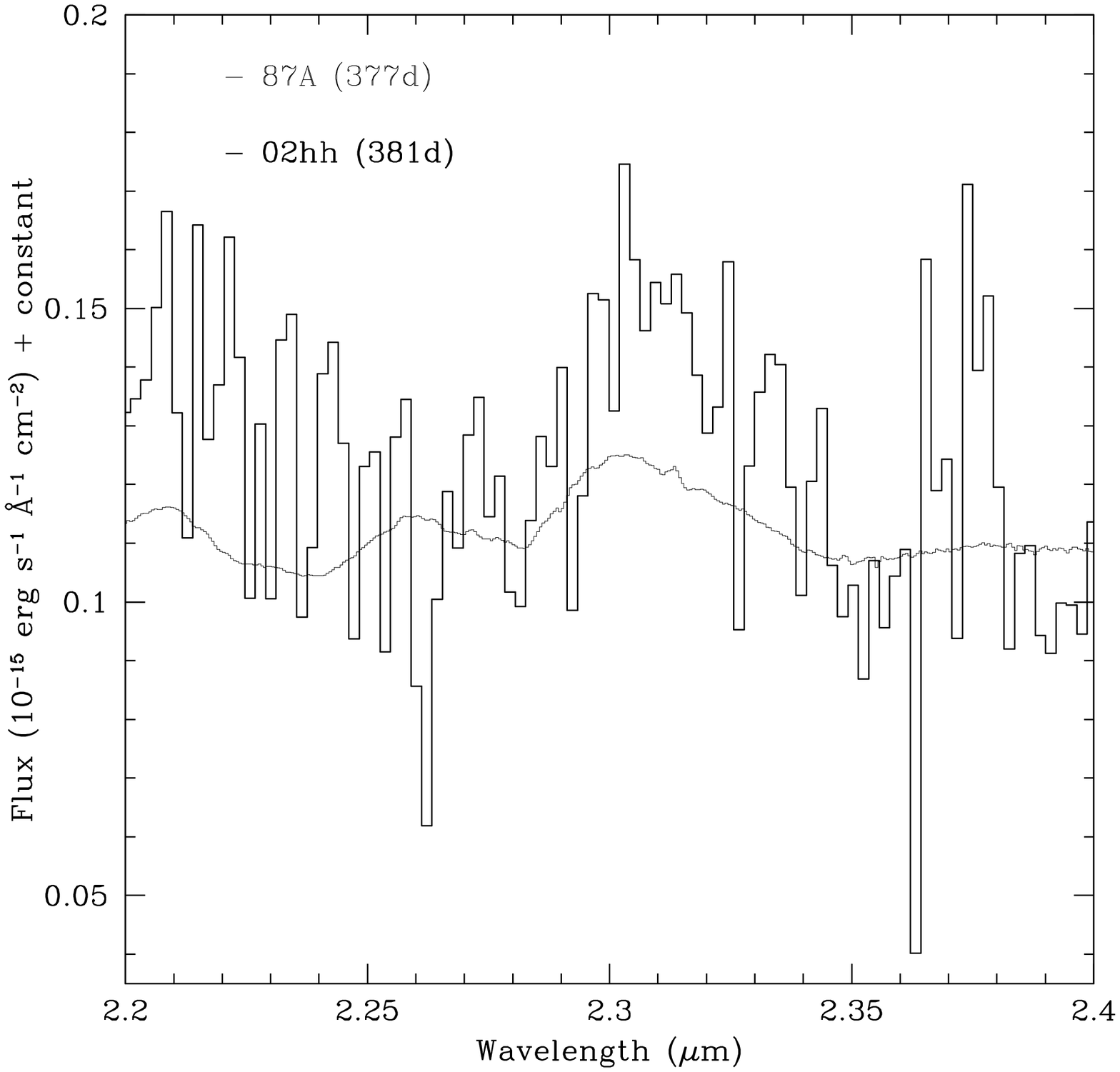}
\caption[] {As for Fig.~25 but for SN~2002hh at 381~d and SN~1987A at
377~d.}
%\label{}
\end{figure}

\noindent  {\it L-band spectra}\\ 

\noindent In spite of the lower S/N of the SN~2002hh spectra, we can
see that Br$\alpha$ and Pf$\gamma$ have strengths similar to those in
SN~1987A
at 200~d.  By 266~d Br$\alpha$ is still of roughly comparable strength
in the two SNe, but there is no longer any sign of Pf$\gamma$ in
SN~2002hh.  As in the $K$ band, a striking aspect of the $L$-band
spectrum is the much stronger continuum in SN~2002hh.  The
(SN~2002hh):(SN~1987A) continuum ratios in the $L$ band are similar to
those seen in the $K$ band (1.8 at 200~d and 3.6 at 266~d).
The continuum is discussed below.

\subsubsection{Carbon monoxide}\label{CO}
In Figures~25--28 we compare the CO first overtone emission spectra of
SNe~1987A and 2002hh.  The SN~1987A spectra have been scaled to the
distance of SN~2002hh. In addition, in order to effect a better
comparison between the two supernovae, the SN~1987A spectra have been
displaced vertically so that the underlying continuum (judged from the
spectrum to the blue of the CO emission) matches that of
SN~2002hh. The excess continuum in SN~2002hh is discussed below. Both
spectra in each figure have been corrected for extinction and
redshift.  \\

In Fig.~25 we compare the 137~d spectrum of SN~2002hh with
the 110~d spectrum of SN~1987A. In both cases CO emission is barely
visible.  In Fig.~26 we compare the 200~d spectrum of SN~2002hh with
the 192~d spectrum of SN~1987A.  First-overtone emission is clearly
present in both spectra, but is significantly weaker in SN~2002hh.
The unidentified 2.265~$\mu$m feature prominent in SN~1987A is
completely absent in SN~2002hh (see below).  While the relative
weakness of the SN~2002hh CO emission persists to 266~d (Fig.~27), the
supernovae exhibit a similar relative decline in the P~branch
($\sim$2.37~$\mu$m).  Local thermodynamic equilibrium (LTE) 
modelling of the CO emission profile in
SN~1987A (e.g., Spyromilio et al. 1988) showed that by 255~d the
decline in the P~branch ($\sim$2.37~$\mu$m) flux implies that the
temperature of the CO emitting region had fallen below 2000 K. We
conclude that similar cooling took place in SN~2002hh. This is
important as it suggests that some regions of the ejecta may cool
sufficiently for dust condensation to take place.  A quasi-periodic
structure is produced by the R-branch bandheads of the 0--2, 1--3,
2--4 (etc.) vibrational transitions.  The visibility of this structure
in the 266~d SN~2002hh spectrum is similar to that of SN~1987A at
255~d, implying a similar CO velocity, 1800--2000 km s $^{-1}$. \\

In Figure~28, we compare the CO spectrum of SN~1987A at 377~d with
that of SN~2002hh at 381~d. In spite of the very low S/N of the
latter, it seems that by this epoch the R-branch intensity of
SN~2002hh was about twice as strong as in SN~1987A.

\subsection{The unidentified 2.265~$\mu$m feature}\label{unid}
An emission feature at 2.265~$\mu$m is present in the spectra of
SN~1987A from as early as 192~d to at least 2~years post-explosion.
Spyromilio, Leibundgut \& Gilmozzi (2001) also claim detection of this
feature in the Type~IIP SN~1999em at $\sim$170~d, albeit at a much
lower S/N.  However, in other CCSNe $K$-band spectra the feature is at
best marginally present [SN~1996ad (105~d), Spyromilio \& Leibundgut
1996; SN~1998dl ($\sim$150~d), Spyromilio et al. 2001] or absent
[SN~1998S (109~d), Fassia et al. 2001; SN~1999gi (126~d), SN~2000ew
(97~d), Gerardy et al. 2002].  It might be inferred from this that the
2.265~$\mu$m feature is actually quite ubiquitous, but does not
strongly appear until after $\sim$150~d. However, our SN~2002hh
observations (see Figs.~25--28) show that the feature was certainly
absent up to 200~d and, perhaps, even up to 266~d.  While the origin
of this feature has been debated for many years (Spyromilio et
al. 1988; Meikle et al. 1993; Liu \& Dalgarno 1995; Spyromilio \&
Leibundgut 1996; Gearhart, Wheeler \& Swartz 1999; Gerardy et
al. 2000, 2002; Spyromilio et al. 2001), we must conclude that as yet
there is no consensus as to its true identity nor why it seems to
exhibit such a large range of intensities from one supernova event to
another.

\subsection{The late-time near-IR continuum}

As already mentioned, one of the most notable differences between the
late-time coeval NIR spectra of SN~1987A and SN 2002hh is the much
stronger $KL$~continuum in the latter event.  This was illustrated
for epochs 200~d and 266~d, for which both $K$ and $L$-band
spectra were available. Inspection of Figs.~23 and 24 shows that
this excess in SN~2002hh probably persisted to beyond 314~d.
We would like to understand the origin of this continuum, and in
particular, why it is so much stronger than in SN~1987A.\\

In SN~1987A, Meikle et al. (1993) found that the late-time $JHK$
continuum could be well represented by a $\lambda^{-2.5}$ law between
192 and 574~days.  They found that this continuum was several times
stronger than would be expected from optically thin, ionised hydrogen,
as well as having a steeper spectrum than that of pure free-free
emission. They suggested that the $JHK$ continuum was due to a mixture
of emission from optically thick and optically thin gas, comprising
many contributing species and subject to spatial variations in
temperature and ionisation.  Different behaviour was seen in the
$L$-band continuum. While the $\lambda^{-2.5}$ law provided a fair
representation of the 192~d continuum to about 3.65~$\mu$m, at longer
wavelengths the slope flattened. Moreover, by 349~d, the break in
slope had moved back to $\sim$3.3~$\mu$m. Meikle et al. found that
between 192~d and 349~d the $L$-band continuum faded at the same rate as
did the CO first overtone emission in the $K$ band. They therefore
suggested that the excess $L$-band continuum was caused by a blend of
unidentified molecular emission bands.\\

\begin{figure*}
\vspace{9.5cm} 
\includegraphics{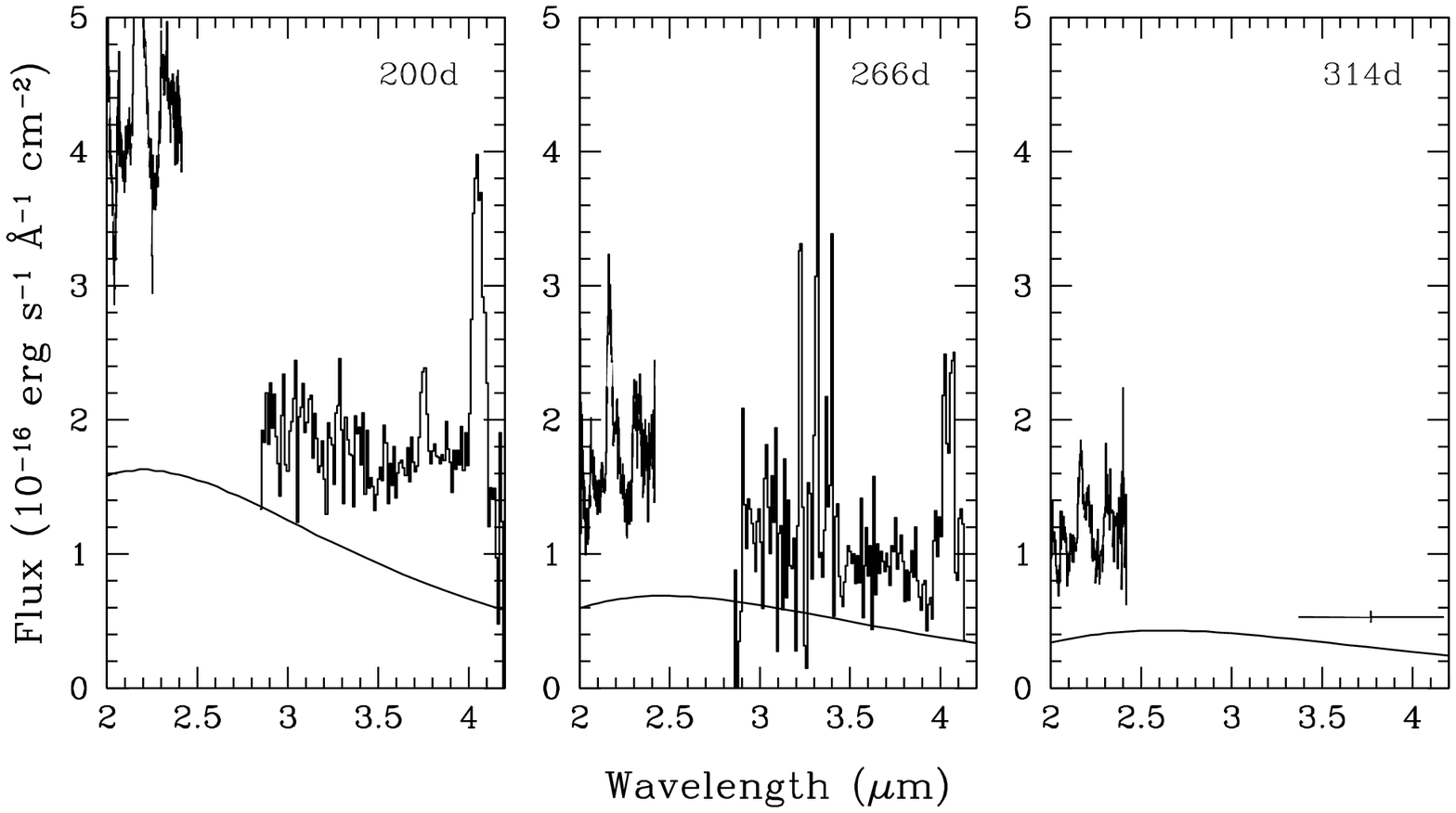}
\caption[] {Infrared spectra of SN~2002hh compared with IR-echo model
spectra for three epochs, illustrating that much of the $KL$ excess
could be due to an IR echo (see text for details of the model).  No
$L$-band spectrum was available for 314~d, so we show the
corresponding $L'$ photometry point instead.}
%\label{}
\vspace{0.5cm}
\end{figure*}

\begin{figure*}
%\vspace{9.5cm} \special{psfile=FIGS/plot_KLp_dered2dust_02hh_allSNeII.ps 
%hoffset=40 voffset=-68 hscale=70 vscale=47 angle=0}
\vspace{10.5cm} \includegraphics{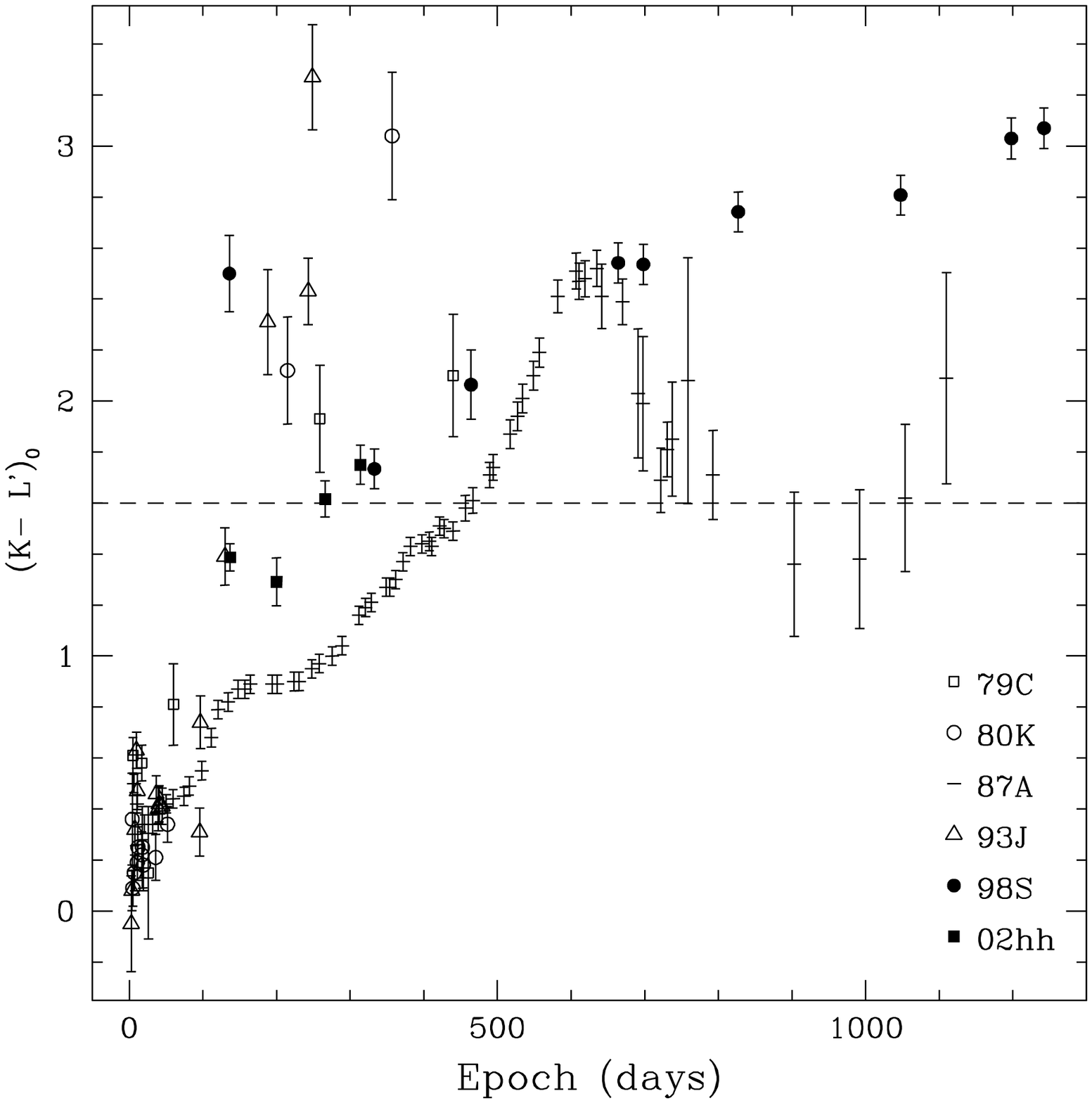}
\vspace{-0.5cm}
\caption[]{The temporal evolution of the extinction-corrected colour
($K-L'$)$_0$ for core-collapse supernovae, including SN~2002hh (filled
squares).  The epoch gives the approximate number of days since
outburst.  Pozzo et al. (2004) suggested that colours lying above the
dashed line at ($K-L'$)$_0 = 1.6$ indicate a contribution to the IR
emission from hot dust.  The ($K-L'$)$_0$ colour of SN~2002hh is redder
than that of SN~1987A by about 0.5 mag, although the temporal trend is
similar.  However, even the latest SN~2002hh point lies only slightly
above the $K-L'=1.6$ line and so provides little (if any) support for dust
condensation.}
%\label{plot_kl_CIT}
\end{figure*}

Wooden (1989; see also Wooden et al. 1993 and Wooden 1997)
hypothesized that at 60~d and 260~d, part of the 3--12~$\mu$m continuum
of SN~1987A was due to warm dust in the circumstellar environment,
heated by the early-time luminosity --- an IR echo. She derived dust
temperatures of $\sim$1300--1600~K at 60~d, falling to $\sim$560--700~K
at 260~d, and suggested that the dust lay in the circumstellar ring
(Wooden 1997). However, as Meikle et al. (1993) point out, the Wooden
dust model does not account for much of the $L$-band continuum which
appears to be in {\it excess} of the Wooden dust-model spectra.
Nevertheless, even if we accept the Meikle et al. ``blended molecular
bands'' hypothesis for SN~1987A, it is not clear that this solution
can be extended to the significantly larger continuum excess in
SN~2002hh. Two points argue against doing so.  One is that there is
convincing evidence that the excess is at least as strong in the
$K$~window, where strong, blended, and unidentified molecular emission
is perhaps less likely. The other point is that, if we attribute all
of the SN~2002hh continuum to molecular emission, then why is the CO
emission so much {\it weaker} than in SN~1987A, at least for epochs
200~d and 266~d?  We are therefore persuaded to examine the possibility
that at least some of the excess continuum in SN~2002hh is due to warm
dust. \\

Comparison of the SN~2002hh 1-4~$\mu$m spectra with blackbody
functions shows that much of the excess continuum in the $K$ and $L$
windows could be reproduced with blackbodies at 1100--1300~K and
having an expansion velocity of $\sim$1700~km s$^{-1}$. Such temperatures and
velocities may be consistent with an origin in newly condensing dust
or in an IR echo.  However, the presence of the $KL$ spectral excess as
early as 200~d plus the rather high blackbody temperature tend to
argue against the condensing dust hypothesis.  In addition, the 200~d
CO emission is a factor of 3 weaker in SN~2002hh than in
SN~1987A, further reducing the prospect that sufficiently cold
dust-forming conditions might have occurred around this epoch.  The
alternative possibility is that much of the excess could have been due
to an IR echo from pre-existing dust in the progenitor CSM.  We have
therefore constructed an IR-echo model and have used this to explore
the parameters of a dusty CSM that could produce an echo of about the
right spectral shape and intensity to account for the excess. \\

The IR-echo model follows those of Bode \& Evans, (1980), Dwek (1983)
and Graham \& Meikle (1986). The input UV-optical luminosity is a
parameterised version of the peak and plateau parts of the SN~1999em
$UBVRI$ light curve (Elmhamdi et al. 2003) scaled to the Cepheid distance
of 11.7~Mpc (Leonard et al. 2003).  To allow for radiation shortward
of the $U$ band, we assumed that the supernova spectrum could be roughly
described by a blackbody at 12,000~K during the peak, and 5,500~K
during the plateau. Therefore, we scaled the adopted light curve by
factors of $\sim$1.9 in the peak and 1.05 in the plateau.  At wavelengths
longer than the $I$~band the fractions of the total blackbody
radiation are about 5\% at 12,000~K and 30\% at 5500~K. However, at
these wavelengths the absorptivity of the dust grains (size
$\sim$0.1~$\mu$m) is likely to fall quite rapidly.  Therefore, to
simplify the IR-echo calculation, we ignored the IR contribution to
the bolometric luminosity. We also ignored the radioactive tail since
this is (a) much fainter than the earlier phases ($<$10\% of the peak
flux), and (b) dominated by infrared radiation. Thus, the bolometric light
curve used for the model is
$47.0\times10^{41}$e$^{-t(d)/18.7}$~erg s$^{-1}$ to 24~days, and
$11.1\times10^{41}$e$^{-t(d)/171}$~erg s$^{-1}$ for 24--118~days.  A grain
UV-optical absorptivity of 1 was assumed.  The dust grain size was
0.05~$\mu$m, the grain material density was 3~g~cm$^{-3}$, and the
pre-explosion grain number density at 6700~A.U. was
$9\times10^{-9}$~cm$^{-3}$ with an $r^{-2}$ (steady wind) dependence.
The resulting optical depth to the input UV-optical luminosity was
$\tau=0.05$. A dust grain emissivity proportional $\lambda^{-2}$ was
assumed for the IR emission. The evaporated dust-free cavity was $R_v
= 8750$~A.U.  This was obtained by scaling from the SN~1979C value
(Dwek 1983) assuming $r_v\propto L_{peak}^{0.5}$ (Dwek 1985).  The
paraboloid corresponding to the peak SN luminosity lay beyond the
dust-free cavity at all three epochs.  The CSM outer radius was
133,000~A.U. and the distance of SN~2002hh was 5.9~Mpc.  The derived
vertex dust temperatures were 1195~K, 1075~K and 1025~K on days~200,
266 and 314, respectively.  The total mass of dust is
$1.6\times10^{-3}$~M$_{\odot}$. Assuming a standard grain-to-gas ratio
(by mass) of 0.006, the corresponding CSM mass was 0.26~M$_{\odot}$,
or 0.28~M$_{\odot}$ if we include the gas in the dust-free cavity.\\

In Fig.~29 we compare the IR echo model spectra with the observed $KL$
spectra for 200~d, 266~d and 314~d. In the last case, no $L$-band
spectrum was available and so we show the corresponding $L'$
photometric point instead.  We conclude that much of the $KL$ excess
can be reproduced with a CSM mass of $\sim$0.3~M$_{\odot}$.  Only
small optical depth ($\sim$0.05) is required to reproduce the
$KL$ excess. In Section~3.2 we obtained $A_V=1.9$ mag, i.e.,
$\tau_V=1.6$ for the ``dust pocket.''  We conclude that most of the
dust-pocket extinction occurred well beyond the progenitor CSM.\\

Pozzo et al. (2004) suggested that if the $K-L'$ colour of a CCSN
exceeds $\sim$1.6 within a year after explosion it indicates the
presence of circumstellar dust heated by the supernova's early-time
luminosity (i.e., an ``IR echo''), although there may also be a
contribution from newly condensed dust in the ejecta. In Fig.~30 we
update Fig.~15 of Pozzo et al. where we plot the $K-L'$ evolution of
SN~2002hh compared with other CCSNe. Even the latest SN~2002hh point
lies only slightly above the $K-L'=1.6$ line.  We therefore conclude
that the $K-L'$ analysis does not provide additional support for
either hypothesis.  As indicated in Section~4.1.4, the line-profile
behaviour up to 381~d also yields no evidence for dust condensation in the
ejecta. \\

The lack of evidence for dust condensation during the first year does
not rule out the possibility of grain formation at later epochs.  Dust
formation happened only after the elapse of one year in SN~1987A,
SN~1998S and SN~1999em. \\ 

\section{Conclusions}

We have presented optical/IR photometric and spectroscopic
observations of SN~2002hh covering a period of $\sim 400$ days during
the photospheric and nebular phases. We confirm it to be a normal
Type~IIP event.  We show the first-ever $L$-band spectra for such an
event. The only other supernova for which $L$-band spectra have been
obtained was the peculiar (and very nearby) Type~II SN~1987A. \\

SN~2002hh is one of the most highly extinguished supernovae ever
investigated. By assuming a similar early-time spectroscopic behaviour
to the well-studied ``template'' Type~IIP SN~1999em, we first derived
a total extinction of $A_V=5.2$ mag. This value was checked against
the K~I 7699~\AA\ absorption line, from which we infer $A_V$ values of
1.55 and 1.77 mag ($R_V=3.1$) in the H~I ISMs of the Milky Way and
NGC~6946, respectively. This suggests that there is an additional
$A_V=1.7$ mag of extinction which is not traced by the K~I absorption
line.  However, we find that a good match between coeval early-time
spectra of SNe~1999em and 2002hh is obtainable using a 2-component
extinction model for SN~2002hh, with $A_V=3.3$ mag and $R_V=3.1$
(Component~1), and $A_V=1.7$ mag with $R_V=1.1$ (Component~2).  An
$R_V$ value as low as 0.7 has been invoked in the past in
multi-component extinction models (see Wang et al. 2004 and
references therein); it probably implies a distinct line-of-sight ``dust
pocket'' where the mean dust grain size is smaller than in the H~I
ISM.  We suggest that this dust pocket may be associated with the H~II
region and dust lane which lies close to the SN~2002hh line-of-sight.
We do {\it not} associate most of the second extinction component with
either newly condensed dust or the progenitor CSM (see below).\\

The early-time optical light curves of SNe~1999em and 2002hh are
generally well-matched, as are the radioactive tails of these two SNe
and SN~1987A.  This suggests a certain robustness in the adopted
extinction in spite of it being derived using only one epoch of
spectroscopic data.  The late-time similarity of SN~2002hh to
SN~1987A implies that $0.07 \pm 0.02$~M$_{\odot}$ of $^{56}$Ni was
ejected by SN~2002hh.  This is confirmed by our derivation of the
OIR luminosity at 200~d and 266~d and by the similar coeval
luminosities of the [Fe~II]~1.257~$\mu$m line in SNe~1987A and 2002hh.
In the nebular phase, however, the $HKL'$ luminosities of
SN~2002hh exhibit a growing excess with respect to those of
SN~1987A. We attribute much of this excess to an IR echo from a
pre-existing, dusty CSM. The NIR light curves of SN~1999em are
systematically fainter than those of SN~2002hh by about 0.6 mag. We
suspect an error in the calibration of the SN~1999em magnitudes. \\

During the nebular phase, line widths (FWHM) of around 3800--4600~km s$^{-1}$
are seen in the hydrogen lines.  There is weak evidence of a modest
decline in the widths of the other lines with time.  During the
earlier part of the nebular phase both the Balmer and Paschen lines
tend to be asymmetric with enhanced red wings.  The presence of strong
He~I~1.083~$\mu$m emission as early as 137~d (Fig.~8b) implies
upward mixing of radioactive iron-group elements (Graham 1988, Fassia
et al. 1998, Fassia \& Meikle 1999).  [O~I]~1.129~$\mu$m is also
prominent during the nebular phase, indicating microscopic mixing of
hydrogen and oxygen. Further evidence of mixing is provided by the
profiles of [O~I]~6300, 6364~\AA\ and Mg~I~1.503~$\mu$m which exhibit
pronounced irregularities during the earlier part of the nebular
phase.  This suggests clumping in the SN ejecta as was inferred for
SNe~1985F, 1987A, and 1993J. In 
contrast, the lack of well-defined small-scale structures in
[Ca~II]~7291,7323~\AA\ points to a more uniform distribution in the SN
ejecta. \\

Persistent blueshifts are seen in a number of prominent nebular-phase 
lines.  The
pattern of blueshifts is probably due to variations in relative
abundancies and conditions across the ejecta.  However, apart from
Pa$\beta$, no lines show evidence of an {\it increasing} blueshift
with time nor progressive attenuation of the red wing, suggesting
that, up to 381~d, there was no significant dust condensation in the
ejecta. \\

We have presented a detailed comparison of the late time optical and
NIR spectra of SNe~1987A and 2002hh. While the overall impression is
one of similarity between the spectra of the two events, there are
notable differences. The Mg~I~1.503~$\mu$m luminosity in SN~2002hh is
much stronger while emission from silicon and calcium is much
weaker. Yet, a number of pieces of evidence point to similar
iron-group masses having been produced in the two events. The
explosive nucleosynthesis models of Thielemann, Nomoto \& Hashimoto
(1996) suggest that the magnesium mass in particular is a sensitive
indicator of progenitor mass, increasing by a factor of about
15 as we go from a progenitor mass of 13~M$_{\odot}$ to
20~M$_{\odot}$. For this progenitor mass range, the same models also predict 
a silicon mass increase of a factor of 2 while the calcium abundance 
decreases slightly as the progenitor mass increases up to 15~M$_{\odot}$ 
and then increases slowly thereafter.  In contrast, due to the rising mass cut, 
the mass of ejected iron {\it falls} by a factor of about 2 over the same
progenitor mass range. Clearly the overall abundance trends in
SN~1987A and SN~2002hh are not consistent with the explosion model
predictions. It appears that during the burning to intermediate-mass
elements, the nucleosynthesis did not progress as far as might have
been expected given the mass of iron ejected.\\

Elmhamdi et al. 2003 (and references therein) made use of the
[O~I]~6300,6364~\AA\ doublet luminosity to place constraints on the
progenitor mass of SN~1999em, by scaling from the [O~I] luminosity and
$^{56}$Ni mass of SN~1987A. Following a similar procedure for
SN~2002hh, at +397~d the coeval [O~I] doublet luminosity is about
a factor of 2.2 lower than in SN~1987A, while the $^{56}$Ni masses are
similar.  We therefore deduce an oxygen mass of about one half of that
found in SN~1987A (i.e., $\sim$0.7--1.0~M$_{\odot}$ in SN~2002hh). This
suggests a massive progenitor with a main-sequence stellar mass of
16--18 M$_{\odot}$ (Woosley \& Weaver 1995), somewhat less than the 20
M$_{\odot}$ of SN~1987A. Within the context of the Thielemann et
al. models, this would be consistent with the observed lower silicon
mass and possibly also with the lower calcium abundance in SN~2002hh.
However, the apparent conflict with the strong Mg~I~1.503~$\mu$m
luminosity remains. \\
 
A striking difference between SN~1987A and SN~2002hh lies in the
velocity shifts of some of the more prominent, isolated lines. While
blueshifts of order --500~km s$^{-1}$ were seen in SN~2002hh, {\it redshifts}
of about the same size were seen in the same lines and phases in
SN~1987A.  These shifts were discussed by Spyromilio et al. (1990) and
Meikle et al. (1993). It was concluded that asymmetry in the ejecta
was the most likely explanation. As indicated above, we propose that
the blueshifts in SN~2002hh have a similar origin.\\

As already mentioned, during the first year of SN~2002hh we find no
line profile-based evidence of dust condensation in the ejecta.  This
is consistent with an SN~1987A-like behaviour, where the first sign of
dust condensation was at about 1~year (Meikle et al. 1993) and
substantial dust-induced line attenuation did not occur until about
500-530~days (Danziger et al. 1991).  The clear presence of CO
emission by 200~d indicates that conditions may eventually allow grain
condensation.  However, the CO emission was generally weaker than seen
in SN~1987A and so the mass of condensing dust may be
less. Nevertheless, Meikle et al. (2005a,b and in prep.) report mid-IR
evidence for the condensation of $\sim10^{-3}$~M$_{\odot}$ of dust by
day~590 --- a very similar mass and condensation epoch to those of
SN~1987A.  Modelling of the CO first-overtone emission using the
constraint of the $^{56}$Ni mass derived from the bolometric
light curve, might provide an additional constraint on the progenitor
core mass as was done for SN~1998S by Fassia et al. (2001). However,
such modelling is beyond the scope of this paper. \\

Based on the IR-echo interpretation of the late-time $KL$ excess we
deduce that the progenitor of SN~2002hh underwent a mass-loss of
$\sim$0.3~M$_{\odot}$.  This is similar to the CSM mass derived for
the Type~IIL SN~1980K by Dwek (1983), also based on IR-echo analysis.
While this may be less than that of SN~1987A (e.g. Sugerman et
al. 2005), the strong and relatively early IR excess points to a
substantial proportion lying within $\sim$100 light-days of the
supernova - much closer than for SN~1987A. Likewise, the strong radio
luminosity of SN~2002hh (Beswick et al. 2005) as early as 381~d points
to recent substantial mass-loss.  Indeed, the radio luminosities
reported by Beswick et al. are very similar to those of SN~1980K from
which Weiler et al. (1992) derived a CSM mass exceeding
0.34M$_{\odot}$.  Extrapolating our IR~echo model to the earliest
epoch of the {\it Spitzer Space Telescope} observations {\it viz.} 590~days,
it predicts a flux of $\sim$0.3~mJy at 5.8~$\mu$m and 8.0~$\mu$m.
This is less than 10\% of the total fluxes quoted by Barlow et al.,
but it is closer to the {\it decline} in flux found by Meikle et
al. (2005a,b and in prep) between days 590 and 994. Thus, if we
attribute only the declining component of the flux to the supernova, a
much smaller CSM mass is needed to account for it. This is consistent
with the finding of Meikle et al (2005a,b and in prep.) that the total
mid-IR emission could not have arisen from an IR-echo within a
dusty CSM ejected by the progenitor (see Introduction).  It is,
however, less consistent with the extremely massive
(10--15~M$_{\odot}$) CSM claimed by Barlow et al. (2005).
Furthermore, a mass loss of 10--15~M$_{\odot}$ is about 100 times
larger than the value recently estimated by Chevalier, Fransson \&
Nymark by modelling the radio data of Chandra et al. (2003) and
Stockdale et al. (2004). \\

Given that the progenitor was probably massive and that the light
curves exhibit a significant plateau phase, we conclude that the
progenitor of SN~2002hh was a red supergiant as expected for normal
type~IIP SNe (Popov 1993; Li et al. 2006, and references
therein). There is also evidence that a substantial, dusty wind was
emitted by the progenitor star and that the mass outflow was ongoing
almost up to the epoch of explosion. However, in spite of the
progenitor differences between SN~2002hh and SN~1987A, much of their
nebular spectroscopic behaviour was remarkably similar.\\

\section*{Acknowledgments}

We would like to thank R. Stathakis for having provided us with the
SN~1987A optical spectra in digitized format. We are also grateful
to M. Gustafsson, 
P. Jakobsson, J. Melinder and G. {\O}ye for observing and reducing the 
NOT spectrum of SN~2002hh; and to R. Chornock, S. Jha, M. Papenkova, B. Swift 
and D. Weisz for help with the Lick and Keck/ESI SN~2002hh spectroscopy. 
This work is based on observations collected at the William Herschel Telescope 
(WHT, La Palma), the Nordic Optical Telescope (NOT, La Palma), the Infrared
Telescope Facility (IRTF, Hawaii), the Keck Telescope (Hawaii), and the
Shane Telescope (Lick Observatory, CA). The WHT is operated on the island of 
La Palma by the Isaac Newton Group (ING) in the Spanish Observatorio del Roque 
de los Muchachos of the Instituto de Astrofisica de Canarias.  The NOT is 
operated on the island of La Palma jointly by Denmark, Finland, Iceland, 
Norway, and Sweden, in the Spanish Observatorio del Roque de los Muchachos 
of the Instituto de Astrofisica de Canarias.  IRTF is operated on behalf of 
NASA by the University of Hawaii. The W. M. Keck Observatory is operated
as a scientific
partnership among the California Institute of Technology, the University of
California, and NASA; the Observatory was made possible by the generous
financial support of the W. M. Keck Foundation. 
Lick Observatory is
operated by the University of California.
KAIT was made possible by generous donations from Sun Microsystems, Inc.,
the Hewlett-Packard Company, AutoScope Corporation, Lick Observatory,
the National Science Foundation (NSF), the University of California, and the
Sylvia \& Jim Katzman Foundation.
MP is supported through
PPARC grant PPA/G/S/2001/00512.
AVF's group at UC Berkeley is supported by NSF grant AST-0307894.

%\appendix
%\section[]{}

\bsp

\label{lastpage}

\end{document}